\documentclass[12pt]{article}
\usepackage{amssymb,bm,float,booktabs, diagbox, authblk}
\usepackage[colorlinks,citecolor=blue]{hyperref}
\usepackage{comment}

\usepackage{natbib,setspace,lscape,longtable}
\usepackage{epsfig,graphicx,abstract}
\usepackage{amsmath,amsthm,color}
\usepackage{extarrows,multirow,booktabs}
\usepackage{threeparttable}
\RequirePackage[mathlines, displaymath]{lineno}
\usepackage{fancyhdr}
\usepackage{lastpage}
\usepackage{dsfont}
\usepackage{fancyref}

\usepackage{caption}
\usepackage{subcaption}

\usepackage{enumitem}
\setenumerate[1]{itemsep=0pt,partopsep=0pt,parsep=\parskip,topsep=0pt}
\setitemize[1]{itemsep=0pt,partopsep=0pt,parsep=\parskip,topsep=0pt}
\setdescription{itemsep=0pt,partopsep=0pt,parsep=\parskip,topsep=0pt}
\def\beqr{\begin{eqnarray}}
\def\eeqr{\end{eqnarray}}
\def\beqrs{\begin{eqnarray*}}
\def\eeqrs{\end{eqnarray*}}
\def\bep{\begin{prop}}
\def\eep{\end{prop}}

\numberwithin{equation}{section}

\newcommand{\trans}{^{\mbox{\tiny{T}}}}
\def\mR{\mathbb{R}}
\def\mN{\mathbb{N}}
\newcommand{\vech}{\mathbf{vech}}
\newcommand{\Y}{\mathbf{Y}}
\newcommand{\X}{\mathbf{X}}
\newcommand{\x}{\mathbf{x}}
\newcommand{\Op}{\mathbf{O}}
\newcommand{\op}{\mathbf{o}}
\newcommand{\M}{\mathbf{M}}
\newcommand{\T}{\mathbf{T}}
\newcommand{\V}{\mathbf{V}}
\newcommand{\B}{\mathbf{B}}
\newcommand{\C}{\mathbf{C}}
\newcommand{\Q}{\mathbf{Q}}
\newcommand{\G}{\mathbf{G}}
\newcommand{\R}{\mathbf{R}}
\newcommand{\W}{\mathbf{W}}
\newcommand{\Z}{\mathbf{Z}}
\newcommand{\E}{\mathbf{E}}
\newcommand{\N}{\mathbf{N}}
\newcommand{\U}{\mathbf{U}}
\newcommand{\I}{\mathbf{I}}

\newcommand{\bfH}{\mathbf{H}}
\newcommand{\proba}{\mathbf{P}}
\newcommand{\J}{\mathbf{J}}
\newcommand{\A}{\mathbf{A}}
\newcommand{\Span}{\text{Span}}
\newcommand{\w}{\mathbf{w}}
\newcommand{\m}{\mathbf{m}}
\newcommand{\be}{\bm{\beta}}

\newcommand{\1}{\mathds{1}}
\newcommand{\di}{\rm{dim}}
\def\eop{\hfill $\Box$ \\}
\newcommand{\Cov}{\rm{Cov}}
\newcommand{\et}{\bm{\eta}}
\newcommand{\la}{\bm{\lambda}}

\newcommand{\pr}{{\rm{pr}}}
\def \calS{\mbox{$\mathcal{S}$}}
\def \calD{\mbox{$\mathcal{D}$}}
\def \calF{\mbox{$\mathcal{F}$}}
\newcommand{\SIR}{\rm{SIR}}
\newcommand{\SAVE}{\rm{SAVE}}
\newcommand{\DR}{\rm{DR}}
\newcommand{\EE}{\mathrm {E}}

\newcommand{\indep}{\;\, \rule[0em]{.03em}{.67em} \hspace{-.25em}
\rule[0em]{.65em}{.03em} \hspace{-.25em}
\rule[0em]{.03em}{.67em}\;\,}
\def\calS{\mathcal{S}}
\newcommand{\bfSigw}{\mathbf{\Sigma_{w}}}
\newcommand{\bfSig}{\mathbf{\Sigma}}

\newtheorem{theo}{\bf Theorem}
\newtheorem{lemm}{\bf Lemma}
\newtheorem{prop}{\bf Proposition}
\newtheorem{remark}{\bf Remark}

\newcommand{\syx}{\mathcal{S}_{\Y|\X}}
\newcommand{\syxW}{\mathcal{S}_{\Y|\X}^{(\W)}}

\DeclareMathOperator{\rank}{rank}

\allowdisplaybreaks

\title{Dynamic Partial Sufficient Dimension Reduction}

\author[1]{Lu Li}
\author[2]{Kai Tan}
\author[3]{Xuerong Meggie Wen}
\author[4]{Zhou Yu}
\affil[1,2,4]{School of Statistics, East China Normal University}
\affil[3]{Department of Mathematics and Statistics, Missouri University of Science and Technology}

\date{}
\begin{document}
	\maketitle
	\begin{abstract}
Sufficient dimension reduction aims for reduction of dimensionality of a regression without loss of information by replacing the original predictor with its lower-dimensional subspace. Partial (sufficient) dimension reduction arises when the predictors naturally fall into two sets, $\X$ and $\W$, and we seek dimension reduction on $\X$ alone while considering all predictors in the regression analysis. Though partial dimension reduction is a very general problem, only very few research results are available when $\W$ is continuous. To the best of our knowledge, these methods generally perform poorly when $\X$ and $\W$ are related, furthermore, none can deal with the situation where the reduced lower-dimensional subspace of $\X$ varies dynamically with  $\W$. In this paper, We develop a novel {\bf dynamic} partial dimension reduction method, which could handle the dynamic dimension reduction issue and also allows the dependency of $\X$ on $\W$. The asymptotic consistency of our method is investigated. Extensive numerical studies and real data analysis show that our {\it Dynamic Partial Dimension Reduction} method has superior performance comparing to the existing methods.
\end{abstract}


\section{Introduction}
The rapid developments of brain imaging, microarray data analysis, computer vision, network analysis, econometrics, and many other applications call for the analysis of high-dimensional data. Sufficient dimension reduction (SDR)  \citep{Li1991, Cook1998}  is arguably one of the most important tools in analyzing high-dimensional data. Let $\Y$ be a univariate response, $\X=(X_{1},\ldots,X_{p})\trans\in\mR^{p}$ be a $p$-dimensional predictors, sufficient dimension reduction methods, aim to find a lower-dimensional subspace of $\X$ without loss of information on the conditional distribution of $\Y|\X$, and without pre-specifying a model for the regression. 
This subspace is then called a dimension reduction subspace for the regression. The goal of SDR is to search for the smallest dimension reduction subspace, the central subspace (CS, $\syx$) and its dimension $d$, which is called the structural dimension of the regression.
We refer readers to \cite{Cook1998} for more details. Many methods have been developed 
in the past two decades due to the ubiquity of large high-dimension data sets which are now more readily available than in the past.
To name a few: 
sliced inverse regression (SIR; \cite{Li1991}),  sliced average variance estimation (SAVE; \cite{Cook1991}), minimum average variance estimation (MAVE; \cite{Xia:2002}),  the kth moment estimation  \citep{Yin2002,Yin2003}, inverse regression \citep{Cook2005}, directional regression (DR;  \cite{LiW2007}),  sliced regression (SR; \citet{Wang2008}), likelihood acquired directions (LAD;  \citet{Cook2009}), and semiparametric approaches of \cite{Ma2012,Ma2013a,Ma2013b,Ma2014}.
More detailed discussion can be found in \cite{Xue2018}.

Partial dimension reduction (PDR) \citep{Chiaromonte2002, Wen2007, Feng2013} arises when the predictors naturally fall into two groups, 
$\X=(X_1, \ldots, X_p)$ and $\W=(W_1, \ldots, W_q)$, and we seek dimension reduction on $\X$ alone while
considering all predictors in the regression analysis. This might happen when $\W$ plays a particular role in the regression and
must, therefore, be shielded from the reduction process.  Considering
the Boston Housing dataset (Feng et al., 2013), which was collected by the U.S. Census Service concerning housing in the
18 area of Boston, where the goal was to study how the house prices are affected by certain given attributes regarding those houses.
Among all those features, it is well known that {\it Crime rate} ($\W$), plays an important role in the housing price, hence it should be treated discriminately and the dimension reduction should focus on the remaining features ($\X$).

To be specific, PDR performs regression of $\Y$ on $(\X, \W)$ by seeking a projection $\mathbf{P}_{\mathcal{S}}\X$ of $\X$ that preserves information on $\Y \mid (\X,\W)$, where $\mathbf{P}_{\mathcal{S}}$ indicates the projection onto the subspace $\mathcal S$ in the usual inner product. If  the intersection of all subspaces $\mathcal S\subseteq \mR^p$ such that
\begin{equation}\label{PDR}
\Y\indep \X|(\mathbf{P}_{\mathcal S}\X,\W),  
\end{equation}
also satisfies condition \eqref{PDR}, we call it the partial central subspace, and denote it by $\calS_{\Y|\X}^{(\W)}$.  And ${\di}\{\calS_{\Y|\X}^{(\W)}\}=d$ is called the structural dimension of the partial central subspace.
The concept of partial central subspace was first proposed by \cite{Chiaromonte2002} to deal with regressions with a mixture of continuous ($\X$) and categorical predictors ($\W$). \cite{Feng2013} developed a method called PDEE to incorporate the continuous $\W$ scenario via a dichotomization transformation. Though PDEE widens the application of PDR, it could not deal with the situation where $\syxW$ varies dynamically with continuous $\W$, which is often the case in real-world applications.

In this article, we propose the concept of {\it Dynamic Partial Dimension Reduction}, where the partial CS, $\syxW$ is allowed to vary with  $\W$.
Hence the aim of dynamic partial dimension reduction  is to find a matrix  of smooth functions of $\W$ with minimum rank,  $\B(\W) \in\mR^{p\times d(\W)}$, such that
\begin{equation}\label{dps}
\begin{aligned}
\Y \indep \X |(\B \trans(\W) \X, \W).  
 \end{aligned}
\end{equation}
 Then $d(\W)\triangleq \rank \{B(\W)\}$ is the structural dimension function  of the dynamic partial CS. It is worth noting that the covariance matrix $\Cov(\X|\W)$, the column space of $\B(\W)$ and the structural  dimension $d(\W)$ may all vary as $\w$ changes, which poses a great challenge for the  estimation procedure.

The contribution of this paper is threefold. First, to the best of our knowledge, the proposed dynamic partial dimension reduction is the first attempt to perform partial dimension reduction for the dynamic case. Second, we adapt three classical SDR methods into the new framework to develop dynamic partial dimension reduction methods and establish the corresponding asymptotic normality and consistency properties rigorously for our methods. Last but not the least, we propose to determine the structural dimension by a nonparametric version of the ladle estimator\citep{Luo2016}, and also derive the consistency property for our nonparametric ladle estimator as well.

The rest of this paper is organized as follows. In Section 2, we first introduce the principles of dynamic partial
dimension reduction, then develop dynamic partial SIR, dynamic partial SAVE and DR,  and also propose the estimation schemes along with
the large sample theories for each method.  In Section 3,  we develop the nonparametric ladle estimator to determine the structural
dimension of dynamic partial CS. Section 4 focuses on how to conduct the bandwidth selection involved in the kernel estimation. Section 5 presents the finite sample performance of our proposed methods via extensive simulation studies. To illustrate the efficiency of our proposed methods, four real data analysis are conducted in section 6. For the ease of exposition, we defer all proofs to the Appendix.

\section{The Principle of Dynamic Partial Dimension Reduction}  \label{pdr}
Let $(\X_{\w},\Y_{\w})$ denote a generic pair distributed like $(\X,\Y)|(\W=\w)$, and
$\calS_{\Y|(\X,\W=\w)}=\calS_{\Y_{\w}|\X_{\w}}$. As we discussed in Section~1, the aim of  dynamic PDR is to find a subspace spanned
by the columns of matrix $\B(\w)\in\mR^{p\times d(\w)}$ such that
\begin{equation}\label{partial dynamic subspace}
\begin{aligned}
\Y_{\w}\indep \X_{\w}|\B\trans(\w)\X_{\w},
\end{aligned}
\end{equation}where $\B(\w)$ is a matrix of smooth functions of $\w$. 

We employ nonparametric covariance models to analyze this dynamic scenario. Let $\m(\w)=\left( \m_{1}(\w),\ldots,\m_{p}(\w) \right)\trans$ and $\bfSigw=\{\sigma_{ij}(\w)\}_{p\times p}$ denote the mean and covariance of $\X_{\w}$, where both $\m(\w)$ and $\bfSigw$ are smooth functions of $\w$. Equation \eqref{partial dynamic subspace} implies the reduction of the predictor from $\X_{\w}$
to $\B\trans(\w)\X_{\w}$. 
If $\bfSigw$ is invertible, one can also work with the standardized data $\Z_{\w}=\bfSigw^{-1/2}\{\X-\m(\w)\}$
to obtain $\calS_{Y_{\w}|\Z_{\w}}$ and then recover the dynamic partial  CS, $\calS_{Y_{\w}|\X_{\w}}$ by
the well-known invariance property $\calS_{Y_{\w}|\X_{\w}}=\bfSigw^{-1/2}\calS_{Y_{\w}|\Z_{\w}}$ (see \cite{Cook1998} Proposition 6.1).
Note that the estimation of $\calS_{Y_{\w}|\X_{\w}}$ consists of two parts, the order determination for $d(\w)$ and the basis estimation for $\calS_{Y_{\w}|\X_{\w}}$. We first consider the basis estimation assuming
$d(\w)$ is known, then 
propose an order determination method for $d(\w)$.
 
\subsection{Implementation of Dynamic Partial  Dimension Reduction via SIR}
We adopt three popular sufficient dimension reduction approaches,
SIR \citep{Li1991}, SAVE \citep{Cook1991}, DR \citep{LiW2007}, to perform dynamic PDR.  Let $\B(\w)\in\mR^{p\times d(\w)}$ be a matrix such that $\Span(\B(\w)) = \mathcal{S}_{\X_{\w}|\Y_{\w}}$.
Prior to the main development of our dynamic partial dimension reduction methods, we first present the following two assumptions,
\begin{description}
	\item  (A1)  (Linear Conditional Mean) $\EE\{\X_{\w}|\B\trans(\w)\X_{\w}\}$ is a linear function of  $\B\trans(\w)\X_{\w}$.
	\item  (A2)  (Constant Conditional Variance) $\Cov\{\X_{\w}|\B\trans(\w)\X_{\w}\}$ is a nonrandom matrix.
\end{description}
 Condition (A1) has been commonly assumed in sufficient dimension reduction literature and it is indispensable for almost all inverse-regression based methods. Condition (A2) is similar to (A1) in nature, and is important for all the second-order sufficient dimension reduction methods. Both conditions (A1) and (A2) are guaranteed when $\X_{\w}$ is normally distributed. More discussions about conditions (A1) and (A2) can be found in \cite{Chiaromonte2002}, \cite{Li2003}, \cite{Li2009} and \cite{Dong2010}, \cite{Li2018}, etc.
 
Firstly, we briefly review the development of SIR \citep{Li1991}. The main idea of SIR is to work with the
inverse regression, 
the conditional distribution of $\X|\Y$, and in particular by examining the kernel matrix $\Cov \{E(\X|\Y)\}$.
SIR procedure starts with the partition of the response $\Y$. Let $\{J_1,J_2, \ldots, J_H\}$ be a measurable
partition of the sample space of $\Y$, consider the discretized version
$\widetilde{\Y} = \sum_{l=1}^{H}l\cdot \1(\Y\in J_l)$. If $\Y$ is categorical or $H$
is sufficiently large ($H\geq d+1$), \cite{Bura2001} and \cite{Cook2009}
verified that there is no loss of information for identifying $\mathcal{S}_{\Y|\X}$
when $\Y$ is replaced with $\widetilde{\Y}$.

Let $\mathbf P_{\B(\w)}$ be the projection on to $\calS_{\Y_{\w}|\X_{\w}}$ with respect to the inner product
$\langle a,b \rangle := a\trans\bfSigw b$, assuming (A1), the following proposition states that the random vector $\bfSig_{\w}^{-1}\{\mathrm{E}(\X_{\w}|Y_{\w})-\m(\w)\}$ belongs to $\calS_{\Y_{\w}|\X_{\w}}$ almost surely.

\begin{prop}\label{partial least squares}
   Given  $\W=\w$, suppose that the linear conditional mean condition (A1) holds, then
	\begin{align*}
	\bfSig_{\w}^{-1}(\EE\{\X|(\Y,\W=\w)\} - \m(\w)) =\mathbf P_{\B(\w)}\bfSig_{\w}^{-1}(\EE\{\X|(\Y,\W=\w)\}-\m(\w)).
	\end{align*}
 \end{prop}
Proposition \ref{partial least squares} indicates that the random vector $\bfSig_{\w}^{-1}(\EE(\X_{\w}|\Y_{\w})- \m(\w))$ belongs to the range of the projection operator $\mathbf P_{\B(\w)}$, which is actually $\calS_{\Y_{\w}|\X_{\w}}$. Consequently,
the column space of the matrix $\bfSig_{\w}^{-1}{\Cov}\{\EE(\X_{\w}|\Y_{\w})\}$ is a subspace of $\calS_{Y_{\w}|\X_{\w}}$.
The proof of Proposition \ref{partial least squares} relies on \cite{Li2009}, and is provided in the Appendix.

Motivated by this finding, we now construct the following kernel matrix for dynamic  partial SIR,
\begin{equation}\label{Dynamic SIR}
	\begin{aligned}
	\M_{\SIR}(\w) &\triangleq \Cov \{ E(\X|\widetilde{\Y},\w)\}\\
	& = \sum_{l=1}^{H} \proba_{l,\w}(\V_{l,\w} - \m(\w))(\V_{l,\w} - \m(\w))\trans\\
	&=\sum_{l=1}^{H}\proba_{l,\w}\V_{l,\w}\V_{l,\w}\trans-\m(\w)\m(\w)\trans,
	\end{aligned}
\end{equation}
where $\proba_{l,\w}=\pr(\widetilde{\Y}=l|\w)$, $\V_{l,\w}=\mathrm{E}\{\X|\widetilde{\Y}=l,\w\}$ and $\m(\w)=\mathrm{E}(\X_{\w})$. Note that the term $\V_{l,\w}=\mathrm{E}\{\X|\widetilde{\Y}=l,\w\}$ contains two conditional variables, and thus makes it hard to deal with. However, this difficulty can be overcome by using the following proposition, whose proof is provided in the Appendix.
\begin{prop}\label{equal inquality}
	Given  $\W=\w$, for each $l = 1, 2,\ldots, H$, we have
	\begin{equation}\label{identity}
		\begin{aligned}
		\EE(\X|\widetilde{\Y}=l,\w)=\frac{\EE\{\X\1(\widetilde{\Y}=l)|\w\}}{\EE\{\1(\widetilde{\Y}=l)|\w\}}, 
		\end{aligned}
	\end{equation}
\end{prop}
where $\1(.)$ is the indicator function. Proposition \ref{equal inquality} shows that the term  $\V_{l,\w}$ in \eqref{Dynamic SIR} can be written as a fraction of  simple conditional expectations. 
We can rewrite the kernel matrix $\M_{\SIR}(\w)$ in the following form:
\begin{align}\label{eq:PDSIR}
  \M_{\SIR}(\w)=\sum_{l=1}^{H}\frac{\U_{l,\w}\U_{l,\w}\trans}{\proba_{l,\w}}-\m(\w)\m(\w)\trans,
\end{align}
where $\U_{l,\w}=\EE\{\X\1(\widetilde{\Y}=l)|\w\}$.

Let $\{( \Y_i, \X_i, \W_i), i=1, \ldots, n\}$ be random samples from $(\Y,\X, \W)$.  Since the structure of $\bfSig_{\w}^{-1}\M_{\SIR}(\w)$ is dynamic,
we employ the nonparametric covariance
model \cite{Yin2010} for estimation. Specifically, we adopt the following Nadaraya-Watson estimator \cite{Nadaraya1964}, \cite{Watson1964} of $\m(\w)$
\begin{align*}
\widehat{\m}(\w)=\frac{\sum\nolimits_{i=1}^{n}\X_{i}K_{h}(\W_{i}-\w)}{\sum\nolimits_{i=1}^{n}K_{h}(\W_{i}-\w)}.
\end{align*}
Similarly, the Nadaraya-Watson estimators of $\U_{l,\w}$ and $\proba_{l,\w}$ are given by
\begin{align*}
  \widehat{\U}_{l,\w}=\frac{\sum\nolimits_{i=1}^{n}\X_{i}\1(\widetilde{\Y}_{i}=l)K_{h}(\W_{i}-\w)}{\sum\nolimits_{i=1}^{n}
    K_{h}(\W_{i}-\w)},
\quad\widehat{\proba}_{l,\w}=\frac{\sum\nolimits_{i=1}^{n}\1(\widetilde{\Y}_{i}=l)K_{h}(\W_{i}-\w)}{\sum\nolimits_{i=1}^{n}K_{h}(\w_{i}-\w)}.
\end{align*}
Then it's straightforward to obtain the sample estimator of  $\M_{\SIR}(\w)$ by substituting $\widehat{\m}(\w), \widehat{\U}_{l,\w}$, and $\widehat{\proba}_{l,\w}$ into equation \eqref{eq:PDSIR}, that is,
\begin{equation}\label{M estimate}
\begin{aligned}
  \widehat{\M}_{\SIR}(\w)=\sum\nolimits_{l=1}^{H}\frac{\widehat{\U}_{l,\w}\widehat{\U}\trans_{l,\w}}{\widehat{\proba}_{l,\w}}-
  \widehat{\m}(\w)\widehat{\m}(\w)\trans.
\end{aligned}
\end{equation}

\begin{remark}
	As for the estimation of the conditional covariance matrix, one may use different bandwidths for different elements of $\bfSigw$. However, the resulting estimate with different bandwidth is not guaranteed to be positive definite \citep{Li2007}, which is the desired property in practice. Thus, we suggest using the same bandwidth for all elements. And the selection of bandwidth will be discussed in Section \ref{bandwidth selection}.
\end{remark}

Recall that $\calS_{\Y_{\w}|\X_{\w}}$ is a $d(\w)$-dimensional subspace of $\mR^{p}$. Proposition \ref{partial least squares} leads us to consider the singular value decomposition of $\widehat{\bfSig}_{\w}^{-1}\widehat{\M}_{\SIR}(\w)$.
Let
\begin{align*}
	\bfSig_{\w}^{-1}\M_{\SIR}(\w) &= \sum_{k=1}^{p} \la_{k}^{\SIR}(\w) \be_{k}^{\SIR}(\w) \et_{k}^{\SIR}(\w), \quad \la^{\SIR}_1(\w)\ge\cdots\ge\la^{\SIR}_d(\w)=0=\cdots =\la^{\SIR}_p(\w),\\
	\widehat\bfSig_{\w}^{-1}\widehat\M_{\SIR}(\w) &= \sum_{k=1}^{p} \widehat\la_{k}^{\SIR}(\w) \widehat\be_{k}^{\SIR}(\w) \widehat\et_{k}^{\SIR}(\w),\quad  \widehat\la^{\SIR}_1(\w)\ge\cdots\ge\widehat\la^{\SIR}_d(\w)\ge\cdots \ge\widehat\la^{\SIR}_p(\w),
\end{align*}
be the singular value decomposition of $\bfSig_{\w}^{-1}\M_{\SIR}(\w)$ and
$\widehat{\bfSig}_{\w}^{-1}\widehat{\M}_{\SIR}(\w)$, respectively.

Then we can use $\Span\{\widehat{\be}_{1}^{\SIR}(\w),\ldots,\widehat{\be}_{d(\w)}^{\SIR}(\w)\}$ to estimate $\calS_{\Y_{\w}|\X_{\w}}$. By large sample theory and singular value decomposition, the asymptotic normality of $\widehat{\M}_{\SIR}(\w)$ and the asymptotic expansion of $\widehat{\be}_{k}^{\SIR}(\w)$ are presented in Theorem \ref{generalized eigenvalue decomposition}.
\begin{theo}\label{generalized eigenvalue decomposition}
  Let $G\subset\{\w:f(\w)>0\}$ be a compact subset on the support of $\W$, where $f(\w)$ is the density of
  $\W$. Under the condition (A1) and assumptions (C1)-(C8) listed in the Appendix, we have  
	\begin{align*}
	\sqrt{nh}\left(\vech\{\widehat{\M}_{\SIR}(\w)\}-\vech\{\M_{\SIR}(\w)\}-\vech\{\B_{\SIR}(\w)\}\right) \xrightarrow{d}  {\bf{N}} \left(\bm{0},f^{-1}(\w)\omega_{0}\C^{\SIR}(\w)\right).
	\end{align*}
	Assume that $\X_{\w}$ has finite fourth moment and all the nonzero eigenvalues of  $\M_{\SIR}(\w)$ are distinct, then for $k=1,\ldots,d(\w)$, we have
	\begin{align*}
	\sqrt{nh}\left(\widehat{\be}_{k}^{\SIR}(\w)-\be_{k}^{\SIR}(\w)-\B_{k}^{\SIR}(\w)\right)\xrightarrow{d}{\bf{N}}(\bm{0},\bfSig_{k}^{\SIR}(\w)),
	\end{align*}
	where $\vech(\cdot)$ is the vectorization of the upper triangular part of a matrix, and the closed forms of $\B_{\SIR}(\w)$, $\omega_{0}$, $\C^{\SIR}(\w)$, $\B_{k}^{\SIR}(\w)$ and $\bfSig_{k}^{\SIR}(\w)$ are provided by \eqref{BSIR}, \eqref{w}, \eqref{CSIR}, \eqref{be} and \eqref{bfsig} in the Appendix, respectively.
\end{theo}

\subsection{ Dynamic Partial SAVE}
Though SIR has received much attention, it cannot recover any vector in the central subspace $\calS_{\Y|\X}$ if the regression function is symmetric about the origin because SIR is based on the estimation of the conditional mean. To address this, SAVE \citep{Cook1991} was proposed to estimate the central space by
utilizing the conditional variance function of the covariates when the response is given. For partial dimension reduction, \cite{Shao2009} also developed partial SAVE and showed that partial SAVE is more comprehensive than partial SIR. As an extension of partial SAVE, we now develop the dynamic partial  SAVE, which is based on the following proposition \ref{nonrandom variance condition 1}.
\begin{prop}\label{nonrandom variance condition 1}
	Conditional on $\W=\w$, suppose that linear conditional mean condition (A1) and constant conditional variance  condition (A2) hold, then
	\begin{align*}
	\bfSig_{\w}^{-1}\{\bfSig_{\w}-\Cov(\X|\Y,\W=\w)\}=\mathbf P_{\B(\w)}\{\I_{p}-\bfSig_{\w}^{-1}\Cov(\X|\Y,\W=\w)\}\mathbf P_{\B(\w)}.
	\end{align*}
\end{prop}
The proof of Proposition \ref{nonrandom variance condition 1} is based on the Theorem 1 of \cite{Cook1999}, and the detailed proof is provided in the Appendix. Parallel to SAVE, we define the following kernel matrix for  dynamic partial SAVE.
\begin{align*}
\M_{\SAVE}(\w)&\triangleq\EE_{\widetilde{\Y}}\{{\Cov}(\X_{\w})-{\Cov}(\X|\widetilde{\Y}=l,\w)\}^{2}\\
&=\sum\nolimits_{l=1}^{H}\proba_{l,\w}\left(\bfSigw-\R_{l,\w}+\V_{l,\w}\V_{l,\w}\trans\right)^{2},
\end{align*}
where $\R_{l,\w}=\EE(\X\X\trans|\widetilde{\Y}=l,\w)$ and $\V_{l,\w}$, $\proba_{l,\w}$ are defined as previously. $\EE_{\widetilde{\Y}}$ represents expectation with respect to $\widetilde{\Y}$.

Proposition \ref{nonrandom variance condition 1} implies that $\bfSig_{\w}^{-1}\M_{\SAVE}(\w)\subseteq \calS_{\Y_{\w}|\X_{\w}}$ almost surely. Similar to Proposition \ref{equal inquality}, the following proposition \ref{equal inquality 1} shows that the term $\R_{l,\w}=\EE(\X\X\trans|\widetilde{\Y}=l,\w)$ can be expressed as a fraction, whose numerator and denominator are easy to be estimated.
\begin{prop}\label{equal inquality 1}
	Conditional on $\W = \w$, for each $l=1,2,\ldots,H$, we have
	\begin{equation}\label{identity1}
	\begin{aligned}
	\EE(\X\X\trans|\widetilde{\Y}=l,\w)=\frac{\EE\{\X\X\trans\1(\widetilde{\Y}=l)|\w\}}{\EE\{\1(\widetilde{\Y}=l)|\w\}}.
	\end{aligned}
	\end{equation}
\end{prop}
Thus, the kernel matrix of partial dynamic SAVE can be rewritten as
\begin{equation}\label{SAVE}
\begin{aligned}
\M_{\SAVE}(\w) =\sum\nolimits_{l=1}^{H}\proba_{l,\w}\left\{\bfSigw-\frac{\N_{l,\w}}{\proba_{l,\w}}+\frac{\U_{l,\w}\U_{l,\w}\trans}{\proba_{l,\w}\proba_{l,\w}}\right\}^{2},
\end{aligned}
\end{equation}
where $\N_{l,\w}=\EE\{\X\X\trans\1(\widetilde{\Y}=l)|\w\}$. The proof of Proposition \ref{equal inquality 1} is
similar to that  of Proposition \ref{equal inquality}, and is omitted.

Recall that the goal of  dynamic partial SAVE is to estimate the $\calS_{\Y_{\w}|\X_{\w}}$ by the estimates of $\bfSig_{\w}^{-1}\M_{\SAVE}(\w)$. Similar to dynamic partial SIR, we have the following NW kernel estimator of $\N_{l,\w}$:
\begin{align*}
\widehat{\N}_{l,\w}=\frac{\sum\nolimits_{i=1}^{n}\x_{i}\x_{i}\trans\1(\widetilde{\Y}_{i}=l)K_{h}(\w_{i}-\w)}{\sum\nolimits_{i=1}^{n}K_{h}(\w_{i}-\w)}.
\end{align*}
Then it is easy for us to get the sample estimator of $\M_{\SAVE}(\w)$: 
\begin{equation}\label{M-SAVE}
\begin{aligned}
\widehat{\M}_{\SAVE}(\w)=\sum\nolimits_{l=1}^{H}\widehat{\proba}_{l,\w}\left\{\widehat{\bfSig}_{\w}-\frac{\widehat{\N}_{l,\w}}{\widehat{\proba}_{l,\w}}+\frac{\widehat{\U}_{l,\w}\widehat{\U}_{l,\w}\trans}{\widehat{\proba}_{l,\w}\widehat{\proba}_{l,\w}}\right\}^{2}.
\end{aligned}
\end{equation}
Now we can use $\widehat{\bfSig}_{\w}^{-1}\widehat{\M}_{\SAVE}(\w)$  to estimate $\calS_{\Y_{\w}|\X_{\w}}$. Proposition \ref{nonrandom variance condition 1} also leads us to consider the singular value decomposition. Let
\begin{align*}
\bfSig_{\w}^{-1}\M_{\SAVE}(\w) &= \sum_{k=1}^{p} \la_{k}^{\SAVE}(\w) \be_{k}^{\SAVE}(\w) \et_{k}^{\SAVE}(\w), \\ \la^{\SAVE}_1(\w)\ge &\cdots\ge\la^{\SAVE}_d(\w)=0=\cdots =\la^{\SAVE}_p(\w),\\
\widehat\bfSig_{\w}^{-1}\widehat\M_{\SAVE}(\w) &= \sum_{k=1}^{p} \widehat\la_{k}^{\SAVE}(\w) \widehat\be_{k}^{\SAVE}(\w) \widehat\et_{k}^{\SAVE}(\w),\\  \widehat\la^{\SAVE}_1(\w)\ge&\cdots\ge\widehat\la^{\SAVE}_d(\w)\ge\cdots \ge\widehat\la^{\SAVE}_p(\w),
\end{align*}
be the singular value decomposition of $\bfSig_{\w}^{-1}\M_{\SAVE}(\w)$ and  $\widehat{\bfSig}_{\w}^{-1}\widehat{\M}_{\SAVE}(\w)$, respectively.

Note that $\Span\{\be_{1}^{\SAVE}(\w),\ldots,\be_{d(\w)}^{\SAVE}(\w)\} = \calS_{\Y_{\w}|\X_{\w}}$, naturally, we propose to use  $\Span\{\widehat{\be}_{1}^{\SAVE}(\w),\ldots,\widehat{\be}_{d(\w)}^{\SAVE}(\w)\}$ to estimate $\calS_{\Y_{\w}|\X_{\w}}$. Using large sample theory and singular value decomposition, we get the asymptotic normality of $\widehat{\M}_{\SAVE}(\w)$ and the asymptotic expansion of $\widehat{\be}_{k}^{\SAVE}(\w)$ as follows.
{\theo\label{generalized eigenvalue decomposition 1}
	Let $G\subset\{\w:f(\w)>0\}$ be a compact subset on the support of $\W$, where $f(\w)$ is the density
	of  $\W$. Under the condition (A1), (A2) and assumptions (C1)-(C8) listed in Appendix, we have
	\begin{align*}
	\sqrt{nh}\left(\vech\{\widehat{\M}_{\SAVE}(\w)\}-\vech\{\M_{\SAVE}(\w)\}-\vech\{\B_{\SAVE}(\w)\}\right)\\
	\xrightarrow{d}  {\bf{N}} (\bm{0},f^{-1}(\w)\omega_{0}\C^{\SAVE}(\w)).
	\end{align*}
	Assume that $\X_{\w}$ has finite fourth moment and all the nonzero eigenvalues of $\M_{\SAVE}(\w)$ are distinct, then for $k=1,\ldots,d(\w)$, we have
	\begin{align*}
	\sqrt{nh}\left(\widehat{\be}_{k}^{\SAVE}(\w)-\be_{k}^{\SAVE}(\w)-\B_{k}^{\SAVE}(\w)\right)\xrightarrow{d}{\bf{N}}(\bm{0},\bfSig_{k}^{\SAVE}(\w)),
	\end{align*}
	where the closed form of $\omega_{0}$, $\B_{\SAVE}(\w)$, $\C^{\SAVE}(\w)$, $\B_{k}^{\SAVE}(\w)$ and $\bfSig_{k}^{\SAVE}(\w)$ are provided by \eqref{w}, \eqref{B SAVE}, \eqref{CSAVE}, \eqref{be1} and \eqref{bfsig1} in the Appendix, respectively.
}
\subsection{ Dynamic Partial DR}
Another popular method of sufficient dimension reduction is Directional Regression (DR) \citep{LiW2007}, which implicitly synthesizes sliced inverse regression and sliced average variance estimation. DR enjoys the advantage of high accuracy and convenient computation, and has received substantial attention in the literature of sufficient dimension reduction \citep{Yu2014, Yu2016}. Parallel to  dynamic partial SIR and  dynamic partial SAVE, we now propose dynamic  partial DR approach to
perform dynamic  partial dimension reduction.
\begin{prop}\label{nonrandom variance condition 2}
  Conditional on $\W=\w$, assume that linear conditional mean condition (A1) and
  constant conditional variance  condition (A2) hold, then 
	\begin{align*}
	&\bfSig_{\w}^{-1}[2\bfSig_{\w}-\EE\{(\X-\breve{\X})(\X-\breve{\X})\trans|\Y,\breve{\Y},\W=\w\}]\\
	=&\proba_{\B(\w)}[2\I_{p}-\bfSig_{\w}^{-1}\EE\{(\X-\breve{\X})(\X-\breve{\X})\trans|\Y,\breve{\Y},\W=\w\}]\proba_{\B(\w)},
	\end{align*}
	where $(\breve{\X},\breve{\Y})$ is an independent copy of $(\X,\Y)$.
\end{prop}
The proof of Proposition \ref{nonrandom variance condition 2} is provided in the Appendix. Now we can define the  kernel matrix
for dynamic partial DR,
\begin{align*}
\M_{\DR}(\w)\triangleq\EE_{\Y,\breve{\Y}}\{2\bfSig_{\w}-\EE\{(\X-\breve{\X})(\X-\breve{\X})\trans|\Y,\breve{\Y},\W=\w\}\}^{2},
\end{align*}
Where $\EE_{\Y,\breve{\Y}}$ represents expectation with respect to $\Y$ and $\breve{\Y}$. Proposition \ref{nonrandom variance condition 2} implies that $\bfSig_{\w}^{-1}\M_{\DR}(\w)\subseteq \calS_{Y_{\w}|\X_{\w}}$ almost surely.

According to Proposition 1 of \cite{LiW2007}, the kernel matrix $\M_{\DR}(\w)$ can be rewritten  as
\begin{align*}
\M_{\DR}(\w)=&2\sum\limits_{l=1}^{H}\proba_{l,\w}\left(\R_{l,\w}-\bfSig_{\w}\right)^{2}+2\left(\sum\limits_{l=1}^{H}\proba_{l,\w}\V_{l,\w}\V_{l,\w}\trans\right)^{2}\\
&+2\left(\sum\limits_{l=1}^{H}\proba_{l,\w}\V_{l,\w}\trans\V_{l,\w}\right)\left(\sum\limits_{l=1}^{H}\proba_{l,\w}\V_{l,\w}\V_{l,\w}\trans\right),
\end{align*}
where $\proba_{l,\w}$, $\R_{l,\w}$ and  $\V_{l,\w}$ are defined as previously. And the sample estimator of $\M_{\DR}(\w)$ can be easily found by
\begin{equation}\label{M-SAVE}
\begin{aligned}
\widehat{\M}_{\DR}(\w)=&2\sum\limits_{l=1}^{H}\widehat{\proba}_{l,\w}\left(\widehat{\R}_{l,\w}-\widehat{\bfSig}_{\w}\right)^{2}+2\left(\sum\limits_{l=1}^{H}\widehat{\proba}_{l,\w}\widehat{\V}_{l,\w}\widehat{\V}_{l,\w}\trans\right)^{2}\\
&+2\left(\sum\limits_{l=1}^{H}\widehat{\proba}_{l,\w}\widehat{\V}_{l,\w}\trans\widehat{\V}_{l,\w}\right)\left(\sum\limits_{l=1}^{H}\widehat{\proba}_{l,\w}\widehat{\V}_{l,\w}\widehat{\V}_{l,\w}\trans\right).
\end{aligned}
\end{equation}
Then we use $\widehat{\bfSig}_{\w}^{-1}\widehat{\M}_{\DR}(\w)$  to estimate $\calS_{\Y_{\w}|\X_{\w}}$. Proposition \ref{nonrandom variance condition 2} also leads us to consider the singular value decomposition. Let
\begin{align*}
\bfSig_{\w}^{-1}\M_{\DR}(\w) &= \sum_{k=1}^{p} \la_{k}^{\DR}(\w) \be_{k}^{\DR}(\w) \et_{k}^{\DR}(\w), \quad \la^{\DR}_1(\w)\ge\cdots\ge\la^{\DR}_d(\w)=0=\cdots =\la^{\DR}_p(\w),\\
\widehat\bfSig_{\w}^{-1}\widehat\M_{\DR}(\w) &= \sum_{k=1}^{p} \widehat\la_{k}^{\DR}(\w) \widehat\be_{k}^{\DR}(\w) \widehat\et_{k}^{\DR}(\w),\quad  \widehat\la^{\DR}_1(\w)\ge\cdots\ge\widehat\la^{\DR}_d(\w)\ge\cdots \ge\widehat\la^{\DR}_p(\w),
\end{align*}
be the singular value decomposition of $\bfSig_{\w}^{-1}\M_{\DR}(\w)$ and  $\widehat{\bfSig}_{\w}^{-1}\widehat{\M}_{\DR}(\w)$, respectively.

Then $\Span\{\be_{1}^{\DR}(\w),\ldots,\be_{d(\w)}^{\DR}(\w)\} = \calS_{\Y_{\w}|\X_{\w}}$, and the final sample estimator of $\calS_{\Y_{\w}|\X_{\w}}$ is $\Span\{\widehat{\be}_{1}^{\DR}(\w),\ldots,\widehat{\be}_{d(\w)}^{\DR}(\w)\}$. Using large sample theory and singular value decomposition, we get the asymptotic normality of $\widehat{\M}_{\DR}(\w)$ and the asymptotic expansion of $\widehat{\be}_{k}^{\DR}(\w)$
as follows.

{\theo\label{generalized eigenvalue decomposition 2}
	Let $G\subset\{\w:f(\w)>0\}$ be a compact subset on the support of $\W$, where $f(\w)$ is the density
	function  of $\W$. Under the condition (A1), (A2) and assumptions (C1)-(C8) listed in Appendix, we have
	\begin{align*}
	\sqrt{nh}\left(\vech\{\widehat{\M}_{\DR}(\w)\}-\vech\{\M_{\DR}(\w)\}-\vech\{\B_{\DR}(\w)\}\right)
	\xrightarrow{d}  {\bf{N}} (\bm{0},f^{-1}(\w)\omega_{0}\C^{\DR}(\w)).
	\end{align*}
	Assume that $\X_{\w}$ has finite fourth moment and all the nonzero eigenvalues of $\M_{\DR}(\w)$ are distinct, then for $k=1,\ldots,d(\w)$, we have
	\begin{align*}
	\sqrt{nh}\left(\widehat{\be}_{k}^{\DR}(\w)-\be_{k}^{\DR}(\w)-\B_{k}^{\DR}(\w)\right)\xrightarrow{d}{\bf{N}}(\bm{0},\bfSig_{k}^{\DR}(\w)),
	\end{align*}
	where the closed form of $\omega_{0}$, $\B_{\DR}(\w)$, $\C^{\DR}(\w)$, $\B_{k}^{\DR}(\w)$ and $\bfSig_{k}^{\DR}(\w)$
	are provided by \eqref{w}, \eqref{B DR} \eqref{CDR}, \eqref{be2} and \eqref{bfsig2} in the Appendix, respectively.}

\section{Determination of Dimensionality $d(\w)$}
Recall that when we estimate the $\calS_{\Y_{\w}|\X_{\w}}$ in Section \ref{pdr}, we assume that the dynamic structural dimension $d(\w)$ is known. However, $d(\w)$ is usually unknown in practice, and its estimation is of independent interest. In this section, we extend the state-of-the-art ladle estimator in \cite{Luo2016} into a nonparametric version and establish its consistency property as well.

Let $F$ be the distribution function of $(\X_\w,\Y_\w)$, and let $F_{n}$ be the empirical distribution based on $(\X_{1},\Y_{1}),\ldots,(\X_{n},\Y_{n})$ conditional on $\W=\w$. Let $\{(\X^{*}_{1,i},\Y^{*}_{1,i}),\ldots,(\X^{*}_{n,i},\Y^{*}_{n,i})\}_{i=1}^n$ be $n$ independent and identically distributed bootstrap sample from $F_{n}$, and let $F^{*}_{n}$ be the empirical distribution based on the bootstrap sample.

 Let $\M(\w)$ denote the kernel matrix of a specific dimension reduction approach,
 let  $\G(\w) = \bfSig_{\w}^{-1}\M(\w)$, and $d(\w)$ be the rank of $G(\w)$.  
Rearrange the eigenvalues of $\G(\w)$ as $\la_{1}(\w)\geq\ldots\geq\la_{d}(\w)>0=\la_{d+1}(\w)=\ldots=\la_{p}(\w)$,
 and denote the corresponding eigenvectors by $\be_{1}(\w),\ldots,\be_{p}(\w)$. Let $\widehat{\G}(\w)$ be the sample kernel matrix based on the sample $\{ Y_i, \X_i, \W_i \}_{i=1}^n$, and $\G^*(\w)$ be the sample kernel matrix based on the bootstrap sample. In parallel, we can define $\{\widehat{\la}_{1}(\w),\ldots,\widehat{\la}_{p}(\w),\widehat{\be}_{1}(\w),\ldots,\widehat{\be}_{p}(\w)\}$ and  $\{\la^{*}_{1}(\w),\ldots,\la^{*}_{p}(\w),\be^{*}_{1}(\w),\ldots,\be^{*}_{p}(\w)\}$ for $\widehat{\G}(\w)$ and $\G^{*}(\w)$. For each $k<p$, let
\begin{align*}
  \T_{k}(\w)&=\Big(\be_{1}(\w),\ldots,\be_{k}(\w)\Big),\\ \widehat{\T}_{k}(\w)&=\Big(\widehat{\be}_{1}(\w),\ldots,\widehat{\be}_{k}(\w)\Big),\\
  \T^{*}_{k}(\w)&=\Big(\be^{*}_{1}(\w),\ldots,\be^{*}_{k}(\w)\Big).
\end{align*}
Since $\T^{*}_{k}(\w)$ is repeatedly calculated for $n$ bootstrap samples, we denote its realization at the $i$th bootstrap sample by $\T^{*}_{k,i}(\w)$.

Conditional on $\W = \w$, define a function from $\{0, 1, \ldots, p-1\}$ to $\mathbb{R}$ as
\begin{align*}
f^{0}_{n}(\w,k)=\left\{
\begin{array}{lc}
{0}, & k=0; \\
{n^{-1}\sum_{i=1}^{n}[1-|\det\{\widehat{\T}_{k}\trans(\w) \T_{k,i}^{*}(\w)\}|]}, &k=1,\ldots,p-1.
\end{array}
\right.
\end{align*}
As in \cite{Ye2003}, $1-|\det\{\widehat{\T}_{k}\trans(\w) \T_{k,i}^{*}(\w)\}|$ is a number between 0 and 1 that measures the discrepancy between column spaces of  $\widehat{\T}_{k}(\w)$ and $\T_{k,i}^{*}(\w)$, with 1 representing the largest discrepancy. Therefore, $f^{0}_{n}(\w,k)$ measures the variability of the bootstrap estimates $\T_{k,1}^{*}(\w),\ldots,\T_{k,n}^{*}(\w)$ around the full sample estimate $\widehat{\T}_{k}(\w)$. We then normalize $f^{0}_{n}(\w,k)$ to be $f_{n}(\w,k)=f^{0}_{n}(\w,k)/\{1+\sum_{i=0}^{p-1}f^{0}_{n}(\w,i)\}$.
The asymptotic behavior of $f_{n}(\w,\cdot)$ is presented in Lemma \ref{asymptotic behavior of eigenvector}, whose
proof is similar to that of Theorem 1 in \cite{Luo2016}.
{\lemm\label{asymptotic behavior of eigenvector}
	Let $c_{n}=\{\log(\log n)\}^{-1}$. If conditions (C9), (C10), (C11) and (C12) hold, and $\G(\w)\in\mR^{p\times p}$ is a positive semi-definite matrix of rank $d(\w)\in\{0,\ldots,p-1\}$, then for any $k=1,\ldots,p-1$, the following relation holds for almost every sequence $S=\{(Y_{1},\X_{1},\W_{1}),(Y_{2},\X_{2},\W_2),\ldots\}:$
	\begin{align*}
	f_{n}(\w,k)=\left\{
	\begin{array}{lcl}
	{\Op_{P}(\frac{1}{nh})},  & \la_{k}(\w)>\la_{k+1}(\w); \\
	{\Op_{P}^{+}(c_{n})},  &\la_{k}(\w)=\la_{k+1}(\w);
	\end{array}
	\right.
	\end{align*}
	where $\Op_{P}^{+}$ is defined in \cite{Luo2016}.} 

Similarly, we normalize the sample eigenvalues and define the function $\phi_{n}(\w, \cdot):\{0,\ldots,p\}\rightarrow\mR$ as
$$ \phi_{n}(\w,k)=\widehat{\la}_{k+1}(\w)/\{1+\sum_{i=0}^{p-1}\widehat{\la}_{i+1}(\w)\},$$
where the constant $1$ in the denominator is introduced to stabilize the performance of the criterion when $d(\w) = 0$. The technique here is to shift the eigenvalues so that $\phi_{n}(\w,\cdot)$ takes a small value at $k = d(\w)$ instead of 
at $k = d(\w)+1$.  Lemma \ref{asymptotic behavior of eigenvalue} gives the asymptotic property of $\phi_{n}(\w,\cdot)$. 
{\lemm\label{asymptotic behavior of eigenvalue}
	Let $c_{n}=\{\log(\log n)\}^{-1}$ and $r=\lfloor p/\log(p) \rfloor$. For any positive semi-definite candidate matrix $\G(\w)\in\mR^{p\times p}$ of rank $d(\w)$, and for each $k\in \{0,1,\ldots,r\}$, we have
	\begin{align*}
	\phi_{n}(\w,k)=\left\{
	\begin{array}{lcl}
	{\Op_{P}^{+}(1)}, &\text{if} & k<d(\w); \\
	{\Op_{P}(\frac{1}{\sqrt{c_{n}nh}})}, &\text{if} &k\geq d(\w);
	\end{array}
	\right. \text{ almost surely } \proba_{\calS}.
	\end{align*}
}

Now, we can define the objective function of our estimator as
\begin{equation}\label{eq:11}
	\begin{aligned}
		g_{n}(\w, \cdot):\{0,\ldots,p-1\}\rightarrow\mR, \quad g_{n}(\w,k)=f_{n}(\w,k)+\phi_{n}(\w,k),
	\end{aligned}
\end{equation}
which collects information from both the eigenvectors and the eigenvalues. The reason for using this
objective function is that, the eigenvalue term $\phi_n(\w,\cdot)$ is small when $k<d(\w)$, while the eigenvector term $f_n(\w,\cdot)$
is large when $k>d(\w)$, and they are both small when  $k=d(\w)$.

As a rule of thumb, in most applications it is reasonable to assume $d(\w) \leq \lfloor p/\log(p) \rfloor$, where $\lfloor a \rfloor$ stands for the greatest integer less than or equal to $a$, thus it suffices to minimize $g_n(\w,
\cdot)$ over $\{0, 1, \ldots, \lfloor p/\log(p) \rfloor\}$, which yields
\begin{equation}\label{eq:12}
	\begin{aligned}
	g_{n}(\w,k)&=f_{n}(\w,k)+\phi_{n}(\w,k)\\
	&= \frac{f_{n}^{0}(\w,k)}{1+\sum_{i=0}^{\lfloor p/\log(p) \rfloor}f_{n}^{0}(\w,i)}+\frac{\widehat{\la}_{k+1}(\w)}{1+\sum_{i=0}^{\lfloor p/\log(p) \rfloor}\widehat{\la}_{i+1}(\w)}.
	\end{aligned}
\end{equation}
Let $\calD(f)$ denote the domain of a function $f$. Since the sample estimator of $\G(\w)$ is a nonparametric estimator, which is similar to the ladle estimator in \cite{Luo2016}, we define the nonparametric ladle estimator for $d(\w)$ by
\begin{equation}\label{eq:10}
\begin{aligned}
\widehat{d}(\w)=\arg \min\{g_{n}(\w,k):k\in\calD[g_{n}(\w,\cdot)]\},
\end{aligned}
\end{equation}
where $g_{n}(\w,\cdot)$ is defined by \eqref{eq:11} if $p\leq 10$ or by \eqref{eq:12} if $p>10$.

The following theorem \ref{consistency property} establishes the consistency of the nonparametric ladle estimator for dynamic partial dimension reduction.
\begin{theo}\label{consistency property}
	Under assumptions (C9), (C10), (C11) and (C12), for positive semi-definite matrix $\G(\w)\in\mR^{p\times p}$ of rank $d(\w)\in\left\{0,\ldots,p-1\right\}$, the nonparametric ladle estimator \eqref{eq:10} enjoys the following property: 
	\begin{align*}
	\proba\left\{\lim\limits_{n\rightarrow\infty}\proba(\widehat{d}(\w)=d(\w)|\calS)=1\right\}=1.
	\end{align*}
\end{theo}

\section{The Bandwidth Selection}\label{bandwidth selection}
Bandwidth selection for both kernel regression estimator and local estimator has been well studied.
Since \cite{Cook2001} showed that SIR can be viewed as linear discriminant analysis, we can
choose the bandwidth in a way similar to the tuning parameter selection based on linear discriminant analysis.

Assuming that  $\X|(\widetilde{\Y}=l,\W=\w)\sim N(\m_{l}(\w),\bfSig_{l\w})$, with density function
$f_{\X|\widetilde{\Y}=l,\W=\w}(\x)$, thus, $\X|\w$ follows a mixture multivariate normal distribution, and its likelihood function is given by
\begin{equation}\label{likelihood}
	\begin{aligned}
		L(\m_l(\w), \bfSig_{l\w}|\x) &= \prod_{i=1}^{n}\sum_{l=1}^{H}\proba_{l,\w} f_{\X|\widetilde{\Y}=l,\W=\w}(\mathbf{x_{i}})\\
	&=\prod_{i=1}^{n} \sum_{l=1}^{H}\proba_{l,\w}\left[\frac{e^{-\frac 12\{\x_{i}-\m_{l}(\w)\}\trans\bfSig_{l,\w}^{-1}\{\x_{i}-\m_{l}(\w)\}}}{\sqrt{(2\pi)^p|\bfSig_{l,\w}|}}\right],
	\end{aligned}
\end{equation}
where $\proba_{l,\w} $ is defined as before. Recall that $\m_l(\w)$ and $\bfSig_{l\w}$ are both of conditional structure, so we propose to estimate them by the following Nadaraya-Watson (NW) kernel estimators
\begin{align*}
&\widehat{\m}_{l}(\w)=\frac{\sum_{i=1}^{n}\x_{i}\1(\widetilde{\Y}_{i}=l)K_{h}(\w_{i}-\w)}{\sum_{i=1}^{n}\1(\widetilde{\Y}_{i}=l)K_{h}(\w_{i}-\w)},\\
  &\widehat{\bfSig}_{l\w}=\frac{\sum_{i=1}^{n}\x_{i}\x_{i}\trans\1(\widetilde{\Y}_{i}=l)K_{h}(\w_{i}-\w)}
  {\sum_{i=1}^{n}\1(\widetilde{\Y}_{i}=l)K_{h}(\w_{i}-\w)}-\widehat{\m}_{l}(\w)\widehat{\m}_{l}\trans(\w),
  \quad \widehat{\bfSig}_{\w}=\sum_{l=1}^{H}\widehat{\proba}_{l,\w}\widehat{\bfSig}_{l\w}.
\end{align*}
Since it's too hard to calculate the log-likelihood type of \eqref{likelihood} directly, we consider finding out the optimal bandwidth in each slice instead of targeting at the overall bandwidth, which is much more reasonable and computationally feasible. Following the argument of leave-one-out cross validation in \cite{Jiang2017}, we propose to find the $h_{l}$ which is the bandwidth for $l$-th slice such that
\begin{equation}\label{function}
\begin{aligned}
CV(h_{l})=\frac{1}{n_{l}}\sum_{i=1}^{n_{l}}\left[\{\x_{i}-\widehat{\m}_{l}^{(-i)}(\w)\}\trans\widehat{\bfSig}_{\w(-i)}^{-1}(\w)\{\x_{i}-\widehat{\m}_{l}^{(-i)}(\w)\}+\log|\widehat{\bfSig}_{\w(-i)}(\w)|\right]
\end{aligned}
\end{equation}
is minimized, where $\widehat{\m}_{l}^{(-i)}(\w)$ and $\widehat{\bfSig}_{\w(-i)}$ are estimators of the mean and covariance matrix of $\X|\widetilde{\Y}=l,\W=\w$, computed without the $i$-th observation, $n_{l}$ is the total number of observations in
the $l$-th slice. We choose the value of $h_{l}$ which maximizes  \eqref{likelihood} as $h_{opt}$, which is the bandwidth selected 
for dynamic  partial SIR.

For dynamic partial SAVE, since \cite{Cook2001} has shown that SAVE is closely related to quadratic
discriminant analysis, the bandwidth selection can be conducted by first minimizing \eqref{function 1},
assuming that
$\X|(\widetilde{\Y}=l,\W=\w) \sim N(\m_{l}(\w),\bfSig_{l\w})$, where $\bfSig_{l\w}$ is different in every slice, 
\begin{equation}\label{function 1}
\begin{aligned}
  CV(h_{l})=\frac{1}{n_{l}}\sum_{i=1}^{n_{l}}\left[\{\x_{i}-\widehat{\m}_{l}^{(-i)}(\w)\}\trans\widehat{\bfSig}_{l\w(-i)}^{-1}(\w)\{\x_{i}-\widehat{\m}_{l}^{(-i)}(\w)\}+\log|\widehat{\bfSig}_{l\w(-i)}(\w)|\right].
\end{aligned}
\end{equation}
The optimal bandwidth $h_{opt}$ for  dynamic partial SAVE is then selected by choosing the value of $h_l$ which
maximizes \eqref{likelihood}. 
For dynamic partial  DR, we use the same bandwidth selection procedure as dynamic partial  SAVE since
DR synthesizes the dimension reduction methods based on the first two conditional moments.

\section{Simulation studies}
In this section, we conduct simulation studies to evaluate our dynamic PDR methods. We consider the following six
models:

Model I: $\Y=X_{1}|W|+3 X_{2}\cos W+0.2\varepsilon,$

Model II: $\Y=2\exp\{X_{1}\exp(W)-X_{2}\cos W +1\}\cdot\text{sign}\{0.01X_1\cos W+2 (W+1)^{2} X_{2}\}+0.2\varepsilon,$

Model III: $\Y= \{X_{1}\sin(W) + 5 X_{2}\cos(W)\}^{2}+0.2\varepsilon,$

Model IV: $\Y=\exp\{(X_{1} |W|+X_{2})^{2}\}\log\{(X_{3}\cos W)^{2}\}+0.2\varepsilon,$

Model V: $\Y=10\frac{\exp\{X_1\sin W+5 X_{2} |W|\}}{X_{1} \exp(W) - X_{2}\cos W}+0.2\varepsilon,$

Model VI: $\Y=\frac{X_{1} (W_{2}+10)+X_{2} \sin W_{1}+7}{X_{1} \exp (W_{1})+10 X_{2}\cos W_{2} }+0.2\varepsilon, $\\
where $\text{sign}(\cdot)$ is the sign function, and $\varepsilon\sim N(0,1)$.
For Models I-V, $W$ is univariate and has a uniform distribution $U(-1,1)$, $\X|\W
\sim N_p(\m(W),\mathbf{\Sigma}_{W})$, where $\m(W) = \frac{\sin(W)}{2}\mathbf{1}_p$, $\mathbf{\Sigma}_{W} = (\sigma_{ij})_{p\times p}$ with $\sigma_{ij} = 1$ for $i=j$, $\sigma_{ij} =\frac 12\sin(W)$ for $i\neq j$.
For Model VI, $\W=(W_1,W_2)\trans$ is a two-dimensional vector with $W_1, W_2\overset{\text{iid}}\sim U(-1,1)$, $\X|\W \sim N_p(\m(\W),\mathbf{\Sigma}_{\W})$, where $\m(\W) = \frac{\sin(W_1) + \cos(W_2)}{2}\mathbf{1}_p$, and $\mathbf{\Sigma}_{\W} = (\sigma_{ij})_{p\times p}$ with $\sigma_{ij} = 1$ for $i=j$, $\sigma_{ij} =\frac 12(\sin(W_1) + \cos(W_2))$ for $i\neq j$.
Hence,  $d(\w) = 1$ for Models I and III, and $d(\w) = 2$ for Models II, IV and VI.  For Model V, 
$d(\w) =2$ when $\w\neq 0$,  and 1 when $\w=0$.

Our simulation studies are two folds. We first use the nonparametric ladle estimator in \eqref{eq:10} to determine $d(\w)$ for each model. And then estimate the basis for $\calS_{Y_{\w}|\X_{\w}}$ on the estimated structural dimension.
Since the inverse conditional mean is symmetric about 0 for Models III and IV, we expect that dynamic partial SIR
may provide a poor estimate for these two models. However, partial dynamic SIR might have advantages over dynamic partial SAVE and DR
for the rest of simulation models when the sample size is small since SIR is based on first inverse moments.

\subsection{Estimation of Structural Dimension}

\begin{table}[H]
	\centering
	\caption{Correct order determinations among 100 runs for Models I-V}\label{od1}
	\begin{tabular}{|c|c|ccc|ccc|}
		\hline
		\multirow{2}{*}{Model} & \multirow{2}{*}{$\w$} & \multicolumn{3}{c}{$(n,p) = (150, 5)$} & \multicolumn{3}{|c}{$(n,p) = (300, 10)$}\\
		\cline{3-5}\cline{6-8}
		&& {DPSIR} & {DPSAVE} & {DPDR} & {DPSIR} & {DPSAVE} & {DPDR} \\
		\hline
		\multirow{5}{*}{I}
		& 0  & 100 & 97 & 100 &  100 & 90 & 100\\
		& -1 & 100 & 96 & 100 & 99 & 91 & 99\\
		& 1 & 100 & 96 & 100 & 100 & 89 & 99\\
		& -0.5 & 100 & 96 & 100 & 98 & 93 & 99\\
		& 0.5 & 100 & 96 & 100 & 99 & 92 & 99\\
		\hline
		\multirow{5}{*}{II}
		& 0 & 100 & 86 & 100 & 100 & 87 & 100\\
		& -1 & 100 & 84 & 100 & 100 & 86 & 100\\
		& 1 & 100 & 85 & 100 & 100 & 87 & 100\\
		& -0.5 & 100 & 86 & 99 & 100 & 86 & 100\\
		& 0.5 & 100 & 86 & 100 & 100 & 87 & 100\\
		\hline
		\multirow{5}{*}{III}
		& 0 &7 & 99 & 95 & 17 & 90 & 84\\
		& -1 &8 & 96 & 94 & 17 & 87 & 84\\
		&1 & 8 & 98 & 95 & 18 & 88 & 85\\
		& -0.5 &7 & 97 & 96 & 18 & 88 & 86\\
		& 0.5 &9 & 99 & 96 & 14 & 88 & 82\\
		\hline
		\multirow{5}{*}{IV}
		& 0 & 0 & 97 & 97 &  1 & 99 & 99\\
		& -1 &  0 & 97 & 97 &  1 & 99 & 99\\
		& 1 & 0 & 98 & 98 &  3 & 99 & 99\\
		& -0.5 & 0 & 99 & 98 &  1 & 99 & 99\\
		& 0.5 & 0 & 98 & 97 &  3 & 99 & 99\\
		\hline
		\multirow{4}{*}{V}
		& -1 & 100 & 72 & 73 & 99 & 70 & 70\\
		& 1 & 100 & 74 & 83 & 99 & 71 & 79\\
		& -0.5 & 100 & 71 & 79 & 98 & 70 & 77\\
		& 0.5 & 100 & 76 & 87 & 99 & 72 & 82\\
		\hline
	\end{tabular}%
\end{table}
Based on the nonparametric ladle estimator, we use  dynamic partial SIR, dynamic partial SAVE and dynamic  partial
DR to estimate $d(\w)$ for different $\w$. The percentages of correct order estimates in 100 runs are presented in Table \ref{od1}.
It  shows that our proposed nonparametric ladle estimator works pretty well, with the percentage of correct order estimation approaches $100\%$. Also, as we expected, dynamic partial SIR fails for Models III and IV, and outperforms the other two dynamic
PDR methods for the remaining models due to the small sample sizes we have ($n=150$ or $300$)
in our simulation studies.

\begin{table}[H]
	\centering
	\caption{Order determination for Model V with $\w$ in a neighborhood of 0}\label{od2}
	\begin{tabular}{|c|c|ccc|ccc|}
		\hline
		\multirow{2}{*}{Model} & \multirow{2}{*}{$\w$} & \multicolumn{3}{c}{$(n,p) = (150, 5)$} & \multicolumn{3}{|c}{$(n,p) = (300, 10)$}\\
		\cline{3-5}\cline{6-8}
		&& {DPSIR} & {DPSAVE} & {DPDR} & {DPSIR} & {DPSAVE} & {DPDR} \\
		\hline
		\multirow{11}{*}{V}
	        & 0.01 & 100 & 72 & 83 & 99 & 70 & 80\\
		& 0.02 & 100 & 74 & 82 & 98 & 71 & 79\\
		& 0.03 & 100 & 75 & 85 & 100 & 72 & 81\\
		& 0.04 & 100 & 71 & 83 & 99 & 69 & 80\\
		& 0.05 & 100 & 72 & 82 & 99 & 71 & 79\\
		& -0.01 & 100 & 72 & 83 & 98 & 69 & 80\\
		& -0.02 & 100 & 75 & 83 & 99 & 72 & 81\\
		& -0.03 & 100 & 76 & 80 & 99 & 73 & 77\\
		& -0.04 & 100 & 71 & 80 & 98 & 69 & 78\\
		& -0.05 & 100 & 69 & 84 & 99 & 67 & 81\\
		\hline
	\end{tabular}%
\end{table}
Table \ref{od2} shows that, for Model V, under the scenario $(n,p)=(150,5)$, if $\w \neq 0$, the structural dimension
can be accurately estimated by dynamic partial SIR approach all the times. When $(n,p)=(300,10)$, at least $98\%$ of
the times, dynamic partial SIR provides with the correct estimates of $d=2$.

Table \ref{od3} suggests that the percentage of correct estimation for Model V when $\w=0$ is somewhat 
unsatisfactory when the sample size $n$ is small. 
However, as $n$ increases, the correct estimation percentage improves steadily,
which is consistent with large sample theory.
\begin{table}[H]
	\centering
	\caption{Order determination for Model V with $\w=0$} \label{od3}
	\begin{tabular}{|cccc|cccc|}
		\hline
		$(n,p)$ & DPSIR & DPSAVE & DPDR & $(n,p)$ & DPSIR & DPSAVE & DPDR\\
		\hline
                (150, 5) & 62 & 21 & 25 & (300, 10) & 59 & 19 & 27\\
		(500, 5) & 73& 51 & 60& (500, 10) & 65& 42 & 53\\
		(800, 5) & 86& 74 & 82& (800, 10) & 75& 57 & 68\\
		(1000, 5) & 93& 88 & 90& (1000, 10) & 81& 70 & 78\\
		\hline
	\end{tabular}
\end{table}

For Model VI, since $\W = (W_1, W_2)$ is two-dimensional, we present the order determination results for different $\w$
values in Tables \ref{od4} and \ref{od5}. The entries are the counts of correct estimates of the structural dimension out of 100
repetitions by dynamic partial SIR, dynamic partial SAVE, dynamic partial DR, respectively.
\begin{table}[H]
	\centering
	\caption{Order Determination for Model VI with $(n,p) = (150, 5)$} \label{od4}
	\begin{tabular}{|@{}c|ccc|ccc|ccc|ccc|ccc|}
		\hline
		\scriptsize
		\diagbox[width=6em,trim=l]{$\w_1$}{$\w_2$} & \multicolumn{3}{c|}{-1} & \multicolumn{3}{c|}{-0.5} & \multicolumn{3}{c|}{0} & \multicolumn{3}{c|}{0.5} & \multicolumn{3}{c|}{1}\\
		\hline
		-1 & 88 & 76 & 85  & 88 & 76 & 86 & 92 & 78 & 86 & 90 & 84 & 86 & 91 & 78 & 86\\
		\hline
		-0.5 & 90 & 77 & 87  & 90 & 77 & 84 & 89 & 80 & 87 & 89 & 82 & 86 & 93 & 77 & 87\\
		\hline
		0 & 89 & 77 & 85  & 89 & 84 & 85 & 91 & 75 & 87 & 89 & 78 & 87 & 88 & 77 & 88 \\
		\hline
		0.5 & 91 & 77 & 87  & 90 & 75 & 87 & 91 & 76 & 87 & 91 & 80 & 87 & 88 & 80 & 89\\
		\hline
		1 & 92 & 78 & 83  & 91 & 79 & 87 & 89 & 76 & 84 & 88 & 79 & 86 & 89 & 85 & 86\\
		\hline
	\end{tabular}

\end{table}

\begin{table}[H]
	\centering
	\caption{Order Determination for Model VI with $(n,p) = (300, 10)$} \label{od5}
        \begin{tabular}{|@{}c|ccc|ccc|ccc|ccc|ccc|}
	  		\hline
		\scriptsize
		\diagbox[width=6em,trim=l]{$\w_1$}{$\w_2$} & \multicolumn{3}{c|}{-1} & \multicolumn{3}{c|}{-0.5} & \multicolumn{3}{c|}{0} & \multicolumn{3}{c|}{0.5} & \multicolumn{3}{c|}{1}\\
		\hline
		-1 & 97 & 80 & 98  & 97 & 81 & 98 & 97 & 78 & 98 & 96 & 81 & 98 & 95 & 78 & 98\\
		\hline
		-0.5 & 96 & 84 & 98  & 95 & 78 & 98 & 95 & 77 & 99 & 97 & 81 & 98 & 97 & 86 & 98\\
		\hline
		0 & 97 & 82 & 99  & 97 & 78 & 98 & 95 & 80 & 98 & 95 & 77 & 98 & 97 & 77 & 98\\
		\hline
		0.5 & 95 & 79 & 98  & 96 & 83 & 98 & 96 & 86 & 98 & 95 & 80 & 99 & 96 & 82 & 98\\
		\hline
		1 & 96 & 79 & 99  & 98 & 79 & 98 & 97 & 79 & 99 & 96 & 80 & 99 & 97 & 80 & 98\\
		\hline
	\end{tabular}

\end{table}

Table \ref{od4} and \ref{od5} suggest that our nonparametric ladles works pretty well for
estimation of structural dimension when $\W$ is two dimensional.

We also use  PDEE methods to estimate the dimension of $\B$ based on the ladle estimator.
Since PDEE methods cannot estimate the dimension of $\B(\w)$ dynamically, it could not deal with situations
such as Model V, the following table gives the number of correct estimates of structural dimensions among 100 simulation
runs for all models except Model V. 
It is easy to see that PDEE methods perform poorly regarding the order determination, especially for complex models. Further,
simulation results (unreported here) also show that, when PDEE fails to select the correct structural dimension, it tends to under-select the dimension
than to over-select it,  
which makes PDEE even less desirable in practice.

\begin{table}[h]
	\centering
	\caption{{Estimated structural dimension based on 100 replications for PDEE methods}}\label{odpdee} 
		\begin{tabular}{|c|c|ccc|}
			\hline
			&	  & PDEE-SIR & PDEE-SAVE & PDEE-DR  \\
			\hline
			\multirow{4}{*}{(n,p)=(150,5)} &	\text{Model I}   & 63 &  5 & 100\\
			\cline{2-5}
			&\text{Model II}  	 & 0 & 0  & 0\\
			\cline{2-5}
			&\text{Model III}  	 & 0 &  6 & 39\\
			\cline{2-5}
			& \text{Model IV}  	 & 0 &  0 & 8\\
			\cline{2-5}
			& \text{Model VI} 	 & 0 & 0  & 1\\
			\hline	
			\multirow{4}{*}{(n,p)=(300,10)} &	\text{Model I}   &100  &  11 & 100\\
			\cline{2-5}
			&\text{Model II}  	 & 0 & 0  & 0\\
			\cline{2-5}
			&\text{Model III}  	 & 0 & 9  & 76\\
			\cline{2-5}
			&\text{Model IV}  	 & 0 & 0  & 18\\
			\cline{2-5}
			& \text{Model VI}  	 & 0 & 0  & 0\\
			\hline	
	\end{tabular}%
\end{table}

\subsection{Estimation of $\calS_{Y_{\w}|\X_{\w}}$}

To assess the accuracy of our dynamic PDR methods, we adopt the trace correlation $r^{2}_{d(\w)}$ proposed by \cite{Ferre1998}. Let $\calS_{1}$ and $\calS_{2}$ be two $d(\w)$-dimensional subspaces of $\mR^{p}$, the distance between subspace $\calS_{1}$ and $\calS_{2}$ can be measured by the following trace correlation coefficient,
\begin{align*}
  r^{2}_{d(\w)}={\rm{Tr}}(\proba_{\calS_{1}}\proba_{\calS_{2}})/d(\w),
\end{align*}
where $\proba_{\calS_{1}}$ and $\proba_{\calS_{2}}$ are orthogonal projections onto $\calS_{1}$ and $\calS_{2}$, and ${\rm{Tr}(\cdot)}$ is the trace of a square matrix. It can be justified that $r^{2}_{d(\w)}\in[0,1]$, $r^{2}_{d(\w)} = 1$ if $\calS_1 = \calS_2$, and $r^{2}_{d(\w)} = 0$ if $\calS_1 \perp \calS_2$ (the two subspaces are perpendicular). Note that a larger value of $r^{2}_{d(\w)}$ implies that $\calS_{1}$ and $\calS_{2}$ are closer. \cite{Li2009} and \cite{Dong2010} applied a similar
criterion to assess the performance of sufficient dimension reduction estimator with non-elliptically distributed predictors.

We first compare the performance among the three dynamic PDR methods under two configurations $(n,p)=(150,5)$
and $(n,p)=(300,10)$. To be fair, we set the number of slices $H$ to be 5 for all three methods. 
Due to space limitations, we only present part of the results.
We present in Table \ref{tab:cor1} the mean of trace correlations between the true and estimated dynamic partial CS
at different values of $\w$ for the first five models based on 100 repetitions.
\begin{table}[H]
	\centering
	\caption{Trace Correlation for the Models 1-5} \label{tab:cor1}
	\begin{tabular}{|c|c|ccc|ccc|}
		\hline
		\multirow{2}{*}{Model} & \multirow{2}{*}{$\w$} & \multicolumn{3}{c}{$(n,p) = (150, 5)$} & \multicolumn{3}{|c}{$(n,p) = (300, 10)$}\\
		\cline{3-5}\cline{6-8}
		&& {DPSIR} & {DPSAVE} & {DPDR} & {DPSIR} & {DPSAVE} & {DPDR} \\
		\hline
		\multirow{5}{*}{I}
		& 0 & 0.978 & 0.962  & 0.945 & 0.980 & 0.958 & 0.933\\
		& -1 & 0.892 & 0.920  & 0.918 & 0.889 & 0.918 & 0.919\\
		& 1 & 0.894 & 0.920  & 0.922 & 0.888 & 0.919 & 0.918\\
		& -0.5 & 0.991 & 0.994  & 0.985 & 0.989 & 0.992 & 0.978\\
		& 0.5 & 0.990 & 0.995  & 0.986 & 0.988 & 0.992 & 0.977\\
		\hline
		\multirow{5}{*}{II}
		& 0 & 0.973 & 0.909 & 0.914 & 0.958 & 0.835 & 0.856\\
		& -1 & 0.940 & 0.910 & 0.911 & 0.927 & 0.831 & 0.854\\
		& 1 & 0.941 & 0.898 & 0.913 & 0.932 & 0.817 & 0.851\\
		& -0.5 & 0.961 & 0.909 & 0.913 & 0.947 & 0.835 & 0.856\\
		& 0.5 & 0.969 & 0.905 & 0.914 & 0.959 & 0.830 & 0.854\\
		\hline
		\multirow{5}{*}{III}
		& 0 & 0.297 & 0.967 & 0.967 & 0.313 & 0.955 & 0.952\\
		& -1 & 0.218 & 0.880 & 0.858 & 0.203 & 0.847 & 0.838\\
		& 1 & 0.492 & 0.941 & 0.945 & 0.536 & 0.939 & 0.945\\
		& -0.5 & 0.230 & 0.956 & 0.950 & 0.242 & 0.938 & 0.935\\
		& 0.5 & 0.389 & 0.965 & 0.969 & 0.418 & 0.957 & 0.955\\
		\hline
		\multirow{5}{*}{IV}
		& 0 & 0.361 & 0.906 & 0.908 & 0.221 & 0.878 & 0.879\\
		& -1 & 0.433 & 0.938 & 0.940 &0.321 & 0.919 & 0.917\\
		& 1 & 0.418 & 0.941 & 0.941 & 0.320 & 0.916 & 0.918\\
		& -0.5 & 0.379 & 0.965 & 0.966 & 0.256 & 0.940 & 0.940\\
		& 0.5 & 0.375 & 0.966 & 0.966 & 0.249 & 0.940 & 0.941\\
		\hline
		\multirow{5}{*}{V}
		& 0 & 0.895 & 0.850  & 0.883 & 0.833 & 0.813 & 0.814\\
		& -1 & 0.833 & 0.827  & 0.857 &0.813 & 0.815 & 0.821\\
		& 1 & 0.904 & 0.847  & 0.884 &0.951 & 0.840 & 0.898\\
		& -0.5 & 0.912 & 0.826  & 0.872 & 0.905 & 0.818 & 0.839\\
		& 0.5 & 0.946 & 0.871  & 0.890 & 0.948 & 0.873 & 0.903\\
		\hline
	\end{tabular}
\end{table}
Table \ref{tab:cor1} shows that dynamic partial SIR works pretty well in most cases except for Models III and IV
just as expected, while both dynamic partial SAVE and DR perform stably for all models, providing with
trace correlations greater than $0.9$ most of the time.

Simulation results for Model VI are provided in Tables \ref{tab:cor2} and \ref{tab:cor3} based on different combinations
of  $\W= (W_1, W_2)$. These results reaffirm the good performance of our dynamic PDR methods even when $\W$ is multivariate.
\begin{table}[H]
	\centering
	\scriptsize
	\caption{Trace Correlation for Model 6 with $(n,p) = (150, 5)$} \label{tab:cor2}
	\begin{tabular}{|@{}c|ccc|ccc|ccc|ccc|ccc|}
		\hline
		\diagbox[width=5em,trim=l]{$\w_1$}{$\w_2$} & \multicolumn{3}{c|}{-1} & \multicolumn{3}{c|}{-0.5} & \multicolumn{3}{c|}{0} & \multicolumn{3}{c|}{0.5} & \multicolumn{3}{c|}{1}\\
		\hline
		-1 & .851 & .829 & .842  & .874 & .868 & .876 & .880 & .890 & .888 & .862 & .892 & .874 & .815 & .872 & .825\\
		\hline
		-0.5 & .884 & .861 & .873  & .906 & .896 & .898 & .912 & .912 & .915 & .894 & .915 & .931 & .864 & .899 & .893\\
		\hline
		0 & .905 & .881 & .907  & .925 & .912 & .919 & .929 & .924 & .931 & .915 & .927 & .92 & .886 & .911 & .921 \\
		\hline
		0.5 & .912 & .893 & .90  & .931 & .919 & .925 & .935 & .924 & .931 & .921 & .920 & .918 & .894 & .903 & .896\\
		\hline
		1 & .900 & .883 & .892  & .925 & .900 & .925 & .930 & .908 & .922 & .916 & .894 & .910 & .883 & .859 & .864\\
		\hline
	\end{tabular}
\end{table}
\begin{table}[H]
	\centering
	\scriptsize
	\caption{Trace Correlation for Model VI with $(n,p) = (300, 10)$} \label{tab:cor3}
	\begin{tabular}{|@{}c|ccc|ccc|ccc|ccc|ccc|}
		\hline
		\diagbox[width=5em,trim=l]{$\w_1$}{$\w_2$} & \multicolumn{3}{c|}{-1} & \multicolumn{3}{c|}{-0.5} & \multicolumn{3}{c|}{0} & \multicolumn{3}{c|}{0.5} & \multicolumn{3}{c|}{1}\\
		\hline
		-1 & .872 & .830 & .880  & .891 & .861 & .875 & .890 & .880 & .885 & .885 & .883 & .878 & .824 & .869 & .837\\
		\hline
		-0.5 & .894 & .859 & .876  & .921 & .889 & .911 & .923 & .910 & .915 & .906 & .909 & .903 & .879 & .896 & .854\\
		\hline
		0 & .915 & .878 & .908  & .931 & .910 & .924 & .945 & .921 & .926 & .923 & .922 & .91 & .894 & .908 & .8778\\
		\hline
		0.5 & .927 & .890 & .921  & .947 & .918 & .932 & .951 & .921 & .945 & .936 & .918 & .931 & .903 & .899 & .900\\
		\hline
		1 & .907 & .876 & .898  & .932 & .892 & .92 & .936 & .903 & .924 & .925 & .890 & .916 & .889 & .856 & .876\\
		\hline
	\end{tabular}
\end{table}

The following Figures give a direct visual presentation of our simulation results, which also agree with
what we discussed previously. It seems that dynamic partial SAVE and DR are very reliable and accurate regarding
the estimation of the dynamic partial CS in all models, while dynamic partial SIR might fail under certain
conditions. We recommend dynamic partial DR, and use dynamic partial SIR as a complementary method.    

\begin{figure}[H]
	\centering
	\begin{subfigure}{0.23\textwidth} 
		\includegraphics[width=\textwidth]{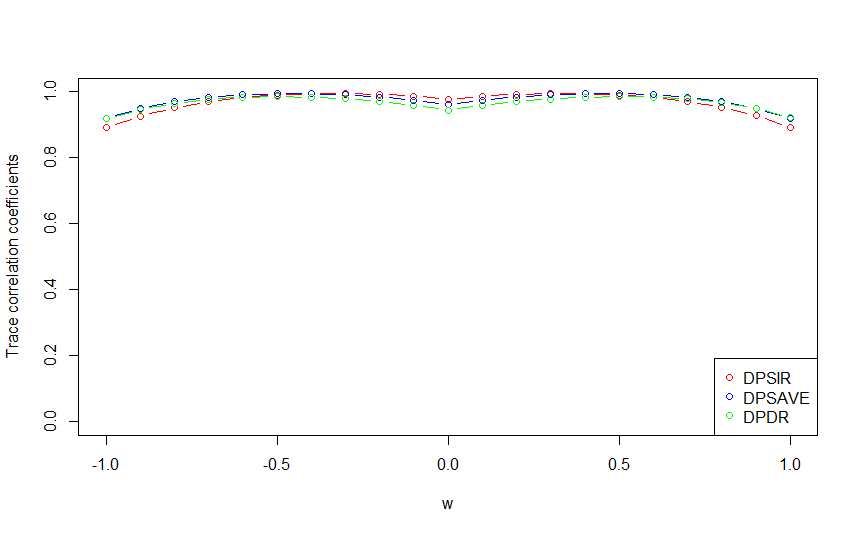}
		\caption{Model 1} 
	\end{subfigure}
	\begin{subfigure}{0.23\textwidth} 
		\includegraphics[width=\textwidth]{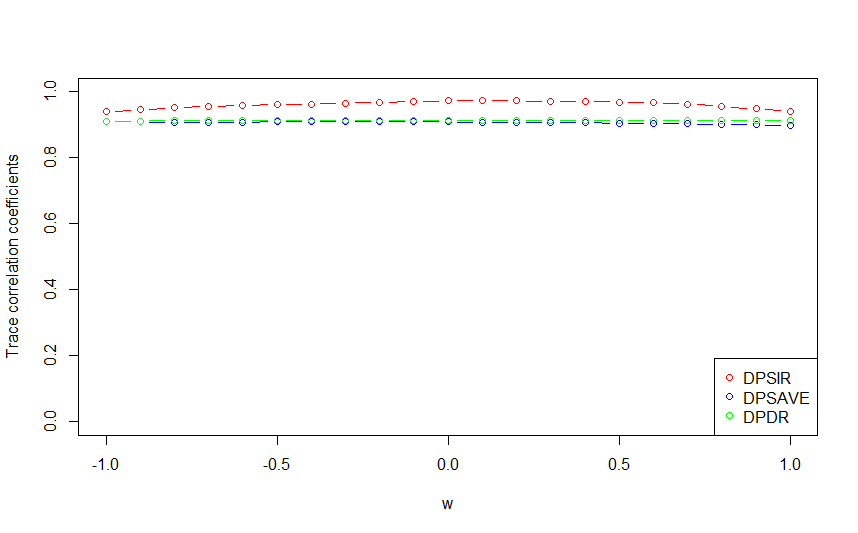}
		\caption{Model 2} 
	\end{subfigure}
\begin{subfigure}{0.23\textwidth} 
	\includegraphics[width=\textwidth]{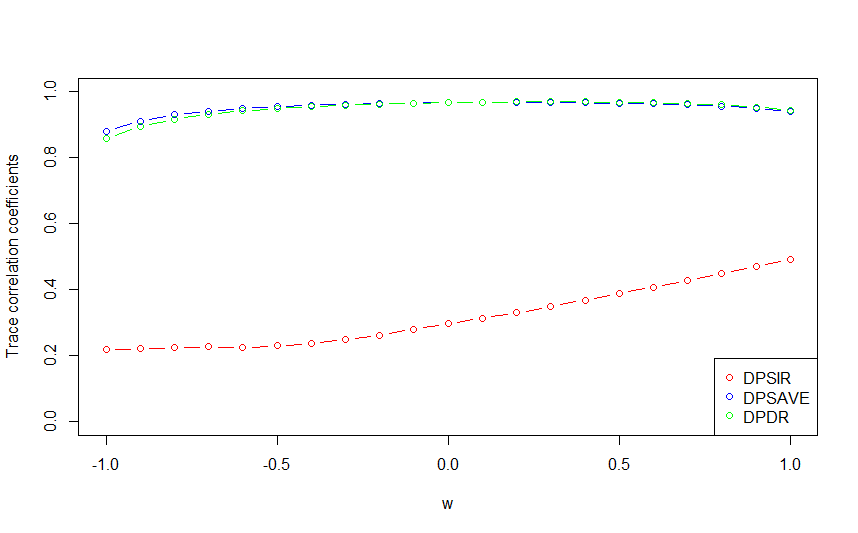}
	\caption{Model 3} 
\end{subfigure}
\begin{subfigure}{0.23\textwidth} 
	\includegraphics[width=\textwidth]{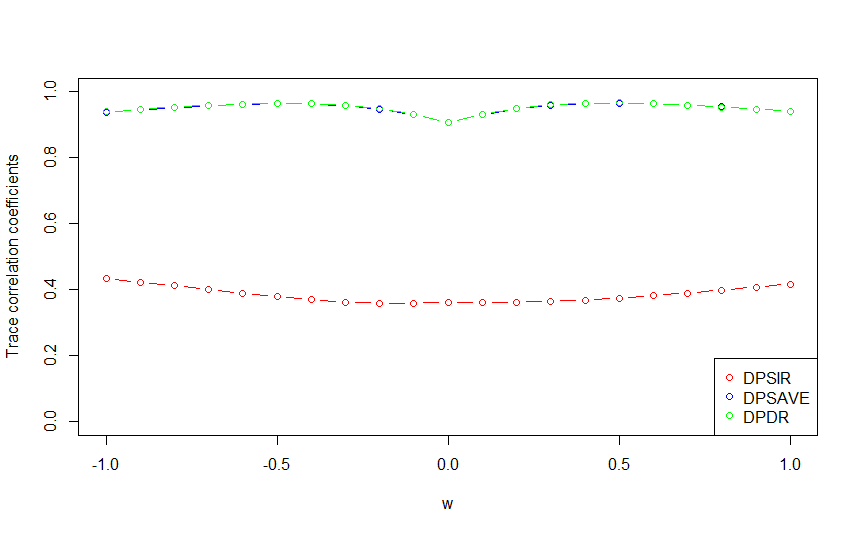}
	\caption{Model 4} 
\end{subfigure}
	\caption{\footnotesize Trace correlation coefficient for Models I-IV with $(n,p)=(150,5)$}
\end{figure}

\begin{figure}[H]
	\centering
\begin{subfigure}{0.23\textwidth} 
	\includegraphics[width=\textwidth]{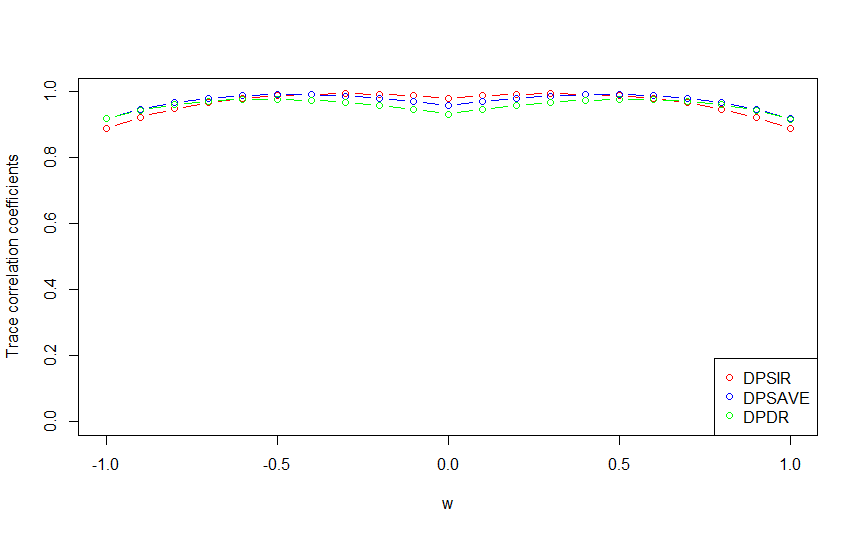}
	\caption{Model 1} 
\end{subfigure}
\begin{subfigure}{0.23\textwidth} 
	\includegraphics[width=\textwidth]{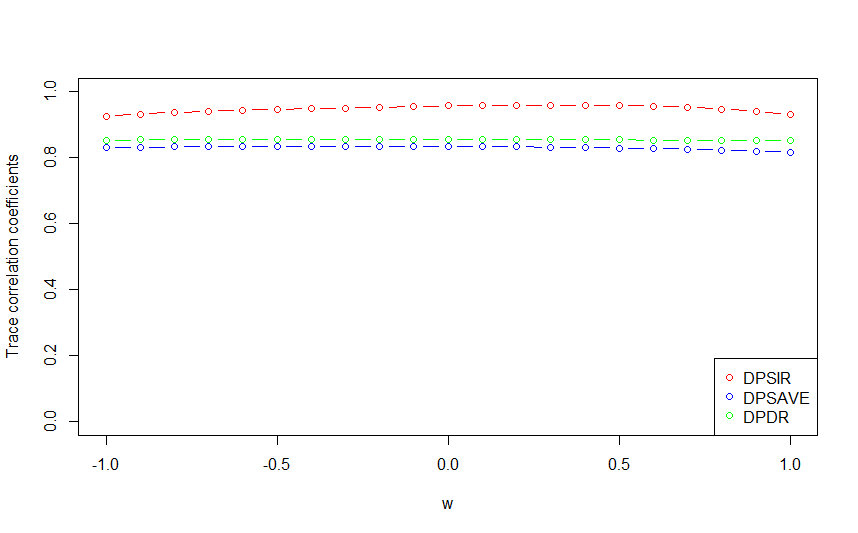}
	\caption{Model 2} 
\end{subfigure}
\begin{subfigure}{0.23\textwidth} 
	\includegraphics[width=\textwidth]{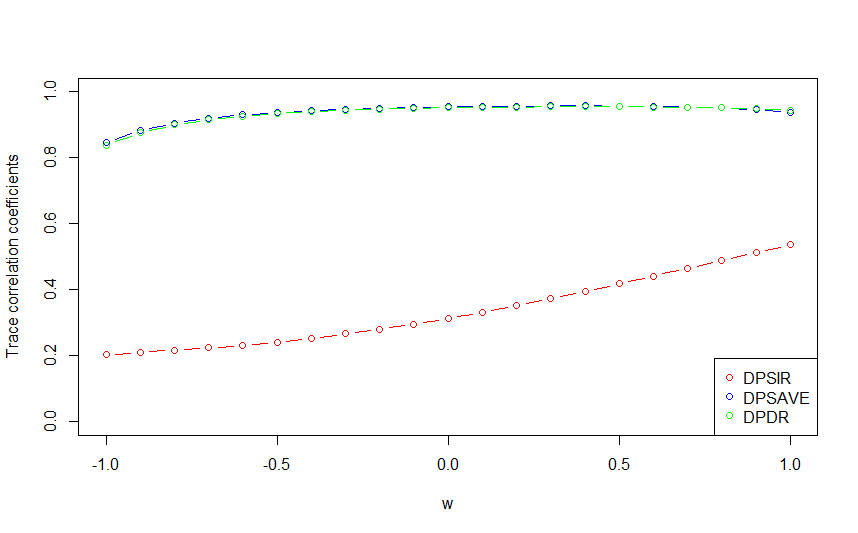}
	\caption{Model 3} 
\end{subfigure}
\begin{subfigure}{0.23\textwidth} 
	\includegraphics[width=\textwidth]{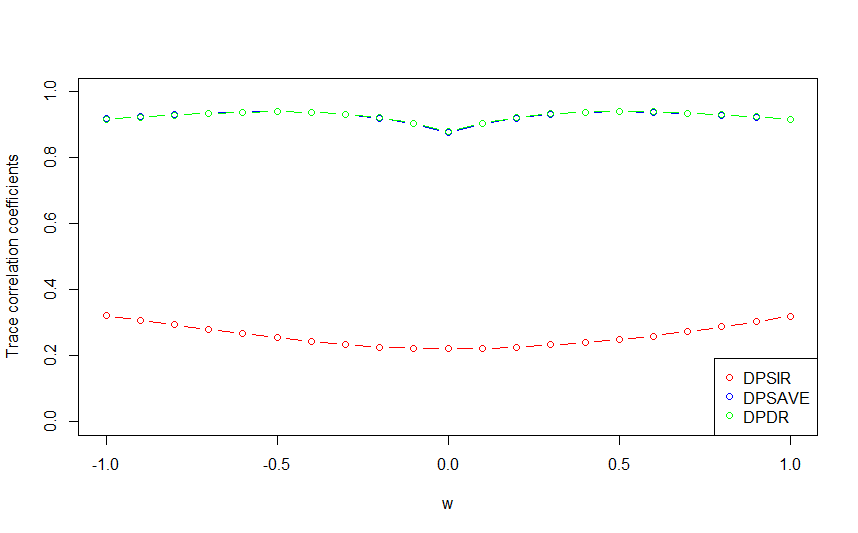}
	\caption{Model 4} 
\end{subfigure}
	\caption{\footnotesize Trace correlation coefficient for Models I-IV with $(n,p)=(300,10)$}
\end{figure}

\begin{figure}[H]
	\centering
	\begin{subfigure}{0.4\textwidth}
		\includegraphics[width=\textwidth]{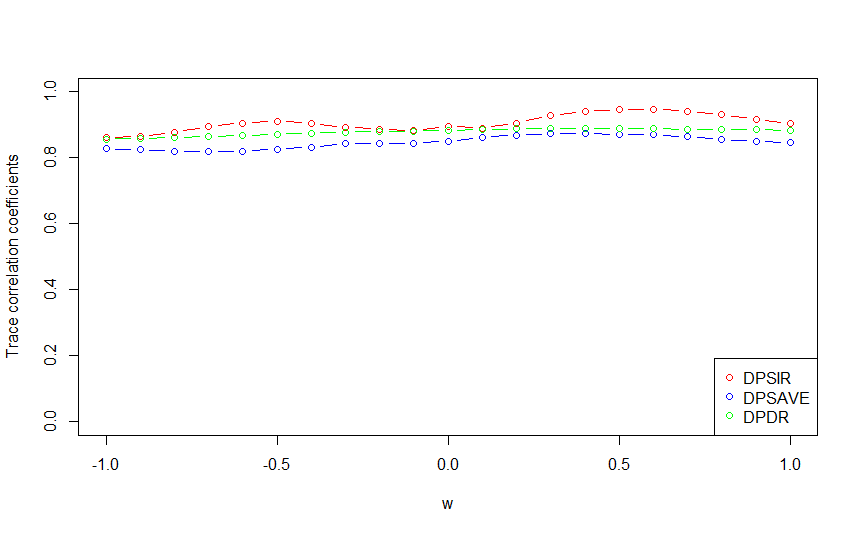}
		\caption{$(n,p)=(150,5)$} 
	\end{subfigure}
\vspace{1em} 
\begin{subfigure}{0.4\textwidth} 
	\includegraphics[width=\textwidth]{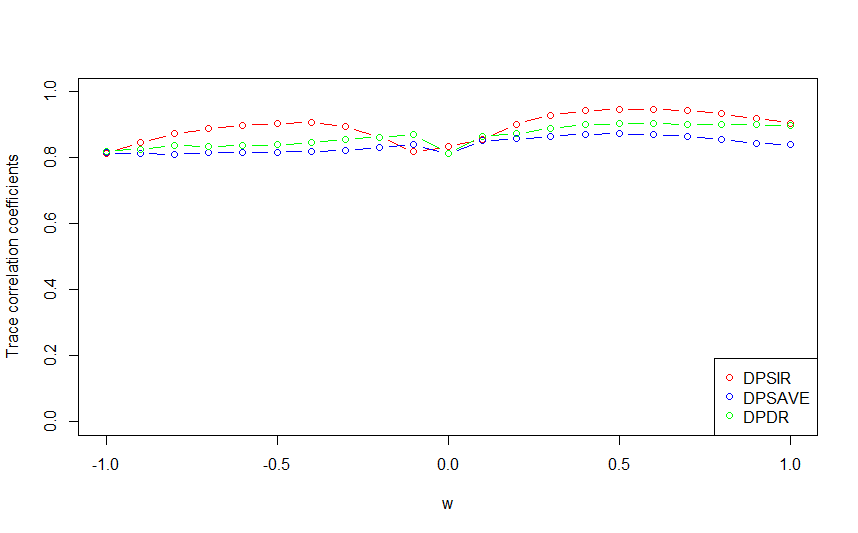}
	\caption{$(n,p)=(300,10)$} 
\end{subfigure}
\caption{Trace correlation coefficient for Model V}
\end{figure}
The 3D plot for Model VI is omitted since it cannot clearly demonstrate the trend of trace
correlation coefficient when $\W$ is two-dimensional.

Furthermore, to illustrate the advantages of our dynamic approach, we
compare the average trace correlation coefficient deriving from those at different $\w$ values,
with the trace correlation coefficient obtained via PDEE \citep{Feng2013} estimation methods. 
Since the PDEE method can not handle dynamic structural dimension scenario,
we only investigate the estimation accuracy in Model I-IV and VI, given a fixed estimated structural dimension. 
The results of the mean trace correlation coefficient from 100 simulation runs are shown in Table \ref{tabcom}.
\begin{table}[h]
	\centering
	\caption{{Mean trace correlation coefficients of 100 replications}}
        \begin{small}
		\begin{tabular}{|c|c|ccc|ccc|}
			\hline
	(n,p)	& Model	& DPSIR  & DPSAVE & DPDR  & PDEE-SIR & PDEE-SAVE & PDEE-DR  \\
			\hline
	\multirow{4}{*}{(150,5)} &	\text{I}  &	0.9723  &  0.9646  &  0.9665 & 0.9483 & 0.7848  & 0.3897\\
			\cline{2-8}
		&\text{II}  &	0.8076  & 0.6940   & 0.7936 & 0.7935 & 0.5222  & 0.5387\\
		\cline{2-8}
		&\text{ III}  &	0.1911  &  0.9147 & 0.9290 & 0.4937 &  0.8638 & 0.8739\\
		\cline{2-8}
		& \text{IV}  &	0.3721  &  0.9033  & 0.9278 & 0.5951 & 0.8722  & 0.8832\\
		\cline{2-8}
		& \text{VI}  &	0.9090  &  0.9104  & 0.9055 & 0.8214 & 0.6437  & 0.7446\\
		\hline	
	\multirow{4}{*}{(300,10)} &	\text{I}  &	0.9728  &  0.9600  & 0.9633  & 0.9553 & 0.6892  & 0.3234\\
		\cline{2-8}
		&\text{ II}  &	0.8096  &  0.5844  & 0.7881 & 0.8014 &  0.3758 & 0.4826\\
		\cline{2-8}
		&\text{III}  &	0.1520  &  0.9218  & 0.9207 & 0.3959 &  0.7600 & 0.8461\\
		\cline{2-8}
		&\text{IV}  &	0.1929  & 0.8948  & 0.9104 & 0.4931 & 0.8490  & 0.8625\\
		\cline{2-8}
		& \text{ VI}  &	0.8929  &  0.8631 & 0.8746 & 0.8659 & 0.5161  & 0.6868\\
		\hline	
	\end{tabular}%
	          \label{tabcom}%
                  \end{small}
\end{table}

It is noticeable that our proposals consistently outperform PDEE approaches. Our methods can estimate $\B(\W)$ at $\W=\w$
dynamically, which changes with the value of $\w$, while the PDEE approach can only estimate $\B$, which is fixed
no matter how $\w$ varies. As a result, the dynamic PDR is a better approach to handle partial dimension reduction with continuous $\W$.

\section{Real Data Analysis}
In this section, we consider analyzing four real-world datasets: Body Fat data, Wage data, Hongkong environmental data, and Boston Housing data. For each dataset, we compare our proposals, dynamic  partial SIR, dynamic partial SAVE, and  dynamic partial
DR with the PDEE methods which include PDEE-SIR, PDEE-SAVE, and PDEE-DR.

To implement each of the methods, we first need to estimate the structural dimension. For our proposed methods, we use the nonparametric ladle estimator in \eqref{eq:10} to estimate $d(\w)$, while using the ladle estimator in \cite{Luo2016} to estimate $d$ for the PDEE methods. Then based on the estimated structural dimension, we obtain the dynamic partial dimension reduction directions $\widehat\B(\w)$ and partial dimension reduction directions $\widehat\B$.
We use the distance correlations \citep{Szekely2007} between $\Y$ and $\widehat\B\trans(\w)\X$ (or $\widehat{\B}\trans\X$) to
evaluate the performance of the estimates $\widehat\B(\w)$ and $\widehat{\B}$.

\subsection{Description of the Datasets}

We illustrate the application of dynamic PDR to the following four data sets in the literature.

\noindent
\textbf{Body Fat dataset.}  we first consider the Body Fat data, which has been analyzed in \cite{Penrose1985}, \cite{Hoeting1999}, \cite{Leng2010}. The Body Fat data contains 252 observations and 14 attributes. Following the analysis of \cite{Leng2010},
we treat {\it{brozek}} as the response $\Y$, {\it{age}} as $W$, and the other 12 predictors (i.e. {\it{weight, height, neck, chest, abdomen, hip, thigh, knee, ankle, biceps, forearm, wrist}}) as $\X$. For the structural determination, we get $\hat{d}(\w) = \hat{d} = 1$. This is consistent with the previous studies in \cite{Leng2010} and \cite{Zhang2013}, both estimated
the structural dimension as 1. 

\noindent
\textbf{Wage dataset.} Wage dataset contains the wage information of 534 workers and their education, living region, gender, union membership, race, occupation, sector, marriage status information and their years of experience. This data set has been investigated in  \cite{Berndt1991}, \cite{Xie2009} and \cite{Zhang2013}. We take {\it{wage}} as the response $\Y$, {\it{years of experience}}
as $W$, and the remaining 8 predictors as  $\X$. The order determination procedure yields
that $\hat{d}(\w)=\hat{d} = 1$.

\noindent
\textbf{Hongkong environmental dataset.} Hongkong environmental dataset has been analyzed in \cite{Li2015}. This data set was collected between January 1 of 1994 and December 31 of 1995. To be more specific, it is a collection of numbers of daily total hospital admissions for circulationary and respirationary problems, measurements of pollutants and many other environmental factors in
Hong Kong. We take {\it{the number of daily total hospital admissions for circulationary and respirationary problems}} as the response $\Y$, {\it{time}} as $W$, and the remaining predictors as $\X$.
The order determination procedure yields that $\hat{d}(\w)=\hat{d} = 1$.

\noindent
\textbf{Boston Housing dataset.} Boston housing data (Harrison and Rubinfeld 1978) has been widely used as a classical dataset in regression study. For example, it has been studied in \cite{Fan2005} and \cite{Chen2010}. It contains information collected by the U.S. Census Service concerning housing in the area of Boston. The original data consist of 14 variables (features) and 506 data points. Following \cite{Chen2010}, we only keep 374 observations with {\it{per capita crime rate by
twon}} smaller than 3.2 in the subsequent analysis. We
take {\it{the median value of the owner-occupied homes in \$1000's}} as the response $\Y$, {\it{ crime rate}}
as $W$, and the remaining predictors as $\X$. The order determination procedure yields that $\hat{d}(\w)=\hat{d} = 2$.

\subsection{Comparison of each method for the data analysis}
The distance correlations between $\Y$ and $\widehat\B(\w)\trans \X$ are reported in columns 2-4 in Table \ref{DC}, where $\widehat\B(\w)$ is the estimate by our proposals (i.e. DPSIR, DPSAVE, DPDR), and the distance correlations between $\Y$ and $\widehat\B\trans \X$ are reported in columns 5-7, where $\widehat\B$ is the estimate by PDEE methods (i.e. PDEE-SIR, PDEE-SAVE, PDEE-DR).
\begin{table}[H]
	\centering
	\scriptsize
	\caption{Distance correlation for each estimation approaches}	\label{DC}
	\begin{tabular}{c|ccc|ccc}
		\hline
		Dataset & DPSIR & DPSAVE & DPDR &  PDEE-SIR & PDEE-SAVE & PDEE-DR\\
		\hline
		\text{Body Fat} & 0.8301 & 0.2149 & 0.5310 &  0.6906 & 0.1664 & 0.2665\\
		\hline
		\text{Wage} & 0.5195 & 0.2215 & 0.5044 &  0.5171 & 0.1512 & 0.1500\\
		\hline
		\text{Hongkong environmental} & 0.4592 & 0.2060 & 0.1942 &  0.2155 & 0.1379 & 0.1193\\
		\hline
		\text{Boston Housing} & 0.9112 & 0.6618 & 0.8688 & 0.8030 & 0.3418 & 0.5408\\
		\hline
	\end{tabular}
\end{table}
It's obvious that our dynamic methods consistently beat the PDEE approaches for each variant of SIR, SAVE, DR
procedure. To be specific, for the Body Fat dataset, the dynamic partial SIR estimation gives
a distance correlation of $0.83$, compared with $0.69$ resulted from PDEE-SIR. For the Boston Housing dataset,
the distance correlation from dynamic partial DR is $0.87$ comparing with $0.54$ from PDEE-DR. The more accurate
dimension reduction estimates we obtained could greatly facilitate further modeling and analysis. 

\section{Discussions}
In this paper, we propose a dynamic approach to better dealing with partial sufficient dimension reduction. 
For the purpose of statistical estimation, a kernel matrix is developed and its asymptotic properties are
thoroughly investigated. We also develop a nonparametric ladle estimator to determine the structural
dimension dynamically.
For future work, we plan to investigate how to apply dynamic PDR to conduct
variable selections, which is of special practical importance since
different $\w$ may lead to different variable selection results.
\clearpage
\section{Appendix}
\subsection{Regularity conditions}
To prove the theoretical results of this paper, we need some regularity conditions. They are not the weakest possible conditions, but they are imposed to facilitate the proofs.
\begin{description}
\item  (C1). (The density of the index variable) We assume that $W$ has a compact support. Then, on the support, we further assume that the probability density function of $W$, denoted by $f(W)$, is bounded away from 0 and has continuous derivatives up to second order.
\item  (C2). (The moment requirement) For any $1\leq j_{1},j_{2}\leq p$, there exists a constant $\delta\in[0,1]$, such that $\sup_{\w}\EE\{|\X_{j_{1}}(\w)\X_{j_{2}}(\w)|\}^{2+\delta}<\infty$.
\item  (C3). (Smoothness of the conditional mean) Assume that the conditional mean $\m_{j}(\cdot)$ has continuous derivatives up to second order.
\item  (C4).  (Smoothness of the conditional variance) We assume that $\EE\{\X_{j_{1}}^{k_{1}}\X_{j_{2}}^{k_{2}}\X_{j_{3}}^{k_{3}}\X_{j_{4}}^{k_{4}}|\W=\w\}$ has continuous derivatives up to second order  in $\w$ for $k_{1},k_{2},k_{3},k_{4}\in\{0,1\}$, where $j_{1},j_{2},j_{3},j_{4}$ are not necessarily different components in the $(\X,\W)$ vectors.
\item  (C5). (Smoothness of the conditional indicator mean) Assume that the conditional indicator mean $\EE\{\X\1(\Y\in\J_{l})|\W=\w\}$ has continuous derivatives up to second order, where $l=1,\ldots,H$.
\item  (C6).  (Smoothness of the conditional indicator variance) We assume that $\EE\{\X_{j_{1}}^{k_{1}}\X_{j_{2}}^{k_{2}}\X_{j_{3}}^{k_{3}}\\\X_{j_{4}}^{k_{4}}\1(\widetilde{\Y}=l)|\W=\w\}$ has continuous derivatives up to second order  in $\w$ for $k_{1},k_{2},k_{3},k_{4}\in\{0,1\}$, $l=1,\ldots,H$, where $j_{1},j_{2},j_{3},j_{4}$ are not necessarily different components in the $(\X,\W)$ vectors.
\item  (C7). (The bandwidth) $h\rightarrow 0$ and $nh^{5}\rightarrow c>0$ for some $c>0$.
\item  (C8). (The kernel function) We assume that $K(\w)$ is a bounded probability density function symmetric about 0. Furthermore, for the $\delta$ in (C2), we assume that $\int K^{2+\delta}(v)v^{j}dv<\infty$, for $j=0,1,2$. Lastly, for two arbitrary indices $\w_{1}$ and $\w_{2}$, we must have $|K(\w_{1})-K(\w_{2})|\leq K_{c}|\w_{1}-\w_{2}|$ for some positive constant $K_{c}$.
\item  (C9)  The bootstrap estimator $\M^{*}$ satisfies
\begin{align}\label{Assumption 2}
  (nh)^{1/2}\{\vech(\M^{*})-\vech(\widehat{\M})-\vech(\B^{*})\}\xrightarrow{d} N(0,\text{Var}_{F}[\vech\{\bfH(\X,\Y)\}]).
\end{align}
\item   (C10)  For any sequence of nonnegative random variables $\{\Z_{n}:n=1,2,\ldots\}$ involved hereafter, if $\Z_{n}=\Op(c_{n})$ for some sequence $\{c_{n}:n\in N\}$ with $c_{n}>0$, then $\EE(c^{-1}_{n}\Z_{n})$ exists for each $n$ and $\EE(c^{-1}_{n}\Z_{n})=\Op(1)$.
\item  (C11)  {\small $\sum\limits_{n=3}^{\infty}(h/\log\log n)\EE\{S^{2}\I(|S|>a_{n})\}<\infty$}, where $S$ represents response variable in nonparametric regression function and $a_{n}=\op\{(nh^{-1}\log\log n)^{1/2}/(\log n)^{2}\}$.
\item  (C12)  {\small $\lim_{\epsilon\rightarrow 0}\lim\sup_{n\rightarrow \infty}\sup_{m\in \Gamma_{n,\epsilon}}|h(m)/h(n)-1|=0$, where $\Gamma_{n,\epsilon}=\{m:|m-n|\leq\epsilon n\}$}.
\end{description}

Both conditions (C1) and (C2) are standard technical assumptions [\cite{Yin2010}], and conditions (C3), (C4), (C5) and (C6) are necessary smoothness constraints [\cite{Fan1993}; \cite{Yao1998}]. By condition (C7) we know that the optimal convergence rate of $n^{-1/5}$ can be used. Condition (C8) is a standard requirement for the kernel function [\cite{Yao1996}] , which is trivially satisfied by both Gaussian kernel and Epanechnikov kernel.

Condition (C9) is quite mild: it is satisfied if the statistical functional $\M$ is Fr$\acute{e}$chet differentiable [\cite{Luo2016}], where $\B^{*}$ represents the bias term of $\M^{*}-\widehat{\M}$. Condition (C10) amounts to asserting that the asymptotic behaviour of  $(nh)^{1/2}(\M^{*}-\widehat{\M}-\B^{*})$ mimics that of $(nh)^{1/2}(\widehat{\M}-\M-\B)$, where $\B$ represents the bias term of $\widehat{\M}-\M$. The validity of this self-similarity has been discussed [\cite{Bickel1981}, \cite{Luo2016}], where $\vech(\cdot)$ is the vectorization of the upper triangular part of a matrix and $\text{var}_{F}[\vech\{\bfH(\X,\Y)\}]$ is positive definite. Condition (C11) and condition (C12) are widely used in law of the iterated logarithm for nonparametric regression \cite{Hardle1984}. These conditions should not be too restrictive on the applicability of our estimator.

\subsection{The proofs of the main results}
\noindent {\sc Proof of Proposition \ref{partial least squares}.} Following \cite{Li2003}, for convenience, we often use the abbreviation
\begin{align*}
\EE\{f(\X,\Y)|g(\X),\W=\w\}\triangleq \EE\{f(\X_{\w},Y_{\w})|g(\X_{\w})\}.
\end{align*}In our case, we define
\begin{align*}
  \EE(\X|\Y,\W=\w)=\EE(\X_{\w}|Y_{\w}),\quad
  \EE\{\X|\Y,\B\trans(\w)\X,\W=\w\}=\EE\{\X_{\w}|Y_{\w},\B\trans(\w)\X_{\w}\}.
\end{align*}Since $\X_{\w}$ satisfies linear conditional mean (A1), given $\W=\w$, it's easy for us to get
\begin{align*}
\EE\{\X_{\w}|\B\trans(\w)\X_{\w}\}=\left[\B(\w)\{\B\trans(\w)\bfSigw\B(\w)\}^{-1}\B\trans(\w)\bfSigw\right]\trans\X_{\w}.
\end{align*}Hence,
\begin{align*}
  &\quad \bfSig_{\w}^{-1}\{\EE(\X|\Y,\W=\w)-\m(\w)\}\\
  &=\bfSig_{\w}^{-1}\{\EE(\X_{\w}|Y_{\w})-\EE(\X_{\w})\}\\
  &=\bfSig_{\w}^{-1}[\EE\{\EE(\X_{\w}|Y_{\w},\B\trans(\w)\X_{\w})|Y_{\w}\}-\EE\{\EE(\X_{\w}|\B\trans(\w)\X_{\w})\}]\\
  &=\bfSig_{\w}^{-1}[\EE\{\EE(\X_{\w}|\B\trans(\w)\X_{\w})|Y_{\w}\}-\EE\{\EE(\X_{\w}|\B\trans(\w)\X_{\w})\}]\\
  &=\bfSig_{\w}^{-1}[\B(\w)\{\B\trans(\w)\bfSigw\B(\w)\}^{-1}\B\trans(\w)\bfSigw]\trans \{\EE(\X_{\w}|Y_{\w})-\EE(\X_{\w})\}\\
  &=\B(\w)\{\B\trans(\w)\bfSigw\B(\w)\}^{-1}\B\trans(\w)\bfSigw\bfSig_{\w}^{-1}\{\EE(\X|\Y,\W=\w)-\m(\w)\}.
\end{align*}
The second equality follows tower property about conditional expectation. Then the third equality is based on the fact that
\begin{align*}
  Y_{\w}\indep\X_{\w}|\B\trans(\w)\X_{\w}.
\end{align*}Denote  $\proba_{\B(\w)}=\B(\w)\{\B\trans(\w)\bfSigw\B(\w)\}^{-1}\B\trans(\w)\bfSigw$. Then we have,
\begin{align*}
  \bfSig_{\w}^{-1}\{\EE(\X|\Y,\W=\w)-\m(\w)\}=\proba_{\B(\w)}\bfSig_{\w}^{-1}\{\EE(\X|\Y,\W=\w)-\m(\w)\}.
\end{align*}
\eop

\noindent {\sc Proof of Proposition \ref{equal inquality}.}
At first, we need to prove the left side of the equation \eqref{identity},
\begin{align*}
  \EE(\X|\widetilde{\Y}=l,\w)&=\int\X\frac{f(\X,\widetilde{\Y}=l,\w)}{f(\widetilde{\Y}=l,\w)}d\X\\
&=\int\X\frac{f(\X,\w|\widetilde{\Y}=l)\proba(\widetilde{\Y}=l)}{f(\w|\widetilde{\Y}=l)\proba(\widetilde{\Y}=l)}d\X\\
&=\frac{\int\X f(\X,\w|\widetilde{\Y}=l)d\X}{f(\w|\widetilde{\Y}=l)}.
\end{align*}
Then we show the right side of the equation \eqref{identity}, since
\begin{equation}\label{r1}
\begin{aligned}
   \EE(\1(\widetilde{\Y}=l)|\w)&=\frac{\1(\widetilde{\Y}=l)\sum\limits_{k=1}^{H}f(\w|\widetilde{\Y}=k)\proba(\widetilde{\Y}=k)}{f(\w)}=\frac{f(\w|\widetilde{\Y}=l)\proba(\widetilde{\Y}=l)}{f(\w)},
\end{aligned}
\end{equation}where $f(\w)$ is the density function of $\w$. Hence
\begin{equation}\label{r2}
\begin{aligned}
  \EE(\X\1(\widetilde{\Y}=l)|\w)&=\int\X\1(\widetilde{\Y}=l)\frac{f(\X,\widetilde{\Y}=l,\w)}{f(\w)}d\X d\widetilde{\Y}\\
  &=\int\X\1(\widetilde{\Y}=l)\frac{\sum\limits_{k=1}^{H}f(\X,\w|\widetilde{\Y}=k)\proba(\widetilde{\Y}=k)}{f(\w)}d\X\\
&=\frac{\int\X f(\X,\w|\widetilde{\Y}=l)\proba(\widetilde{\Y}=l)}{f(\w)}.
\end{aligned}
\end{equation}
Thus, we get
\begin{align*}
  \frac{\EE\{\X\1(\widetilde{\Y}=l)|\w\}}{\EE\{\1(\widetilde{\Y}=l)|\w\}}&=\frac{\int\X f(\X,\w|\widetilde{\Y}=l)P(\widetilde{\Y}=l)d\X/f(\w)}{f(\w|\widetilde{\Y}=l)P(\widetilde{\Y}=l)/f(\w)}\\
&=\frac{\int f(\X,\w|\widetilde{\Y}=l)d\X}{f(\w|\widetilde{\Y}=l)}=\EE(\X|\widetilde{\Y}=l,\w).
\end{align*}
\eop

\noindent {\sc Proof of Proposition \ref{nonrandom variance condition 1}.} Use the similar abbreviation in Proposition \ref{partial least squares},
\begin{align*}
  \Cov(\X|\Y,\W=\w)=\Cov(\X_{\w}|Y_{\w}).
\end{align*}Assume that $\X_{\w}$ satisfies linear conditional mean (A1) and constant conditional variance (A2), furthermore, constant conditional variance (A2) implies {\footnotesize $\Cov\{\X_{\w}|\B\trans(\w)\X_{\w}\}=\bfSig_{\w}\{\I_{p}-\proba_{\B(\w)}\}$}. Conditional on $\W=\w$, it's easy for us to get
\begin{align*}
  &\quad\bfSig_{\w}^{-1}\Cov(\X|\Y,\W=\w)=\bfSig_{\w}^{-1}\Cov(\X_{\w}|Y_{\w})\\
  &=\bfSig_{\w}^{-1}\EE[\Cov\{\X_{\w}|\B\trans(\w)\X_{\w},Y_{\w}\}|Y_{\w}]+\bfSig_{\w}^{-1}\Cov[\EE\{\X_{\w}|\B\trans(\w)\X_{\w},Y_{\w}\}|Y_{\w}]\\
  &=\bfSig_{\w}^{-1}\EE[\Cov\{\proba_{\B(\w)}\trans\X_{\w}+\Q_{\B(\w)}\trans\X_{\w}|\B\trans(\w)\X_{\w}\}|Y_{\w}]+\bfSig_{\w}^{-1}\Cov[\EE\{\X_{\w}|\B\trans(\w)\X_{\w}\}|Y_{\w}]\\
  &=\bfSig_{\w}^{-1}\Q_{\B(\w)}\trans \EE[\Cov\{\X_{\w}|\B\trans(\w)\X_{\w}\}|Y_{\w}]\Q_{\B(\w)}+\bfSig_{\w}^{-1}\proba_{\B(\w)}\trans \Cov(\X_{\w}|Y_{\w})\proba_{\B(\w)}\\
  &=\Q_{\B(\w)}\bfSig_{\w}^{-1} \EE[\Cov\{\X_{\w}|\B\trans(\w)\X_{\w}\}|Y_{\w}]\Q_{\B(\w)}+\proba_{\B(\w)}\bfSig_{\w}^{-1} \Cov(\X_{\w}|Y_{\w})\proba_{\B(\w)}\\
  &=\Q_{\B(\w)}+\proba_{\B(\w)}\bfSig_{\w}^{-1} \Cov(\X_{\w}|Y_{\w})\proba_{\B(\w)},
\end{align*}where $\Q_{\B(\w)}=\I_{p}-\proba_{\B(\w)}$. Such that we can derive that
\begin{align*}
\I_{p}-\bfSig_{\w}^{-1}\Cov(\X|\Y,\W=\w)=\proba_{\B(\w)}\{\I_{p}-\bfSig_{\w}^{-1}\Cov(\X|\Y,\W=\w)\}\proba_{\B(\w)}.
\end{align*}The proof is completed.
\eop

\noindent {\sc Proof of Proposition \ref{nonrandom variance condition 2}.}
Assume that $\X_{\w}$ satisfies linear conditional mean (A1) and constant conditional variance (A2), conditional on $\W=\w$, it's easy for us to get
{\small
\begin{align*}
  &\quad\bfSig_{\w}^{-1}\EE\{(\X-\breve{\X})(\X-\breve{\X})\trans|\Y,\breve{\Y},\W=\w\}\\
  &=\bfSig_{\w}^{-1}\EE\{(\X_{\w}-\breve{\X}_{\w})(\X_{\w}-\breve{\X}_{\w})\trans|Y_{\w},\breve{\Y}_{\w}\}\\
  &=\bfSig_{\w}^{-1}\{\EE(\X_{\w}\X_{\w}\trans|Y_{\w})-\EE(\X_{\w}|Y_{\w})\EE(\breve{\X}_{\w}\trans|\breve{\Y}_{\w})-\EE(\breve{\X}_{\w}|\breve{\Y}_{\w})\EE(\X_{\w}\trans|Y_{\w})+\EE(\breve{\X}_{\w}\breve{\X}_{\w}\trans|\breve{\Y}_{\w})\}\\
  &=\bfSig_{\w}^{-1}\Big(\EE\left[\EE\{\X_{\w}\X_{\w}\trans|\B\trans(\w)\X_{\w},Y_{\w}\}|Y_{\w}\right]-\EE[\EE\{\X_{\w}|\B\trans(\w)\X_{\w},Y_{\w}\}|Y_{\w}]\EE[\EE\{\breve{\X}_{\w}\trans|\B\trans(\w)\breve{\X}_{\w},\breve{\Y}_{\w}\}|\breve{\Y}_{\w}]\\
  &-\EE[\EE\{\breve{\X}_{\w}|\B\trans(\w)\breve{\X}_{\w},\breve{\Y}_{\w}\}|\breve{\Y}_{\w}]\EE[\EE\{\X_{\w}\trans|\B\trans(\w)\X_{\w},Y_{\w}\}|Y_{\w}]+\EE\left[\EE\{\X_{\w}\X_{\w}\trans|\B\trans(\w)\X_{\w},Y_{\w}\}|Y_{\w}\right]\Big)\\
  &=\bfSig_{\w}^{-1}\Big(\EE\left[\Cov\{\X_{\w}|\B\trans(\w)\X_{\w}\}|Y_{\w}\right]+\EE[\EE\{\X_{\w}|\B\trans(\w)\X_{\w}\}\EE\{\X_{\w}\trans|\B\trans(\w)\X_{\w}\}|Y_{\w}]\\
  &-\EE[\EE\{\X_{\w}|\B\trans(\w)\X_{\w}\}|Y_{\w}]\EE[\EE\{\breve{\X}_{\w}\trans|\B\trans(\w)\breve{\X}_{\w}\}|\breve{\Y}_{\w}]-\EE[\EE\{\breve{\X}_{\w}|\B\trans(\w)\breve{\X}_{\w}\}|\breve{\Y}_{\w}]\EE[\EE\{\X_{\w}\trans|\B\trans(\w)\X_{\w}\}|Y_{\w}]\\
  &+\EE\left[\Cov\{\breve{\X}_{\w}|\B\trans(\w)\breve{\X}_{\w}\}|\breve{\Y}_{\w}\right]+\EE\left[\EE\{\breve{\X}_{\w}|\B\trans(\w)\breve{\X}_{\w}\}\EE\{\breve{\X}_{\w}\trans|\B\trans(\w)\breve{\X}_{\w}\}|\breve{\Y}_{\w}\right]\Big)\\
  &=\Q_{\B(\w)}+\proba_{\B(\w)}\bfSig_{\w}^{-1}\EE(\X_{\w}\X_{\w}\trans|Y_{\w})\proba_{\B(\w)}-\proba_{\B(\w)}\bfSig_{\w}^{-1}\EE(\X_{\w}|Y_{\w})\EE(\breve{\X}_{\w}\trans|\breve{\Y}_{\w})\proba_{\B(\w)}\\
  &-\proba_{\B(\w)}\bfSig_{\w}^{-1}\EE(\breve{\X}_{\w}|\breve{\Y}_{\w})\EE(\X_{\w}\trans|Y_{\w})\proba_{\B(\w)}+\Q_{\B(\w)}+\proba_{\B(\w)}\bfSig_{\w}^{-1}\EE(\breve{\X}_{\w}\breve{\X}_{\w}\trans|\breve{\Y}_{\w})\proba_{\B(\w)}.
\end{align*}}Recall that $\Q_{\B(\w)}=\I_{p}-\proba_{\B(\w)}$. Then we can derive that
\begin{align*}
&\quad\bfSig_{\w}^{-1}[2\bfSig_{\w}-\EE\{(\X-\breve{\X})(\X-\breve{\X})\trans|\Y,\breve{\Y},\W=\w\}]\\
&=\proba_{\B(\w)}[2\I_{p}-\bfSig_{\w}^{-1}\EE\{(\X-\breve{\X})(\X-\breve{\X})\trans|\Y,\breve{\Y},\W=\w\}]\proba_{\B(\w)}.
\end{align*}
The proof is completed.
\eop

\noindent {\sc Proof of Theorem \ref{generalized eigenvalue decomposition}.}
Denote $k_{i}(\w)=K(\frac{\w_{i}-\w}{h})$, we get
\begin{align*}
  \frac{1}{nh}\sum\limits_{i=1}^{n}k_{i}(\w)=\frac{1}{nh}\sum\limits_{i=1}^{n}K\left(\frac{\w_{i}-\w}{h}\right)=\widehat{f}(\w).
\end{align*}Then under condition (C1) and (C8), by \cite{Yin2010}, we conclude that $|\widehat{f}(\w)-f(\w)|=\Op_{P}\left\{\sqrt{\frac{\log(n)}{nh}}+h^{2}\right\}$. Next under condition (C1), we have uniformly for $\w\in G$,
\begin{align*}
  \left\{\frac{1}{nh}\sum\limits_{i=1}^{n}k_{i}(\w)\right\}^{-1}=f^{-1}(\w)\left\{1+\Op_{P}\left(\sqrt{\frac{\log(n)}{nh}}+h^{2}\right)\right\}.
\end{align*}
Define for $j=1,2$,
\begin{align*}
  s_{j}(\w)=\frac{1}{nh}\sum\limits_{i=1}^{n}\left(\frac{\w_{i}-\w}{h}\right)K\left(\frac{\w_{i}-\w}{h}\right).
\end{align*}Then, by the same method as in \cite{Yin2010}, we have
\begin{equation}\label{A.4}
\begin{aligned}
  s_{1}(\w)-h\dot{\proba}_{l,\w}\omega_{2}=\Op_{p}\left(\sqrt{\frac{\log(n)}{nh}}+\op(h)\right)=\op_{P}(h),
\end{aligned}
\end{equation}where $\omega_{2}=\int_{-\infty}^{\infty}\w^{2}K(\w)d\w$. Next, by Lemma 2 of \cite{Yao1998} we know that
\begin{equation}\label{A.5}
 \begin{aligned}
   \sup\limits_{\w\in\mR}|s_{2}(\w)-f(\w)\omega_{2}|=\op_{P}(1).
 \end{aligned}
\end{equation}By Taylor's expansion, we then have
\begin{equation}\label{eq:p}
  \begin{aligned}
    &\quad \widehat{\proba}_{l,\w}-\proba_{l,\w}\\
    &=\frac{1}{nhf(\w)}\left\{1+\Op_{P}\left(\sqrt{\frac{\log(n)}{nh}}+h^{2}\right)\right\}\sum\limits_{i=1}^{n}K\left(\frac{\w_{i}-\w}{h}\right)\left\{\1(\widetilde{\Y}_{i}=l)-\proba_{l,\w_{i}}\right\}\\
    &+\frac{1}{f(\w)}\left\{1+\Op_{P}\left(\sqrt{\frac{\log(n)}{nh}}+h^{2}\right)\right\}\left\{h\dot{\proba}_{l,\w}s_{1}(\w)+\frac{h^{2}\ddot{\proba}_{l,\w}}{2}s_{2}(\w)+\op(h^{2})\right\}
  \end{aligned}
\end{equation}Then it follows by \eqref{A.4} and \eqref{A.5} that
\begin{equation}
  \begin{aligned}
    &\quad \widehat{\proba}_{l,\w}-\proba_{l,\w}=\C_{\proba,\w}+\B_{\proba,\w}+\Op_{P}\{\R_{1}(\w)\}.
  \end{aligned}
\end{equation}where
\begin{align*}
  \C_{\proba,\w}=\frac{1}{nhf(\w)}\sum\limits_{i=1}^{n}k_{i}(\w)\left\{\1(\widetilde{\Y}_{i}=l)-\proba_{l,\w_{i}}\right\},\quad
  \B_{\proba,\w}=\frac{h^{2}\omega_{2}}{2}\left(2\frac{\dot{f}(\w)}{f(\w)}\dot{\proba}_{l,\w}+\ddot{\proba}_{l,\w}\right),
  \end{align*} and
  \begin{align*}
  \R_{1}(\w)=\frac{1}{nf(\w)}\left[\left|\sum\limits_{i=1}^{n}K\left(\frac{\w_{i}-\w}{h}\right)\{\1(\widetilde{\Y}_{i}=l)-\proba_{l,\w_{i}}\}\right|\right]+\op(h^{2}).
\end{align*}
$\widehat{\proba}_{l,\w}-\proba_{l,\w}$ follows by the similar argument. Then,
\begin{equation}
  \begin{aligned}
    \widehat{\U}_{l,\w}-\U_{l,\w}=\C_{\U,\w}+\B_{\U,\w}+\Op_{P}\{\R_{2}(\w)\},
  \end{aligned}
\end{equation}where
\begin{align*}
  \C_{\U,\w}=\frac{1}{nhf(\w)}\sum\limits_{i=1}^{n}k_{i}(\w)\left\{\x_{i}\1(\widetilde{\Y}_{i}=l)-\U_{l,\w_{i}}\right\},\quad
  \B_{\U,\w}=\frac{h^{2}\omega_{2}}{2}\left(2\frac{\dot{f}(\w)}{f(\w)}\dot{\U}_{l,\w}+\ddot{\U}_{l,\w}\right),
\end{align*}and
\begin{align*}
\R_{2}(\w)=\frac{1}{nf(\w)}\left\{\left|\sum\limits_{i=1}^{n}K\left(\frac{\w_{i}-\w}{h}\right)\left(\x_{i}\1(\widetilde{\Y}_{i}=l)-\U_{l,\w_{i}}\right)\right|\right\}+\op(h^{2}),
\end{align*}\cite{Yin2010} has already shown us that
\begin{align*}
  \widehat{\m}(\w)-\m(\w)=\C_{\m,\w}+\B_{\m,\w}+\Op_{P}\{\R_{3}(\w)\},
  \quad\widehat{\bfSig}_{\w}-\bfSigw=\C_{\bfSig,\w}+\B_{\bfSig,\w}+\Op_{P}\{\R_{4}(\w)\},
\end{align*}where
\begin{align*}
  &\C_{\m,\w}=\frac{1}{nhf(\w)}\sum\limits_{i=1}^{n}k_{i}(\w)\left\{\x_{i}-\m(\w_{i})\right\},\quad
  \B_{\m,\w}=\frac{h^{2}\omega_{2}}{2}\left(2\frac{\dot{f}(\w)}{f(\w)}\dot{\m}(\w)+\ddot{\m}(\w)\right),\\
  &\C_{\bfSig,\w}=\frac{1}{nhf(\w)}\sum\limits_{i=1}^{n}k_{i}(\w)\left[\left\{\x_{i}-\widehat{\m}(\w_{i})\right\}\left\{\x_{i}-\widehat{\m}(\w_{i})\right\}\trans-\bfSigw-(\w_{i}-\w)\dot{\bfSig}_{\w}\right],\\
  &\B_{\bfSig,\w}=h^{2}\dot{\bfSig}_{\w}\omega_{2}\frac{\dot{f}(\w)}{f(\w)},
  \end{align*}and
{\small
\begin{align*}
&\R_{3}(\w)=\frac{1}{nf(\w)}\left\{\left|\sum\limits_{i=1}^{n}K\left(\frac{\w_{i}-\w}{h}\right)\left(\x_{i}-\m(\w_{i})\right)\right|\right\}+\op(h^{2}),\\
&\R_{4}(\w)=\frac{1}{nf(\w)}\left\{\left|\sum\limits_{i=1}^{n}K\left(\frac{\w_{i}-\w}{h}\right)\left[\left\{\x_{i}-\widehat{\m}(\w_{i})\right\}\left\{\x_{i}-\widehat{\m}(\w_{i})\right\}\trans-\bfSigw-(\w_{i}-\w)\dot{\bfSig}_{\w}\right]\right|\right\}+\op(h^{2}),\\
\end{align*}}Thus
  \begin{align*}
    &\quad \widehat{\M}_{\SIR}(\w)-\M_{\SIR}(\w)\\
    &=\sum\limits_{l=1}^{H}\left(\frac{\widehat{\U}_{l,\w}\widehat{\U}\trans_{l,\w}}{\widehat{\proba}_{l,\w}}-\sum\limits_{l=1}^{H}\frac{\U_{l,\w}\U_{l,\w}\trans}{\proba_{l,\w}}\right)+\m(\w)\m(\w)\trans-\widehat{\m}(\w)\widehat{\m}(\w)\trans\\
    &=\sum\limits_{l=1}^{H}\left\{\frac{(\widehat{\U}_{l,\w}-\U_{l,\w})\U\trans_{l,\w}}{\proba_{l,\w}}+\frac{\U_{l,\w}(\widehat{\U}_{l,\w}-\U_{l,\w})\trans}{\proba_{l,\w}}-\frac{1}{\proba^{2}_{l,\w}}(\widehat{\proba}_{l,\w}-\proba_{l,\w})\U_{l,\w}\U\trans_{l,\w}\right\}\\
    &-\left\{\widehat{\m}(\w)-\m(\w)\}\m(\w)\trans-\m(\w)\{\widehat{\m}(\w)-\m(\w)\right\}\trans+\op_{P}\left(\frac{1}{\sqrt{nh}}\right)\\
    &=\B_{\SIR}(\w)+\C_{\SIR}(\w)+\op_{P}\left(\frac{1}{\sqrt{nh}}\right),
    \end{align*}where
\begin{equation}\label{BSIR}
\begin{aligned}
    \B_{\SIR}(\w)&=\sum\limits_{l=1}^{H}\Bigg\{\frac{\B_{\U,\w}\U\trans_{l,\w}+\U_{l,\w}\B_{\U,\w}\trans}{\proba_{l,\w}}-\frac{1}{\proba^{2}_{l,\w}}\B_{\proba,\w}\U_{l,\w}\U\trans_{l,\w}\Bigg\}-\B_{\m,\w}\m(\w)\trans-\m(\w)\B_{\m,\w}\trans,
\end{aligned}
\end{equation}and
\begin{align*}
  \C_{\SIR}(\w)&=\sum\limits_{l=1}^{H}\Bigg\{\frac{\C_{\U,\w}\U\trans_{l,\w}+\U_{l,\w}\B_{\U,\w}\trans}{\proba_{l,\w}}-\frac{1}{\proba^{2}_{l,\w}}\C_{\proba,\w}\U_{l,\w}\U\trans_{l,\w}\Bigg\}-\C_{\m,\w}\m(\w)\trans-\m(\w)\C_{\m,\w}\trans.
\end{align*}Following by the similar argument of Theorem 1 in \cite{Yin2010}, we have
\begin{equation}
  \begin{aligned}
    \sqrt{nh}\left(\vech\{\widehat{\M}_{\SIR}(\w)\}-\vech\{M_{\SIR}(\w)\}-\vech\{\B_{\SIR}(\w)\}\right) \xrightarrow{d}  {\bf{N}} (\bm{0},f^{-1}(\w)\omega_{0}\C^{\SIR}(\w)),
  \end{aligned}
\end{equation}where
\begin{equation}
  \begin{aligned}\label{w}
  \omega_{0}=\int_{-\infty}^{\infty}K^{2}(\w)d\w.
  \end{aligned}
\end{equation}Hence
 \begin{equation}\label{CSIR}
   \begin{aligned}
     \C^{\SIR}(\w)=\Cov\left[\vech\{\C_{\SIR}(\w)\}|\w\right].
   \end{aligned}
 \end{equation}This completes the proof of the part one of Theorem \ref{generalized eigenvalue decomposition}, then we prove the part two. Observe that $\la_{k}(\w)$ and $\be_{k}(\w)$ satisfy the following singular value decomposition equation:
\begin{align*}
  \G\trans(\w)\G(\w)\be_{k}(\w)=\la^{2}_{k}(\w)\be_{k}(\w),\quad k=1,\ldots,p;
\end{align*}where $\G(\w)=\bfSig_{\w}^{-1}\M(\w)$. Hence,
\begin{align*}
  \M\trans(\w)\bfSig_{\w}^{-1}\bfSig_{\w}^{-1}\M(\w)\be_{k}(\w)=\la^{2}_{k}(\w)\be_{k}(\w),\quad k=1,\ldots,p;
\end{align*}where $\be_{k}\trans(\w)\be_{k}(\w)=1$ and $\be_{k}\trans(\w)\be_{\rho}(\w)=0$ for $k\neq \rho$. Similarly, in the sample level, we have
\begin{align*}
\widehat{\M}\trans(\w)\widehat{\bfSig}_{\w}^{-1}\widehat{\bfSig}_{\w}^{-1}\widehat{\M}(\w)\widehat{\be}_{k}(\w)=\widehat{\la}_{k}^{2}(\w)\widehat{\be}_{k}(\w),\quad k=1,\ldots,p;
\end{align*}and $\widehat{\be}_{k}\trans(\w)\widehat{\be}_{k}(\w)=1$ and $\widehat{\be}_{k}\trans(\w)\widehat{\be}_{\rho}(\w)=0$ for $k\neq \rho$.
The singular value decomposition form in the sample level implies that
\begin{equation}\label{A:1}
  \begin{aligned}
  &\{\bfSig_{\w}^{-1}\M(\w)\}\trans\bfSig_{\w}^{-1}\M(\w)\{\widehat{\be}_{k}(\w)-\be_{k}(\w)\}+\M(\w)\trans(\widehat{\bfSig}_{\w}^{-1}-\bfSig_{\w}^{-1})\trans\bfSig_{\w}^{-1}\M(\w)\be_{k}(\w)\\
  +&\{\widehat{\M}(\w)-\M(\w)\}\trans\bfSig_{\w}^{-1}\bfSig_{\w}^{-1}\M(\w)\be_{k}(\w)+\{\bfSig_{\w}^{-1}\M(\w)\}\trans(\widehat{\bfSig}_{\w}^{-1}-\bfSig_{\w}^{-1})\M(\w)\be_{k}(\w)\\
  +&\{\bfSig_{\w}^{-1}\M(\w)\}\trans\bfSig_{\w}^{-1}\{\widehat{\M}(\w)-\M(\w)\}\be_{k}(\w)=\la_{k}(\w)\{\widehat{\la}_{k}(\w)-\la_{k}(\w)\}\be_{k}(\w)\\
  +&\{\widehat{\la}_{k}(\w)-\la_{k}(\w)\}\la_{k}(\w)\be_{k}(\w)+\la_{k}^{2}(\w)\{\widehat{\be}_{k}(\w)-\be_{k}(\w)\}+\op_{p}(\frac{1}{\sqrt{nh}}),
  \end{aligned}
\end{equation}for $k=1,\ldots,d$. Multiply both sides of \eqref{A:1} by $\be\trans_{k}(\w)$ from the left, we get
\begin{align*}
  &  \be\trans_{k}(\w)\Big[\M(\w)\trans(\widehat{\bfSig}_{\w}^{-1}-\bfSig_{\w}^{-1})\trans\bfSig_{\w}^{-1}\M(\w)+\{\widehat{\M}(\w)-\M(\w)\}\trans\bfSig_{\w}^{-1}\bfSig_{\w}^{-1}\M(\w)\\
  &+\{\bfSig_{\w}^{-1}\M(\w)\}\trans(\widehat{\bfSig}_{\w}^{-1}-\bfSig_{\w}^{-1})\M(\w)+\{\bfSig_{\w}^{-1}\M(\w)\}\trans\bfSig_{\w}^{-1}\{\widehat{\M}(\w)-\M(\w)\}\Big]\be_{k}(\w)\\
  &=\{\widehat{\la}_{k}(\w)-\la_{k}(\w)\}\la_{k}(\w)+\la_{k}\{\widehat{\la}_{k}(\w)-\la_{k}(\w)\}+\op_{p}(\frac{1}{\sqrt{nh}}),
\end{align*}which further suggests that
\begin{equation}\label{o:11}
  \begin{aligned}
    \widehat{\la}_{k}(\w)=&\la_{k}(\w)+\frac{\be\trans_{k}(\w)}{2\la_{k}(\w)}\Bigg[\M(\w)\trans(\widehat{\bfSig}_{\w}^{-1}-\bfSig_{\w}^{-1})\trans\bfSig_{\w}^{-1}\M(\w)\\
    +&\{\widehat{\M}(\w)-\M(\w)\}\trans\bfSig_{\w}^{-1}\bfSig_{\w}^{-1}\M(\w)+\{\bfSig_{\w}^{-1}\M(\w)\}\trans(\widehat{\bfSig}_{\w}^{-1}-\bfSig_{\w}^{-1})\M(\w)\\
    +&\{\bfSig_{\w}^{-1}\M(\w)\}\trans\bfSig_{\w}^{-1}\{\widehat{\M}(\w)-\M(\w)\}\Bigg]\be_{k}(\w)+\op_{p}(\frac{1}{\sqrt{nh}}).
  \end{aligned}
\end{equation}By lemma A.2 of \cite{Cook2005} we know that
\begin{align*}
  \widehat{\bfSig}_{\w}^{-1}-\bfSig_{\w}^{-1}=-\bfSig_{\w}^{-1}(\widehat{\bfSig}_{\w}-\bfSig_{\w})\bfSig_{\w}^{-1}+\op_{p}(\frac{1}{\sqrt{nh}}).
\end{align*}Hence, equation \eqref{o:11} becomes
\begin{equation}\label{A:2}
  \begin{aligned}
    \widehat{\la}_{k}(\w)=&\la_{k}(\w)+\frac{\be\trans_{k}(\w)}{2\la_{k}(\w)}\Big[-\M(\w)\trans\bfSig_{\w}^{-1}(\widehat{\bfSig}_{\w}-\bfSig_{\w})\bfSig_{\w}^{-1}\bfSig_{\w}^{-1}\M(\w)\\
    +&\{\widehat{\M}(\w)-\M(\w)\}\trans\bfSig_{\w}^{-1}\bfSig_{\w}^{-1}\M(\w)-\{\bfSig_{\w}^{-1}\M(\w)\}\trans\bfSig_{\w}^{-1}(\widehat{\bfSig}_{\w}-\bfSig_{\w})\bfSig_{\w}^{-1}M(\w)\\
    +&\{\bfSig_{\w}^{-1}\M(\w)\}\trans\bfSig_{\w}^{-1}\{\widehat{\M}(\w)-\M(\w)\}\Big]\be_{k}(\w)+\op_{p}(\frac{1}{\sqrt{nh}}),\\
    =&\la_{k}(\w)+\C_{\la_{k}}(\w)+\B_{\la_{k}}(\w)+\op_{p}(\frac{1}{\sqrt{nh}})
  \end{aligned}
\end{equation}where
\begin{equation}
  \begin{aligned}
    \C_{\la_{k}}(\w)=&\frac{\be\trans_{k}(\w)}{2\la_{k}(\w)}\Big[-\M(\w)\trans\bfSig_{\w}^{-1}\C_{\bfSig,\w}\bfSig_{\w}^{-1}\bfSig_{\w}^{-1}\M(\w)+\C_{\M,\w}\trans\bfSig_{\w}^{-1}\bfSig_{\w}^{-1}\M(\w)\\
    -&\{\bfSig_{\w}^{-1}\M(\w)\}\trans\bfSig_{\w}^{-1}\C_{\bfSig,\w}\bfSig_{\w}^{-1}\M(\w)+\{\bfSig_{\w}^{-1}\M(\w)\}\trans\bfSig_{\w}^{-1}\C_{\M,\w}\Big]\be_{k}(\w),
  \end{aligned}
\end{equation}and
\begin{equation}\label{BLA}
\begin{aligned}
\B_{\la_{k}}(\w)=&\frac{\be\trans_{k}(\w)}{2\la_{k}(\w)}\Big[-\M(\w)\trans\bfSig_{\w}^{-1}\B_{\bfSig,\w}\bfSig_{\w}^{-1}\bfSig_{\w}^{-1}\M(\w)+\B_{\M,\w}\trans\bfSig_{\w}^{-1}\bfSig_{\w}^{-1}\M(\w)\\
    -&\{\bfSig_{\w}^{-1}\M(\w)\}\trans\bfSig_{\w}^{-1}\B_{\bfSig,\w}\bfSig_{\w}^{-1}\M(\w)+\{\bfSig_{\w}^{-1}\M(\w)\}\trans\bfSig_{\w}^{-1}\B_{\M,\w}\Big]\be_{k}(\w).
 \end{aligned}
 \end{equation} Now we turn to the expansion of $\widehat{\be}_{k}(\w)$. Since $\left(\be_{1}(\w),\ldots,\be_{p}(\w)\right)$ is a basis of $\mR^{p}$, then there exists $c_{kj}^{*}$ for $j=1,\ldots,p$, such that $\widehat{\be}_{k}(\w)-\be_{k}(\w)=\sum\nolimits_{j=1}^{p}c_{kj}^{*}\be_{j}(\w)$ and $c_{kj}^{*}=\Op_{p}(\frac{1}{\sqrt{nh}}+h^{2})$. We will derive the explicit form of $c_{kj}^{*}$ in the next step. Note that \eqref{A:1} can be rewritten as
\begin{equation}\label{A:3}
  \begin{aligned}
    &\quad \left[\{\bfSig_{\w}^{-1}\M(\w)\}\trans\bfSig_{\w}^{-1}\M(\w)-\la_{k}^{2}(\w)\right]\sum\nolimits_{j=1}^{p}c_{kj}^{*}\be_{j}(\w)\\
    &=\la_{k}(\w)\left\{\widehat{\la}_{k}(\w)-\la_{k}(\w)\right\}\be_{k}(\w)+\left\{\widehat{\la}_{k}(\w)-\la_{k}(\w)\right\}\la_{k}(\w)\be_{k}(\w)\\
    &+\Big[\M(\w)\trans\bfSig_{\w}^{-1}(\widehat{\bfSig}_{\w}-\bfSig_{\w})\bfSig_{\w}^{-1}\bfSig_{\w}^{-1}\M(\w)-\{\widehat{\M}(\w)-\M(\w)\}\trans\bfSig_{\w}^{-1}\bfSig_{\w}^{-1}\M(\w)\\
    &+\{\bfSig_{\w}^{-1}\M(\w)\}\trans\bfSig_{\w}^{-1}(\widehat{\bfSig}_{\w}-\bfSig_{\w})\bfSig_{\w}^{-1}\M(\w)-\{\bfSig_{\w}^{-1}\M(\w)\}\trans\bfSig_{\w}^{-1}\{\widehat{\M}(\w)-\M(\w)\}\Big]\be_{k}(\w).
    \end{aligned}
\end{equation}Denote
\begin{align*}
  \A(\w)&= \M(\w)\trans\bfSig_{\w}^{-1}(\widehat{\bfSig}_{\w}-\bfSig_{\w})\bfSig_{\w}^{-1}\bfSig_{\w}^{-1}\M(\w)-\{\widehat{\M}(\w)-\M(\w)\}\trans\bfSig_{\w}^{-1}\bfSig_{\w}^{-1}\M(\w)\\
    &+\{\bfSig_{\w}^{-1}\M(\w)\}\trans\bfSig_{\w}^{-1}(\widehat{\bfSig}_{\w}-\bfSig_{\w})\bfSig_{\w}^{-1}\M(\w)-\{\bfSig_{\w}^{-1}\M(\w)\}\trans\bfSig_{\w}^{-1}\{\widehat{\M}(\w)-\M(\w)\}.
\end{align*}Multiply both sides of \eqref{A:3} by $\be\trans_{j}(\w)$ $(j\neq k)$ from the left, we have
\begin{align*}
  c_{kj}^{*}=\frac{\be\trans_{j}(\w)\A(\w)\be_{k}(\w)}{\la_{j}^{2}(\w)-\la_{k}^{2}(\w)},j\neq k;
\end{align*}in addition, $\be\trans_{k}(\w)\be_{k}(\w)=\widehat{\be}\trans_{k}(\w)\widehat{\be}_{k}(\w)=1$ indicates that
 \begin{align*}
  0=\{\sum\limits_{j=1}^{p}c_{kj}^{*}\be_{j}(\w)\}\trans\be_{k}(\w)+\be_{k}(\w)\trans\{\sum\limits_{j=1}^{p}c_{kj}^{*}\be_{j}(\w)\},
\end{align*}which further implies that $c_{kk}^{*}=0$. Let
\begin{align*}
  \A_{1}(\w)&= \M(\w)\trans\bfSig_{\w}^{-1}\B_{\bfSig,\w}\bfSig_{\w}^{-1}\bfSig_{\w}^{-1}\M(\w)-\B_{\M,\w}\trans\bfSig_{\w}^{-1}\bfSig_{\w}^{-1}\M(\w)\\
    &+\{\bfSig_{\w}^{-1}\M(\w)\}\trans\bfSig_{\w}^{-1}\B_{\bfSig,\w}\bfSig_{\w}^{-1}\M(\w)-\{\bfSig_{\w}^{-1}\M(\w)\}\trans\bfSig_{\w}^{-1}\B_{\M,\w},
\end{align*}and
\begin{align*}
  \A_{2}(\w)&= \M(\w)\trans\bfSig_{\w}^{-1}\C_{\bfSig,\w}\bfSig_{\w}^{-1}\bfSig_{\w}^{-1}\M(\w)-\C_{\M,\w}\trans\bfSig_{\w}^{-1}\bfSig_{\w}^{-1}\M(\w)\\
    &+\{\bfSig_{\w}^{-1}\M(\w)\}\trans\bfSig_{\w}^{-1}\C_{\bfSig,\w}\bfSig_{\w}^{-1}\M(\w)-\{\bfSig_{\w}^{-1}\M(\w)\}\trans\bfSig_{\w}^{-1}\C_{\M,\w}.
\end{align*}Hence, we have
\begin{equation}\label{B:1}
  \begin{aligned}
    &\quad \widehat{\be}_{k}(\w)=\be_{k}(\w)+\B_{k}(\w)+\C_{k}(\w)+\op_{P}(\frac{1}{\sqrt{nh}}).
  \end{aligned}
\end{equation}Denote
\begin{equation}\label{BK}
  \begin{aligned}
    \B_{k}(\w)&=\sum\limits_{j\neq k}\frac{\be_{j}(\w)\be\trans_{j}(\w)\A_{1}(\w)\be_{k}(\w)}{\la_{j}^{2}(\w)-\la_{k}^{2}(\w)},
    \end{aligned}
    \end{equation}and
\begin{equation}\label{CK}
\begin{aligned}
  \C_{k}(\w)&=\sum\limits_{j\neq k}\frac{\be_{j}(\w)\be\trans_{j}(\w)\A_{2}(\w)\be_{k}(\w)}{\la_{j}^{2}(\w)-\la_{k}^{2}(\w)}.
  \end{aligned}
  \end{equation}The asymptotic normality is then straightforward via the central limit theorem and
  \begin{align*}
    \bfSig_{k}(\w)={\Cov}\{\C_{k}(\w)|\w\}.
  \end{align*}In partial dynamic SIR, denote $\G(\w)=\bfSig_{\w}^{-1}\M_{\SIR}(\w)$, then substitute it into \eqref{BK} and \eqref{CK},
\begin{equation}\label{be}
  \begin{aligned}
    \B_{k}^{\SIR}(\w)&=\sum\limits_{j\neq k}\frac{\be_{j}^{\SIR}(\w)\{\be^{\SIR}_{j}(\w)\}\trans}{\{\la_{j}^{\SIR}(\w)\}^{2}-\{\la_{k}^{\SIR}(\w)\}^{2}}\Big[\M_{\SIR}(\w)\trans\bfSig_{\w}^{-1}\B_{\bfSig,\w}\bfSig_{\w}^{-1}\bfSig_{\w}^{-1}\M_{\SIR}(\w)\\
    &-\B_{\SIR}(\w)\trans\bfSig_{\w}^{-1}\bfSig_{\w}^{-1}\M_{\SIR}(\w)+\{\bfSig_{\w}^{-1}\M_{\SIR}(\w)\}\trans\bfSig_{\w}^{-1}\B_{\bfSig,\w}\bfSig_{\w}^{-1}\M_{\SIR}(\w)\\
    &-\{\bfSig_{\w}^{-1}M_{\SIR}(\w)\}\trans\bfSig_{\w}^{-1}\B_{\SIR}(\w)
\Big]\be_{k}^{\SIR}(\w),
    \end{aligned}
\end{equation}
\begin{align*}
  \C_{k}^{\SIR}(\w)&=\sum\limits_{j\neq k}\frac{\be_{j}^{\SIR}(\w)\{\be^{\SIR}_{j}(\w)\}\trans}{\{\la_{j}^{\SIR}(\w)\}^{2}-\{\la_{k}^{\SIR}(\w)\}^{2}}\Big[\M_{\SIR}(\w)\trans\bfSig_{\w}^{-1}\C_{\bfSig,\w}\bfSig_{\w}^{-1}\bfSig_{\w}^{-1}\M_{\SIR}(\w)\\
  &-\C_{\SIR}(\w)\trans\bfSig_{\w}^{-1}\bfSig_{\w}^{-1}\M_{\SIR}(\w)+\{\bfSig_{\w}^{-1}\M_{\SIR}(\w)\}\trans\bfSig_{\w}^{-1}\C_{\bfSig,\w}\bfSig_{\w}^{-1}\M_{\SIR}(\w)\\
  &-\{\bfSig_{\w}^{-1}\M_{\SIR}(\w)\}\trans\bfSig_{\w}^{-1}\C_{\SIR}(\w)
\Big]\be_{k}^{\SIR}(\w).
\end{align*}Hence,
\begin{equation}\label{bfsig}
  \begin{aligned}
    \bfSig_{k}^{\SIR}(\w)={\Cov}\{\C_{k}^{\SIR}(\w)|\w\}
  \end{aligned}
\end{equation}
\eop

\noindent {\sc Proof of Theorem \ref{generalized eigenvalue decomposition 1}.}
Following by similar argument in Theorem \ref{generalized eigenvalue decomposition}, we have
\begin{align*}
    \widehat{\N}_{l,\w}-\N_{l,\w}=\C_{\N,\w}+\B_{\N,\w}+\Op_{P}\{\R_{5}(\w)\},
\end{align*}where
\begin{align*}
  \C_{\N,\w}=\frac{1}{nhf(\w)}\sum\limits_{i=1}^{n}k_{i}(\w)\left\{\x_{i}\x_{i}\trans\1(\widetilde{\Y}_{i}=l)-\N_{l,\w_{i}}\right\},\quad\B_{\N,\w}=\frac{h^{2}\omega_{2}}{2}\left(2\frac{\dot{f}(\w)}{f(\w)}\dot{\N}_{l,\w}+\ddot{\N}_{l,\w}\right),
\end{align*}and
\begin{align*}
\R_{5}(\w)=\frac{1}{nf(\w)}\left\{\left|\sum\limits_{i=1}^{n}K(\frac{\w_{i}-\w}{h})\left(\x_{i}\x_{i}\trans\1(\widetilde{\Y}_{i}=l)-\N_{l,\w_{i}}\right)\right|\right\}+\op(h^{2}).
\end{align*}Denote
\begin{align*}
  \E_{l,\w}=\bfSig_{\w}-\frac{\N_{l,\w}}{\proba_{l,\w}}+\frac{\U_{l,\w}\U_{l,\w}\trans}{\proba_{l,\w}\proba_{l,\w}},
\end{align*}then the sample estimator of $\E_{l,\w}$ is
\begin{align*}
  \widehat{\E}_{l,\w}=\widehat{\bfSig}_{\w}-\frac{\widehat{\N}_{l,\w}}{\widehat{\proba}_{l,\w}}+\frac{\widehat{\U}_{l,\w}\widehat{\U}_{l,\w}\trans}{\widehat{\proba}_{l,\w}\widehat{\proba}_{l,\w}},
\end{align*} thus
\begin{align*}
  &\quad\widehat{\M}_{\SAVE}(\w)-\M_{\SAVE}(\w)=\sum\nolimits_{l=1}^{H}\left(\widehat{\proba}_{l,\w}\widehat{\E}_{l,\w}^{2}-\proba_{l,\w}\E_{l,\w}^{2}\right)\\
  &=\sum_{l=1}^{H}\left\{\left(\widehat{\proba}_{l,\w}-\proba_{l,\w}\right)\widehat{\E}_{l,\w}^{2}+\proba_{l,\w}\left(\widehat{\E}_{l,\w}^{2}-\E_{l,\w}^{2}\right)\right\}\\
  &=\sum\nolimits_{l=1}^{H}\left(\widehat{\proba}_{l,\w}-\proba_{l,\w}\right)\E_{l,\w}^{2}+\sum\nolimits_{l=1}^{H}\proba_{l,\w}\left(\widehat{\E}_{l,\w}-\E_{l,\w}\right)\times\left(\widehat{\E}_{l,\w}+\E_{l,\w}\right)+\op_{P}\left(\frac{1}{\sqrt{nh}}\right)\\
  &=\sum\nolimits_{l=1}^{H}\left(\widehat{\proba}_{l,\w}-\proba_{l,\w}\right)\E_{l,\w}^{2}+\sum\nolimits_{l=1}^{H}2\proba_{l,\w}\Bigg\{\left(\widehat{\bfSig}_{\w}-\bfSig_{\w}\right)\bfSig_{\w}\\
  &+\frac{\N_{l,\w}\left(\widehat{\proba}_{l,\w}-\proba_{l,\w}\right)-\left(\widehat{\N}_{l,\w}-\N_{l,\w}\right)\proba_{l,\w}}{\proba_{l,\w}^{2}}\bfSig_{\w}\\
  &+\frac{\left(\widehat{\U}_{l,\w}-\U_{l,\w}\right)\U_{l,\w}\trans\proba_{l,\w}+\U_{l,\w}\left(\widehat{\U}_{l,\w}-\U_{l,\w}\right)\trans\proba_{l,\w}-2\U_{l,\w}\U_{l,\w}\trans\left(\widehat{\proba}_{l,\w}-\proba_{l,\w}\right)}{\proba_{l,\w}^{3}}\bfSig_{\w}\\
  &-\left(\widehat{\bfSig}_{\w}-\bfSig_{\w}\right)\frac{\N_{l,\w}}{\proba_{l,\w}}-\frac{\N_{l,\w}\left(\widehat{\proba}_{l,\w}-\proba_{l,\w}\right)-\left(\widehat{\N}_{l,\w}-\N_{l,\w}\right)\proba_{l,\w}}{\proba_{l,\w}^{2}}\frac{\N_{l,\w}}{\proba_{l,\w}}\\
  &-\frac{\left(\widehat{\U}_{l,\w}-\U_{l,\w}\right)\U_{l,\w}\trans\proba_{l,\w}+\U_{l,\w}\left(\widehat{\U}_{l,\w}-\U_{l,\w}\right)\trans\proba_{l,\w}-2\U_{l,\w}\U_{l,\w}\trans\left(\widehat{\proba}_{l,\w}-\proba_{l,\w}\right)}{\proba_{l,\w}^{3}}\frac{\N_{l,\w}}{\proba_{l,\w}}\\
  &+\left(\widehat{\bfSig}_{\w}-\bfSig_{\w}\right)\frac{\U_{l,\w}\U_{l,\w}\trans}{\proba_{l,\w}^{2}}+\frac{\N_{l,\w}\left(\widehat{\proba}_{l,\w}-\proba_{l,\w}\right)-\left(\widehat{\N}_{l,\w}-\N_{l,\w}\right)\proba_{l,\w}}{\proba_{l,\w}^{2}}\frac{\U_{l,\w}\U_{l,\w}\trans}{\proba_{l,\w}^{2}}\\
  &+\frac{\left(\widehat{\U}_{l,\w}-\U_{l,\w}\right)\U_{l,\w}\trans\proba_{l,\w}+\U_{l,\w}\left(\widehat{\U}_{l,\w}-\U_{l,\w}\right)\trans\proba_{l,\w}-2\U_{l,\w}\U_{l,\w}\trans\left(\widehat{\proba}_{l,\w}-\proba_{l,\w}\right)}{\proba_{l,\w}^{3}}\frac{\U_{l,\w}\U_{l,\w}\trans}{\proba_{l,\w}^{2}}\Bigg\}\\
  &+\op_{P}\left(\frac{1}{\sqrt{nh}}\right)=\B_{\SAVE}+\C_{\SAVE}+\op_{P}\left(\frac{1}{\sqrt{nh}}\right),
\end{align*}where
\begin{equation}\label{B SAVE}
  \begin{aligned}
  \B_{\SAVE}=&\sum\nolimits_{l=1}^{H}\B_{\proba,\w}\E_{l,\w}^{2}+2\sum\nolimits_{l=1}^{H}\proba_{l,\w}\Big\{\B_{\bfSig,\w}\bfSig_{\w}+\frac{\N_{l,\w}\B_{\proba,\w}-\B_{\N,\w}\proba_{l,\w}}{\proba_{l,\w}^{2}}\bfSig_{\w}\\
  &+\frac{\B_{\U,\w}\U_{l,\w}\trans\proba_{l,\w}+\U_{l,\w}\B_{\U,\w}\trans\proba_{l,\w}-2\U_{l,\w}\U_{l,\w}\trans\B_{\proba,\w}}{\proba_{l,\w}^{3}}\bfSig_{\w}-\B_{\bfSig,\w}\frac{\N_{l,\w}}{\proba_{l,\w}}\\
  &-\frac{\N_{l,\w}\B_{\proba,\w}-\B_{\N,\w}\proba_{l,\w}}{\proba_{l,\w}^{2}}\frac{\N_{l,\w}}{\proba_{l,\w}}-\frac{\B_{\U,\w}\U_{l,\w}\trans\proba_{l,\w}+\U_{l,\w}\B_{\U,\w}\trans\proba_{l,\w}-2\U_{l,\w}\U_{l,\w}\trans\B_{\proba,\w}}{\proba_{l,\w}^{3}}\frac{\N_{l,\w}}{\proba_{l,\w}}\\
  &+\B_{\bfSig,\w}\frac{\U_{l,\w}\U_{l,\w}\trans}{\proba_{l,\w}^{2}}+\frac{\N_{l,\w}\B_{\proba,\w}-\B_{\N,\w}\proba_{l,\w}}{\proba_{l,\w}^{2}}\frac{\U_{l,\w}\U_{l,\w}\trans}{\proba_{l,\w}^{2}}\\
  &+\frac{\B_{\U,\w}\U_{l,\w}\trans\proba_{l,\w}+\U_{l,\w}\B_{\U,\w}\trans\proba_{l,\w}-2\U_{l,\w}\U_{l,\w}\trans\B_{\proba,\w}}{\proba_{l,\w}^{3}}\frac{\U_{l,\w}\U_{l,\w}\trans}{\proba_{l,\w}^{2}}\Big\},
  \end{aligned}
\end{equation}and
\begin{equation}\label{C SAVE}
  \begin{aligned}
  \C_{\SAVE}=&\sum\nolimits_{l=1}^{H}\C_{\proba,\w}\E_{l,\w}^{2}+2\sum\nolimits_{l=1}^{H}\proba_{l,\w}\Big\{\C_{\bfSig,\w}\bfSig_{\w}+\frac{\N_{l,\w}\C_{\proba,\w}-\C_{\N,\w}\proba_{l,\w}}{\proba_{l,\w}^{2}}\bfSig_{\w}\\
  &+\frac{\C_{\U,\w}\U_{l,\w}\trans\proba_{l,\w}+\U_{l,\w}\C_{\U,\w}\trans\proba_{l,\w}-2\U_{l,\w}\U_{l,\w}\trans\C_{\proba,\w}}{\proba_{l,\w}^{3}}\bfSig_{\w}-\C_{\bfSig,\w}\frac{\N_{l,\w}}{\proba_{l,\w}}\\
  &-\frac{\N_{l,\w}\C_{\proba,\w}-\C_{\N,\w}\proba_{l,\w}}{\proba_{l,\w}^{2}}\frac{\N_{l,\w}}{\proba_{l,\w}}-\frac{\C_{\U,\w}\U_{l,\w}\trans\proba_{l,\w}+\U_{l,\w}\C_{\U,\w}\trans\proba_{l,\w}-2\U_{l,\w}\U_{l,\w}\trans\C_{\proba,\w}}{\proba_{l,\w}^{3}}\frac{\N_{l,\w}}{\proba_{l,\w}}\\
  &+\C_{\bfSig,\w}\frac{\U_{l,\w}\U_{l,\w}\trans}{\proba_{l,\w}^{2}}+\frac{\N_{l,\w}\C_{\proba,\w}-\C_{\N,\w}\proba_{l,\w}}{\proba_{l,\w}^{2}}\frac{\U_{l,\w}\U_{l,\w}\trans}{\proba_{l,\w}^{2}}\\
  &+\frac{\C_{\U,\w}\U_{l,\w}\trans\proba_{l,\w}+\U_{l,\w}\C_{\U,\w}\trans\proba_{l,\w}-2\U_{l,\w}\U_{l,\w}\trans\C_{\proba,\w}}{\proba_{l,\w}^{3}}\frac{\U_{l,\w}\U_{l,\w}\trans}{\proba_{l,\w}^{2}}\Big\}.
  \end{aligned}
\end{equation}Similar to the proof of Theorem \ref{generalized eigenvalue decomposition}, we have
\begin{equation}
  \begin{aligned}
    \sqrt{nh}\left(\vech\{\widehat{\M}_{\SAVE}(\w)\}-\vech\{\M_{\SAVE}(\w)\}-\vech\{\B_{\SAVE}(\w)\}\right) \xrightarrow{d}  {\bf{N}} (0,f^{-1}(\w)\omega_{0}\C^{\SAVE}(\w)),
  \end{aligned}
\end{equation}where $\omega_{0}$ is defined as follows. Hence
 \begin{equation}\label{CSAVE}
   \begin{aligned}
     \C^{\SAVE}(\w)=\Cov[\vech\{\C_{\SAVE}(\w)\}|\w].
   \end{aligned}
 \end{equation}Then for singular value decomposition, in partial dynamic SAVE, denote  $\G(\w)=\bfSig_{\w}^{-1}\M_{\SAVE}(\w)$, then substitute it into \eqref{BK} and \eqref{CK},
\begin{equation}\label{be1}
  \begin{aligned}
    \B_{k}^{\SAVE}(\w)&=\sum\limits_{j\neq k}\frac{\be_{j}^{\SAVE}(\w)\{\be^{\SAVE}_{j}(\w)\}\trans}{\{\la_{j}^{\SAVE}(\w)\}^{2}-\{\la_{k}^{\SAVE}(\w)\}^{2}}\Big[\M\trans_{\SAVE}(\w)\bfSig_{\w}^{-1}\B_{\bfSig,\w}\bfSig_{\w}^{-1}\bfSig_{\w}^{-1}\M_{\SAVE}(\w)\\
    &-\B\trans_{\SAVE}(\w)\bfSig_{\w}^{-1}\bfSig_{\w}^{-1}\M_{\SAVE}(\w)+\{\bfSig_{\w}^{-1}\M_{\SAVE}(\w)\}\trans\bfSig_{\w}^{-1}\B_{\bfSig,\w}\bfSig_{\w}^{-1}\M_{\SAVE}(\w)\\
    &-\{\bfSig_{\w}^{-1}\M_{\SAVE}(\w)\}\trans\bfSig_{\w}^{-1}\B_{\SAVE}(\w)
\Big]\be_{k}^{\SAVE}(\w),
    \end{aligned}
\end{equation}
\begin{align*}
  \C_{k}^{\SAVE}(\w)&=\sum\limits_{j\neq k}\frac{\be_{j}^{\SAVE}(\w)\{\be^{\SAVE}_{j}(\w)\}\trans}{\{\la_{j}^{\SAVE}(\w)\}^{2}-\{\la_{k}^{\SAVE}(\w)\}^{2}}\Big[\M\trans_{\SAVE}(\w)\bfSig_{\w}^{-1}\C_{\bfSig,\w}\bfSig_{\w}^{-1}\bfSig_{\w}^{-1}\M_{\SAVE}(\w)\\
  &-\C_{\SAVE}(\w)\trans\bfSig_{\w}^{-1}\bfSig_{\w}^{-1}\M_{\SAVE}(\w)+\{\bfSig_{\w}^{-1}\M_{\SAVE}(\w)\}\trans\bfSig_{\w}^{-1}\C_{\bfSig,\w}\bfSig_{\w}^{-1}\M_{\SAVE}(\w)\\
  &-\{\bfSig_{\w}^{-1}\M_{\SAVE}(\w)\}\trans\bfSig_{\w}^{-1}\C_{\SAVE}(\w)
\Big]\be_{k}^{\SAVE}(\w).
\end{align*}Hence,
\begin{equation}\label{bfsig1}
  \begin{aligned}
    \bfSig_{k}^{\SAVE}(\w)={\Cov}\{\C_{k}^{\SAVE}(\w)|\w\}
  \end{aligned}
\end{equation}
\eop

\noindent {\sc Proof of Theorem \ref{generalized eigenvalue decomposition 2}.} Firstly, we have
{\small
\begin{align*}
  &\quad\widehat{\M}_{\DR}(\w)-\M_{\DR}(\w)\\
  &=2\sum\limits_{l=1}^{H}\Bigg\{\widehat{\proba}_{l,\w}\left(\frac{\widehat{\N}_{l,\w}}{\widehat{\proba}_{l,\w}}-\widehat{\bfSig}_{\w}\right)^{2}-\proba_{l,\w}\left(\frac{\N_{l,\w}}{\proba_{l,\w}}-\bfSig_{\w}\right)^{2}\Bigg\}+2\Bigg\{\left(\sum\limits_{l=1}^{H}\frac{\widehat{\U}_{l,\w}\widehat{\U}_{l,\w}\trans}{\widehat{\proba}_{l,\w}}\right)^{2}-\left(\sum\limits_{l=1}^{H}\frac{\U_{l,\w}\U_{l,\w}\trans}{\proba_{l,\w}}\right)^{2}\Bigg\}\\
  &+2\Bigg\{\left(\sum\limits_{l=1}^{H}\frac{\widehat{\U}_{l,\w}\trans\widehat{\U}_{l,\w}}{\widehat{\proba}_{l,\w}}\right)\left(\sum\limits_{l=1}^{H}\frac{\widehat{\U}_{l,\w}\widehat{\U}_{l,\w}\trans}{\widehat{\proba}_{l,\w}}\right)-\left(\sum\limits_{l=1}^{H}\frac{\U_{l,\w}\trans\U_{l,\w}}{\proba_{l,\w}}\right)\left(\sum\limits_{l=1}^{H}\frac{\U_{l,\w}\U_{l,\w}\trans}{\proba_{l,\w}}\right)\Bigg\}\\
  &=2\sum\limits_{l=1}^{H}\Bigg\{\left(\widehat{\proba}_{l,\w}-\proba_{l,\w}\right)\left(\frac{\N_{l,\w}}{\proba_{l,\w}}-\bfSig_{\w}\right)^{2}+\proba_{l,\w}\left(\frac{\widehat{\N}_{l,\w}}{\widehat{\proba}_{l,\w}}-\widehat{\bfSig}_{\w}+\frac{\N_{l,\w}}{\proba_{l,\w}}-\bfSig_{\w}\right)\\
  &\times\left(\frac{\widehat{\N}_{l,\w}}{\widehat{\proba}_{l,\w}}-\widehat{\bfSig}_{\w}-\frac{\N_{l,\w}}{\proba_{l,\w}}+\bfSig_{\w}\right)\Bigg\}+2\sum\limits_{l=1}^{H}\left(\frac{\widehat{\U}_{l,\w}\widehat{\U}_{l,\w}\trans}{\widehat{\proba}_{l,\w}}+\frac{\U_{l,\w}\U_{l,\w}\trans}{\proba_{l,\w}}\right)\\
  &\times\sum\limits_{l=1}^{H}\left(\frac{\widehat{\U}_{l,\w}\widehat{\U}_{l,\w}\trans}{\widehat{\proba}_{l,\w}}-\frac{\U_{l,\w}\U_{l,\w}\trans}{\proba_{l,\w}}\right)+2\sum\limits_{l=1}^{H}\left(\frac{\widehat{\U}_{l,\w}\trans\widehat{\U}_{l,\w}}{\widehat{\proba}_{l,\w}}-\frac{\U_{l,\w}\trans\U_{l,\w}}{\proba_{l,\w}}\right)\left(\sum\limits_{l=1}^{H}\frac{\U_{l,\w}\U_{l,\w}\trans}{\proba_{l,\w}}\right)\\
  &+2\sum\limits_{l=1}^{H}\frac{\U_{l,\w}\trans\U_{l,\w}}{\proba_{l,\w}}\sum\limits_{l=1}^{H}\left(\frac{\widehat{\U}_{l,\w}\widehat{\U}_{l,\w}\trans}{\widehat{\proba}_{l,\w}}-\frac{\U_{l,\w}\U_{l,\w}\trans}{\proba_{l,\w}}\right)\Bigg\}+\op_{P}\left(\frac{1}{\sqrt{nh}}\right)\\
  &=2\sum\limits_{l=1}^{H}\Bigg\{\left(\widehat{\proba}_{l,\w}-\proba_{l,\w}\right)\left(\frac{\N_{l,\w}}{\proba_{l,\w}}-\bfSig_{\w}\right)^{2}+2\N_{l,\w}\left(\frac{\widehat{\N}_{l,\w}}{\widehat{\proba}_{l,\w}}-\frac{\N_{l,\w}}{\proba_{l,\w}}\right)-2\N_{l,\w}\left(\widehat{\bfSig}_{\w}-\bfSig_{\w}\right)\\
  &-2\proba_{l,\w}\bfSig_{\w}\left(\frac{\widehat{\N}_{l,\w}}{\widehat{\proba}_{l,\w}}-\frac{\N_{l,\w}}{\proba_{l,\w}}\right)+2\proba_{l,\w}\bfSig_{\w}\left(\widehat{\bfSig}_{\w}-\bfSig_{\w}\right)\Bigg\}\\
  &+4\Bigg\{\sum\limits_{l=1}^{H}\frac{\U_{l,\w}\U_{l,\w}\trans}{\proba_{l,\w}}\sum\limits_{l=1}^{H}\left(\frac{\widehat{\U}_{l,\w}\widehat{\U}_{l,\w}\trans}{\widehat{\proba}_{l,\w}}-\frac{\U_{l,\w}\U_{l,\w}\trans}{\proba_{l,\w}}\right)\Bigg\}+2\sum\limits_{l=1}^{H}\left(\frac{\widehat{\U}_{l,\w}\trans\widehat{\U}_{l,\w}}{\widehat{\proba}_{l,\w}}-\frac{\U_{l,\w}\trans\U_{l,\w}}{\proba_{l,\w}}\right)\left(\sum\limits_{l=1}^{H}\frac{\U_{l,\w}\U_{l,\w}\trans}{\proba_{l,\w}}\right)\\
  &+2\sum\limits_{l=1}^{H}\frac{\U_{l,\w}\trans\U_{l,\w}}{\proba_{l,\w}}\sum\limits_{l=1}^{H}\left(\frac{\widehat{\U}_{l,\w}\widehat{\U}_{l,\w}\trans}{\widehat{\proba}_{l,\w}}-\frac{\U_{l,\w}\U_{l,\w}\trans}{\proba_{l,\w}}\right)+\op_{P}\left(\frac{1}{\sqrt{nh}}\right)\\
  &=2\sum\limits_{l=1}^{H}\Bigg\{\left(\widehat{\proba}_{l,\w}-\proba_{l,\w}\right)\left(\frac{\N_{l,\w}}{\proba_{l,\w}}-\bfSig_{\w}\right)^{2}+2\N_{l,\w}\frac{\left(\widehat{\N}_{l,\w}-\N_{l,\w}\right)\proba_{l,\w}-\N_{l,\w}\left(\widehat{\proba}_{l,\w}-\proba_{l,\w}\right)}{\proba_{l,\w}^{2}}\\
  &-2\N_{l,\w}\left(\widehat{\bfSig}_{\w}-\bfSig_{\w}\right)-2\bfSig_{\w}\frac{\left(\widehat{\N}_{l,\w}-\N_{l,\w}\right)\proba_{l,\w}-\N_{l,\w}\left(\widehat{\proba}_{l,\w}-\proba_{l,\w}\right)}{\proba_{l,\w}}+2\proba_{l,\w}\bfSig_{\w}\left(\widehat{\bfSig}_{\w}-\bfSig_{\w}\right)\Bigg\}\\
  &+2\sum\limits_{l=1}^{H}\left(2\frac{\U_{l,\w}\U_{l,\w}\trans}{\proba_{l,\w}}+\frac{\U_{l,\w}\trans\U_{l,\w}}{\proba_{l,\w}}\right)\\
  &\times\sum\limits_{l=1}^{H}\frac{\left(\widehat{\U}_{l,\w}-\U_{l,\w}\right)\U_{l,\w}\trans\proba_{l,\w}+\U_{l,\w}\left(\widehat{\U}_{l,\w}-\U_{l,\w}\right)\trans\proba_{l,\w}-\U_{l,\w}\U_{l,\w}\trans\left(\widehat{\proba}_{l,\w}-\proba_{l,\w}\right)}{\proba_{l,\w}^{2}}\\
  &+2\sum\limits_{l=1}^{H}\frac{\left(\widehat{\U}_{l,\w}-\U_{l,\w}\right)\trans\U_{l,\w}\proba_{l,\w}+\U_{l,\w}\trans\left(\widehat{\U}_{l,\w}-\U_{l,\w}\right)\proba_{l,\w}-\U_{l,\w}\trans\U_{l,\w}\left(\widehat{\proba}_{l,\w}-\proba_{l,\w}\right)}{\proba_{l,\w}^{2}}\\
  &\times\left(\sum\limits_{l=1}^{H}\frac{\U_{l,\w}\U_{l,\w}\trans}{\proba_{l,\w}}\right)+\op_{P}\left(\frac{1}{\sqrt{nh}}\right)=\B_{\DR}+\C_{\DR}+\op_{P}\left(\frac{1}{\sqrt{nh}}\right),
\end{align*}}where
\begin{equation}\label{B DR}
\begin{aligned}
  \B_{\DR}=&2\sum\limits_{l=1}^{H}\Bigg\{\B_{\proba,\w}\left(\frac{\N_{l,\w}}{\proba_{l,\w}}-\bfSig_{\w}\right)^{2}+2\N_{l,\w}\frac{\B_{\N,\w}\proba_{l,\w}-\N_{l,\w}\B_{\proba,\w}}{\proba_{l,\w}^{2}}-2\N_{l,\w}\B_{\bfSig,\w}\\
&-2\bfSig_{\w}\frac{\B_{\N,\w}\proba_{l,\w}-\N_{l,\w}\B_{\proba,\w}}{\proba_{l,\w}}+2\proba_{l,\w}\bfSig_{\w}\B_{\bfSig,\w}\Bigg\}+2\sum\limits_{l=1}^{H}\left(2\frac{\U_{l,\w}\U_{l,\w}\trans}{\proba_{l,\w}}+\frac{\U_{l,\w}\trans\U_{l,\w}}{\proba_{l,\w}}\right)\\
&\times \sum\limits_{l=1}^{H}\frac{\B_{\U,\w}\U_{l,\w}\trans\proba_{l,\w}+\U_{l,\w}\B_{\U,\w}\trans\proba_{l,\w}-\U_{l,\w}\U_{l,\w}\trans\B_{\proba,\w}}{\proba_{l,\w}^{2}}\\
  &+2\sum\limits_{l=1}^{H}\frac{\B_{\U,\w}\trans\U_{l,\w}\proba_{l,\w}+\U_{l,\w}\trans\B_{\U,\w}\proba_{l,\w}-\U_{l,\w}\trans\U_{l,\w}\B_{\proba,\w}}{\proba_{l,\w}^{2}}\left(\sum\limits_{l=1}^{H}\frac{\U_{l,\w}\U_{l,\w}\trans}{\proba_{l,\w}}\right)\\
  \end{aligned}
\end{equation}and
\begin{align*}
  \C_{\DR}=&2\sum\limits_{l=1}^{H}\Bigg\{\C_{\proba,\w}\left(\frac{\N_{l,\w}}{\proba_{l,\w}}-\bfSig_{\w}\right)^{2}+2\N_{l,\w}\frac{\C_{\N,\w}\proba_{l,\w}-\N_{l,\w}\C_{\proba,\w}}{\proba_{l,\w}^{2}}-2\N_{l,\w}\C_{\bfSig,\w}\\
&-2\bfSig_{\w}\frac{\C_{\N,\w}\proba_{l,\w}-\N_{l,\w}\C_{\proba,\w}}{\proba_{l,\w}}+2\proba_{l,\w}\bfSig_{\w}\C_{\bfSig,\w}\Bigg\}+2\sum\limits_{l=1}^{H}\left(2\frac{\U_{l,\w}\U_{l,\w}\trans}{\proba_{l,\w}}+\frac{\U_{l,\w}\trans\U_{l,\w}}{\proba_{l,\w}}\right)\\
&\times\sum\limits_{l=1}^{H}\frac{\C_{\U,\w}\U_{l,\w}\trans\proba_{l,\w}+\U_{l,\w}\C_{\U,\w}\trans\proba_{l,\w}-\U_{l,\w}\U_{l,\w}\trans\C_{\proba,\w}}{\proba_{l,\w}^{2}}\\
  &+2\sum\limits_{l=1}^{H}\frac{\C_{\U,\w}\trans\U_{l,\w}\proba_{l,\w}+\U_{l,\w}\trans\C_{\U,\w}\proba_{l,\w}-\U_{l,\w}\trans\U_{l,\w}\C_{\proba,\w}}{\proba_{l,\w}^{2}}\left(\sum\limits_{l=1}^{H}\frac{\U_{l,\w}\U_{l,\w}\trans}{\proba_{l,\w}}\right)\\
  \end{align*}Similar to the proof of Theorem \ref{generalized eigenvalue decomposition} and \ref{generalized eigenvalue decomposition 1}, we have
\begin{equation}
  \begin{aligned}
    \sqrt{nh}\left(\vech\{\widehat{\M}_{\DR}(\w)\}-\vech\{\M_{\DR}(\w)\}-\vech\{\B_{\DR}(\w)\}\right) \xrightarrow{d}  {\bf{N}} (0,f^{-1}(\w)\omega_{0}\C^{\DR}(\w)),
  \end{aligned}
\end{equation}where $\omega_{0}$ is defined as previously. Hence
 \begin{equation}\label{CDR}
   \begin{aligned}
     \C^{\DR}(\w)=\Cov[\vech\{\C_{\DR}(\w)\}|\w].
   \end{aligned}
 \end{equation}Then for singular value decomposition, in partial dynamic DR, denote  $\G(\w)=\bfSig_{\w}^{-1}\M_{\DR}(\w)$, then substitute it into \eqref{BK} and \eqref{CK},
\begin{equation}\label{be2}
  \begin{aligned}
    \B_{k}^{\DR}(\w)&=\sum\limits_{j\neq k}\frac{\be_{j}^{\DR}(\w)\{\be^{\DR}_{j}(\w)\}\trans}{\{\la_{j}^{\DR}(\w)\}^{2}-\{\la_{k}^{\DR}(\w)\}^{2}}\Big[\M\trans_{\DR}(\w)\bfSig_{\w}^{-1}\B_{\bfSig,\w}\bfSig_{\w}^{-1}\bfSig_{\w}^{-1}\M_{\DR}(\w)\\
    &-\B\trans_{\DR}(\w)\bfSig_{\w}^{-1}\bfSig_{\w}^{-1}\M_{\DR}(\w)+\{\bfSig_{\w}^{-1}\M_{\DR}(\w)\}\trans\bfSig_{\w}^{-1}\B_{\bfSig,\w}\bfSig_{\w}^{-1}\M_{\DR}(\w)\\
    &-\{\bfSig_{\w}^{-1}\M_{\DR}(\w)\}\trans\bfSig_{\w}^{-1}\B_{\DR}(\w)
\Big]\be_{k}^{\DR}(\w),
    \end{aligned}
\end{equation}
\begin{align*}
  \C_{k}^{\DR}(\w)&=\sum\limits_{j\neq k}\frac{\be_{j}^{\DR}(\w)\{\be^{\DR}_{j}(\w)\}\trans}{\{\la_{j}^{\DR}(\w)\}^{2}-\{\la_{k}^{\DR}(\w)\}^{2}}\Big[\M\trans_{\DR}(\w)\bfSig_{\w}^{-1}\C_{\bfSig,\w}\bfSig_{\w}^{-1}\bfSig_{\w}^{-1}\M_{\DR}(\w)\\
  &-\C_{\DR}(\w)\trans\bfSig_{\w}^{-1}\bfSig_{\w}^{-1}\M_{\DR}(\w)+\{\bfSig_{\w}^{-1}\M_{\DR}(\w)\}\trans\bfSig_{\w}^{-1}\C_{\bfSig,\w}\bfSig_{\w}^{-1}\M_{\DR}(\w)\\
  &-\{\bfSig_{\w}^{-1}\M_{\DR}(\w)\}\trans\bfSig_{\w}^{-1}\C_{\DR}(\w)
\Big]\be_{k}^{\DR}(\w).
\end{align*}Hence,
\begin{equation}\label{bfsig2}
  \begin{aligned}
    \bfSig_{k}^{\DR}(\w)={\Cov}\{\C_{k}^{\DR}(\w)|\w\}
  \end{aligned}
\end{equation}
\eop

Theorem \ref{consistency property} is closely related to the following two assertions:
\begin{description}
	\item[(a)] if $\la_{k}(\w)>\la_{k+1}(\w)$, then $f_{n}^{0}(k)=\Op_{P}(\frac{1}{nh})$ almost surely $\proba_{\calS}$ given $\W=\w$;
	\item[(b)] if $\la_{k}(\w)=\la_{k+1}(\w)$, then $f_{n}^{0}(k)=\Op_{P}^{+}(c_{n})$ almost surely $\proba_{\calS}$ given $\W=\w$,
\end{description}where $c_{n}=[\log\{\log (n)\}]^{-1}$ and $\Op_{P}^{+}$ symbols has been shown in \cite{Luo2016}. We will prove these assertions, and then prove Theorem 2 based on them.

The nonparametric ladle estimator can pinpoint the rank of a matrix more precisely than the other order-determination methods when they are used for the nonparametric model. We established the consistency of the nonparametric ladle estimator. The next Lemma regulates the order between the eigenvalues and the variability of eigenvectors. In particular, it shows that distant eigenvalues are related to the small variability of eigenvectors. Nonparametric method allows asymmetric matrices, because we can apply it to $\widehat{\G}\trans(\w)\widehat{\G}(\w)$ or $\widehat{\G}(\w)\widehat{\G}\trans(\w)$, which amounts to replacing the eigenvalues and eigenvectors of $\widehat{\G}(\w)$ by its squared singular values and singular vectors. With this modification, all the subsequent results remain valid for asymmetric $\widehat{\G}(\w)$.

{\lemm\label{the osrder between eigenvalue and the variability of eigenvectors} For any $i,j=1,\ldots,p$ with $i\neq j$,
\begin{align*}
  |\be_{i}\trans(\w)\widehat{\be}_{j}(\w)|[\la_{i}(\w)-\la_{j}(\w)-\{\B_{\la_{j}}(\w)+\B(\w)\}]=\Op_{p}(\frac{1}{\sqrt{nh}}),
\end{align*}where the closed form of  $\B_{\la_{j}}(\w)$ and $\B(\w)$ are provided in \eqref{BLA} and \eqref{BW} in the Appendix, respectively.
}

\noindent {\sc Proof of Lemma \ref{the osrder between eigenvalue and the variability of eigenvectors}.}
\cite{Luo2016} has shown that
\begin{equation}\label{l14}
\begin{aligned}
  \G(\w)\widehat{\be}_{j}(\w)=\G(\w)\left\{\sum_{i=1}^{p}r_{ij}\be_{i}(\w)\right\}=\sum_{i=1}^{p}r_{ij}\la_{i}(\w)\be_{i}(\w),
\end{aligned}
\end{equation}and
\begin{equation}\label{l15}
\begin{aligned}
  \G(\w)\widehat{\be}_{j}(\w)=\widehat{\G}(\w)\widehat{\be}_{j}(\w)+\{\G(\w)-\widehat{\G}(\w)\}\widehat{\be}_{j}(\w)=\widehat{\la}_{j}(\w)\widehat{\be}_{j}(\w)+\{\G(\w)-\widehat{\G}(\w)\}\widehat{\be}_{j}(\w),
\end{aligned}
\end{equation}where $r_{ij}=\be_{i}\trans(\w)\widehat{\be}_{j}(\w)$,  $\widehat{\be}_{j}(\w)=\sum_{i=1}^{p}r_{ij}\be_{i}(\w)$ and $\sum_{i=1}^{p}r_{ij}^{2}=1$. $\widehat{\G}(\w)-\G(\w)=\B(\w)+\Op_{P}\left(\frac{1}{\sqrt{nh}}\right)$ and  $\widehat{\la}_{j}(\w)-\la_{j}(\w)=\B_{\la_{j}}(\w)+\Op_{p}(\frac{1}{\sqrt{nh}})$. Here, we denote $\G(\w)=\bfSig_{\w}^{-1}\M(\w)$. Since
\begin{align*}
  \widehat{\G}(\w)-\G(\w)=\bfSig_{\w}^{-1}\{\widehat{\M}(\w)-\M(\w)\}+(\widehat{\bfSig}_{\w}^{-1}-\bfSig_{\w}^{-1})\M(\w)+\op_{p}(\frac{1}{\sqrt{nh}}).
\end{align*}Hence
\begin{equation}\label{BW}
\begin{aligned}
  \B(\w)=\bfSig_{\w}^{-1}\B_{M,\w}+\bfSig_{\w}^{-1}\B_{\bfSig,\w}\bfSig_{\w}^{-1}\M(\w).
\end{aligned}
\end{equation}Combing \eqref{l14} and \eqref{l15}, we have
\begin{align*}
  \sum_{i=1}^{p}r_{ij}\la_{i}(\w)\be_{i}(\w)=&\la_{j}(\w)\widehat{\be}_{j}(\w)+\B_{\la_{j}}(\w)\widehat{\be}_{j}(\w)+\B(\w)\widehat{\be}_{j}(\w)+\Op_{p}(\frac{1}{\sqrt{nh}})\\
  =&\sum_{i=1}^{p}r_{ij}\la_{j}(\w)\be_{i}(\w)+\{\B_{\la_{j}}(\w)+\B(\w)\}\widehat{\be}_{j}(\w)+\Op_{p}(\frac{1}{\sqrt{nh}})\\
  =&\sum_{i=1}^{p}r_{ij}\la_{j}(\w)\be_{i}(\w)+\sum_{i=1}^{p}r_{ij}\{\B_{\la_{j}}(\w)+\B(\w)\}\be_{i}(\w)+\Op_{p}(\frac{1}{\sqrt{nh}}).
\end{align*}Hence
\begin{align*}
  \sum_{i=1}^{p}r_{ij}[\la_{i}(\w)-\la_{j}(\w)-\{\B_{\la_{j}}(\w)+\B(\w)\}]\be_{i}(\w)=\Op_{p}(\frac{1}{\sqrt{nh}}).
\end{align*}Since $\be_{1}(\w),\ldots,\be_{p}(\w)$ are orthogonal, we have, for each $i\neq j$,
\begin{align*}
|\be_{i}\trans(\w)\widehat{\be}_{j}(\w)|[\la_{i}(\w)-\la_{j}(\w)-\{\B_{\la_{j}}(\w)+\B(\w)\}]=\Op_{p}(\frac{1}{\sqrt{nh}}).
\end{align*}
\eop

Since the order of bias term is $h^{2}$, then under condition (C7),
\begin{align*}
\frac{\sqrt{nh}\times h^{2}}{\sqrt{\log(\log n)}}\rightarrow 0, \text { as } n\rightarrow \infty,
 \end{align*}the bias term $\{\B_{\la_{j}}(\w)+\B(\w)\}$ will vanish, such that a direct application of this Lemma leads to the next lemma.
{\lemm\label{A direct application} Under condition (C9) and (C13), for any positive semi-definite candidate matrix $\G(\w)\in\mR^{p\times p}$ and any $i,j\in\{1,\ldots,p\}$, if $\la_{i}(\w)>\la_{j}(\w)$,then
\begin{align*}
  \widehat{\be}_{i}(\w)\trans\be_{j}^{*}(\w)=\Op_{P}(\frac{1}{\sqrt{nh}}),
\end{align*}almost surely $\proba_{\calS}$.
}

\noindent {\sc Proof of Lemma \ref{A direct application}.}
Let $A_{1}\in\calF$ be the event that $\Big\{(nh)^{1/2}\{\widehat{\la}_{i}(\w)-\la_{i}(\w)\}/\sqrt{\log(\log n)}:n\in\mN\Big\}$ is a bounded sequence for each $i=1,\ldots,p$. From Assumption (C12) it follows \cite{Hardle1984} that the bias term of $\widehat{\la}_{i}(\w)-\la_{i}(\w)$ vanishes. By the law of the iterated logarithm and Lemma 3.2 of \cite{Zhao1986}, $\pr(A_{1})=1$. For any $s\in A_{1}$ and $i,j=1,\ldots,p$, we have
\begin{equation}
\begin{aligned}\label{eigenvalues relationship}
  |\widehat{\la}_{i}(\w)-\widehat{\la}_{j}(\w)|=|\la_{i}(\w)-\la_{j}(\w)|+\op(1).
\end{aligned}
\end{equation}Let $A_{2}\in\calF$ be the event in Assumption (C10). Then $\pr(A_{2})=1$. Hence $\pr(A_{1}\cap A_{2})=1$. For any fixed $s\in A_{1}\cap A_{2}$, by Lemma \ref{the osrder between eigenvalue and the variability of eigenvectors},
\begin{align*}
  \widehat{\be}_{i}\trans(\w)\be_{j}^{*}(\w)\{\widehat{\la}_{i}(\w)-\widehat{\la}_{j}(\w)\}=\Op_{P}(\frac{1}{\sqrt{nh}}).
\end{align*}By \eqref{eigenvalues relationship}, we have $\widehat{\be}_{i}\trans(\w)\be_{j}^{*}(\w)=\Op_{P}(\frac{1}{\sqrt{nh}})$.
\eop

\noindent {\sc Proof of Assertion} (a). By the law of iterated logarithm, similar to the proof of Assertion (a) of \cite{Luo2016}, we can show that, when $\la_{k}(\w)>\la_{k+1}(\w)$,
\begin{align*}
  1-|\text{det}\{\G_{11}(\w)\}|=\Op_{P}(\frac{1}{nh}) \text{ almost surely } \proba_{\calS},
\end{align*}where $\G_{11}(\w)=\widehat{\T}_{k}\trans\T_{k}^{*}$. Thus $f_{n}^{0}(\w,k)=\Op_{P}(\frac{1}{nh})$ almost surely $\proba_{\calS}$, as desired.
\eop

Proof of Assertion (b) is omitted here since it's similar in \cite{Luo2016}. We now prove Theorem \ref{consistency property} by combining lemma \ref{asymptotic behavior of eigenvector}, lemma \ref{asymptotic behavior of eigenvalue}, lemma \ref{the osrder between eigenvalue and the variability of eigenvectors}, lemma \ref{A direct application} and Assertion (a) and (b).

\noindent {\sc Proof of Theorem \ref{consistency property}.} It's easy to see that
\begin{align*}
  \Op_{P}(\frac{1}{nh})=\op_{P}(c_{n}),\quad\Op_{P}^{+}(1)=\Op_{P}^{+}(c_{n}),\quad\Op_{P}(\frac{1}{\sqrt{c_{n}nh}})=\op_{P}(c_{n}).
\end{align*}Let $r=p-1$ if $p\leq 10$ and $r=[p/\log(p)]$ otherwise. By assertion (a), assertion (b) and Lemma \ref{asymptotic behavior of eigenvector} and \ref{asymptotic behavior of eigenvalue}, for any $k\in\{0,1,\ldots,r\}$,
\begin{equation*}
  \left\{
  \begin{array}{lclc}
  {f_{n}(\w,k)\geq 0}, &\phi_{n}(\w,k)=\Op_{P}^{+}(c_{n}), &\text{if} &k<d(\w); \\
  {f_{n}(\w,k)=\op_{P}(c_{n})},&\phi_{n}(\w,k)=\op_{P}(c_{n}), &\text{if} &k=d(\w);\\
  {f_{n}(\w,k)=\Op_{P}^{+}(c_{n})},&\phi_{n}(\w,k)>0, &\text{if} &k>d(\w);
  \end{array}
        \right.
\end{equation*}almost surely $\proba_{\calS}$. Since $g_{n}(\w)=f_{n}(\w)+\phi_{n}(\w)$, lemma D (i) of \cite{Luo2016} implies that
\begin{equation*}
  g_{n}(\w,k)=\left\{
  \begin{array}{lcl}
  {\Op_{P}^{+}(c_{n})}, &\text{if} &k\neq d(\w); \\
  {\op_{P}(c_{n})}, &\text{if} &k=d(\w);\\
  \end{array}
        \right. \text{ almost surely } \proba_{\calS}.
\end{equation*}By Lemma D (ii) of \cite{Luo2016}, $g_{n}$ is minimized at $d(\w)$ in probability almost surely $\proba_{\calS}$.
\eop

\newcommand{\enquote}[1]{#1}
\expandafter\ifx\csname
\natexlab\endcsname\relax\def\natexlab#1{#1}\fi



\begin{thebibliography}{999}
	
	\bibitem[{Berndt (1991)}]{Berndt1991}[1]
	Berndt, E. R. (1991) \enquote{The practice of econometrics: classic and contemporary}.
	\textit{Addison Wesley Publishing Company}.
	
	\bibitem[{Bickel and Freedman (1981)}]{Bickel1981}[2]
	Bickel, P. J., and Freedman, D. A. (1981) \enquote{Some asymptotic theory for the bootstrap}.
	\textit{The Annals of Statistics},
	{\bf9} 1196-1217.
	

    \bibitem[{Bura and Cook(2001)}]{Bura2001}[3]
	Bura, E., and Cook, R. D. (2001) \enquote{Estimating the structural dimension of regressions via parametric inverse regression}.
	\textit{Journal of the Royal Statistical Society: Series B (Statistical Methodology)},
	{\bf63(2)} 393-410.
	
	\bibitem[{Chen et al.(2010)}]{Chen2010}[4]
	Chen, X., Zou, C., and Cook, R. D. (2010) \enquote{Coordinate-independent sparse sufficient dimension reduction and variable selection}.
	\textit{The Annals of Statistics},
	{\bf38(6)} 3696-3723.
	
	\bibitem[{Chiaromonte et al.(2002)}]{Chiaromonte2002}[5]
	Chiaromonte, F., Cook, R. D., and Li, B. (2002) \enquote{Sufficient dimension reduction in regressions with categorical predictors}.
	\textit{The Annals of Statistics},
	{\bf} 475-497.
	
	\bibitem[{Cook(1998)}]{Cook1998}[6]
	Cook, R. D. (1998) \enquote{Regression Graphics}.
	\textit{New York: Wiley},
	{\bf237}.
	
	\bibitem[{Cook and Forzani (2009)}]{Cook2009}[7]
	Cook, R. D., and Forzani, L. (2009) \enquote{Likelihood-based sufficient dimension reduction}.
	\textit{Journal of the American Statistical Association},
	{\bf104(485)} 197-208.
	
	\bibitem[{Cook and Lee (1999)}]{Cook1999}[8]
	Cook, R., and Hakbae Lee. (1999) \enquote{Dimension reduction in binary response regression}.
	\textit{Journal of the American Statistical Association},
	{\bf94(448)} 1187-1200.
	
	\bibitem[{Cook and Ni(2005)}]{Cook2005}[9]
	Cook, R. D., and Ni, L. (2005) \enquote{Sufficient dimension reduction via inverse regression: A minimum discrepancy approach}.
	\textit{Journal of the American Statistical Association},
	{\bf100(470)} 410-428.
	
	\bibitem[{Cook and Weisberg(1991)}]{Cook1991}[10]
	Cook, R. D., and Weisberg, S. (1991) \enquote{Comment}.
	\textit{Journal of the American Statistical Association},
	{\bf86(414)} 328-332.
	
	\bibitem[{Cook and Yin (2001)}]{Cook2001}[11]
	Cook, R. D., and Yin, X. (2001) \enquote{Theory and methods: Special invited paper: Dimension reduction and visualization in discriminant analysis (with discussion)}.
	\textit{Australian and New Zealand Journal of Statistics},
	{\bf43(2)} 147-199.
	
	
	\bibitem[{Dong and Li (2010)}]{Dong2010}[12]
	Dong and Li (2010) \enquote{Dimension reduction for non-elliptically distributed predictors: second-order methods}.
	\textit{Biometrika},
	{\bf97(2)} 279-294.
	
	\bibitem[{Fan (1993)}]{Fan1993}[13]
	Fan, J. (1993) \enquote{Local linear regression smoothers and their minimax efficiencies}.
	\textit{The Annals of Statistics},
	{\bf} 196-216.
	
	\bibitem[{Fan and Huang (2005)}]{Fan2005}[14]
	Fan, J., and Huang, T. (2005) \enquote{Profile likelihood inferences on semiparametric varying-coefficient partially linear models}.
	\textit{Bernoulli},
	{\bf11(6)} 1031-1057.
	
	\bibitem[{Feng et al. (2013)}]{Feng2013}[15]
	Feng, Z., Wen, X. M., Yu, Z., and Zhu, L. (2013) \enquote{On partial sufficient dimension reduction with applications to partially linear multi-index models}.
	\textit{Journal of the American Statistical Association},
	{\bf108(501)} 237-246.
	
	\bibitem[{Ferre (1998)}]{Ferre1998}[16]
	Ferr$\acute{e}$, L. (1998) \enquote{Determining the dimension in sliced inverse regression and related methods}.
	\textit{Journal of the American Statistical Association},
	{\bf93(441)} 132-140.
	
	\bibitem[{Hardle (1984)}]{Hardle1984}[17]
	Hardle, W. (1984) \enquote{A law of the iterated logarithm for nonparametric regression function estimators}.
	\textit{The Annals of Statistics},
	{\bf20} 624-635.
	
	\bibitem[{Hoeting et al. (1999)}]{Hoeting1999}[18]
	Hoeting, J. A., Madigan, D., Raftery, A. E., and Volinsky, C. T. (1999) \enquote{Bayesian model averaging: a tutorial}.
	\textit{Statistical Science},
	{\bf} 382-401.
	
	
	\bibitem[{Jiang et al. (2017)}]{Jiang2017}[19]
	Jiang, B., Chen, Z., and Leng, C. (2017) \enquote{Dynamic Linear Discriminant Analysis in High Dimensional Space}.
	
	\bibitem[{Leng (2010)}]{Leng2010}[20]
	Leng, C. (2010) \enquote{Variable selection and coefficient estimation via regularized rank regression}.
	\textit{Statistica Sinica},
	{\bf} 167-181.
	
	\bibitem[{Li (1991)}]{Li1991}[21]
	Li, K. C. (1991) \enquote{Sliced inverse regression for dimension reduction}.
	\textit{Journal of the American Statistical Association},
	{\bf86(414)} 316-327.
	
	\bibitem[{Li (2018)}]{Li2018}[22]
	Li, B. (2018) \enquote{Sufficient Dimension Reduction: Methods and Applications with R}.
	\textit{CRC Press}.
	
	\bibitem[{Li and Dong (2009)}]{Li2009}[23]
	Li, B., and Dong, Y. (2009) \enquote{Dimension reduction for nonelliptically distributed predictors}.
	\textit{The Annals of Statistics},
	{\bf37(3)} 1272-1298.
	
	\bibitem[{Li et al. (2003)}]{Li2003}[24]
	Li, B., Cook, R. D., and Chiaromonte, F. (2003) \enquote{Dimension reduction for the conditional mean in regressions with categorical predictors}.
	\textit{The Annals of Statistics},
	{\bf} 1636-1668.
	
	\bibitem[{Li et al. (2015)}]{Li2015}[25]
	Li, D., Ke, Y., and Zhang, W. (2015) \enquote{Model selection and structure specification in ultra-high dimensional generalised semi-varying coefficient models}.
	\textit{The Annals of Statistics},
	{\bf43(6)} 2676-2705.

	
	\bibitem[{Li and Wang (2007)}]{LiW2007}[26]
	Li, B., and Wang, S. (2007) \enquote{On directional regression for dimension reduction}.
	\textit{Journal of the American Statistical Association},
	{\bf102(479)} 997-1008.
	
	\bibitem[{Li and Zhu (2007)}]{Li2007}[27]
	Li, Y., and Zhu, L. X. (2007) \enquote{Asymptotics for sliced average variance estimation}.
	\textit{The Annals of Statistics},
	{\bf35(1)} 41-69.
	
	\bibitem[{Luo and Li (2016) }]{Luo2016}[28]
	Luo, W., and Li, B. (2016) \enquote{Combining eigenvalues and variation of eigenvectors for order determination}.
	\textit{Biometrika},
	{\bf103(4)} 875-887.
	
	\bibitem[{Ma and Zhu (2012)}]{Ma2012}[29]
	Ma, Y., and Zhu, L. (2012) \enquote{A semiparametric approach to dimension reduction}.
	\textit{Journal of the American Statistical Association},
	{\bf107(497)} 168-179.
	
	\bibitem[{Ma and Zhu (2013a)}]{Ma2013a}[30]
	Ma, Y., and Zhu, L. (2013a) \enquote{A review on dimension reduction}.
	\textit{International Statistical Review},
	{\bf81(1)} 134-150.
	
	\bibitem[{Ma and Zhu (2013b)}]{Ma2013b}[31]
	Ma, Y., and Zhu, L. (2013b) \enquote{Efficient estimation in sufficient dimension reduction}.
	\textit{The Annals of Statistics},
	{\bf41(1)} 250.
	
	\bibitem[{Ma and Zhu (2014)}]{Ma2014}[32]
	Ma, Y., and Zhu, L. (2014) \enquote{On estimation efficiency of the central mean subspace}.
	\textit{Journal of the Royal Statistical Society: Series B (Statistical Methodology)},
	{\bf76(5)} 885-901.

    \bibitem[{Nadaraya (1964)}]{Nadaraya1964}[33]
	Nadaraya (1964) \enquote{On estimating regression}.
	\textit{Theory of Probability $\&$ Its Applications},
	{\bf9(1)} 141-142.
	
	\bibitem[{Penrose et al. (1985)}]{Penrose1985}[34]
	Penrose, K. W., Nelson, A. G., and Fisher, A. G. (1985) \enquote{Generalized body composition prediction equation for men using simple measurement techniques}.
	\textit{Medicine $\&$ Science in Sports $\&$ Exercise},
	{\bf17(2)} 189.
	
	\bibitem[{Shao et al. (2009)}]{Shao2009}[35]
	Shao, Y., Cook, R. D., and Weisberg, S. (2009) \enquote{Partial central subspace and sliced average variance estimation}.
	\textit{Journal of Statistical Planning and Inference},
	{\bf139(3)} 952-961.
	
	\bibitem[{Szekely et al. (2007)}]{Szekely2007}[36]
	Sz$\acute{e}$kely, G. J., Rizzo, M. L., and Bakirov, N. K. (2007) \enquote{Measuring and testing dependence by correlation of distances}.
	\textit{The Annals of Statistics},
	{\bf35(6)} 2769-2794.
	
	\bibitem[{Wang and Xia (2008)}]{Wang2008}[37]
	Wang, H., and Xia, Y. (2008) \enquote{Sliced regression for dimension reduction}.
	\textit{Journal of the American Statistical Association},
	{\bf103(482)} 811-821.

    \bibitem[{Watson (1964)}]{Watson1964}[38]
	Watson, G. S. (1964) \enquote{Smooth regression analysis}.
	\textit{Sankhyi: The Indian Journal of Statistics, Series A},
	{\bf} 359-372.
	
	\bibitem[{Wen and Cook (2007)}]{Wen2007}[39]
	Wen, X., and Cook, R. D. (2007) \enquote{Optimal sufficient dimension reduction in regressions with categorical predictors}.
	\textit{Journal of Statistical Planning and Inference},
	{137(6)} 1961-1978.
	
	\bibitem[{Xia et al.(2002) Xia, Y., Tong, H., Li, W. K., $\&$ Zhu, L. X.}]{Xia:2002}[40]
	\textsc{Xia, Y., Tong, H., Li, W. K., $\&$ Zhu, L. X.}  (2002).
	\newblock An adaptive estimation of dimension reduction space.
	\newblock \textit{Journal of the Royal Statistical Society: Series B (Statistical Methodology)} \textbf{64(3)}, 363-410.
	
	\bibitem[{Xie and Huang (2009)}]{Xie2009}[41]
	Xie, H., and Huang, J. (2009) \enquote{SCAD-penalized regression in high-dimensional partially linear models}.
	\textit{The Annals of Statistics},
	{\bf37(2)} 673-696.
	
	\bibitem[{Xue, Wang and Yin (2018)}]{Xue2018}[42]
	Xue, Y., Wang, Q., and Yin, X. (2018) \enquote{A unified approach to sufficient dimension reduction}.
	\textit{Journal of Statistical Planning and Inference},
	{\bf197} 168-179.
	
	
	\bibitem[{Yao and Tong (1996)}]{Yao1996}[43]
	Yao, Q. and Tong, H. (1996) \enquote{Asymmetric least squares regression estimation: a nonparametric approach}.
	\textit{Journal of Nonparametric Statistics},
	{\bf6(2-3)} 273-292.
	
	\bibitem[{Yao and Tong(1998)}]{Yao1998}[44]
	Yao, Q., and Tong, H. (1998) \enquote{Cross-validatory bandwidth selections for regression estimation based on dependent data}.
	\textit{Journal of Statistical Planning and Inference},
	{\bf68(2)} 387-415.
	
	\bibitem[{Ye and Weiss (2003)}]{Ye2003}[45]
	Ye, Z., and Weiss, R. E. (2003) \enquote{Using the bootstrap to select one of a new class of dimension reduction methods}.
	\textit{Journal of the American Statistical Association},
	{\bf98(464)} 968-979.
	
	\bibitem[{Yin and Cook (2002)}]{Yin2002}[46]
	Yin, X., and Cook, R. D. (2002) \enquote{Dimension reduction for the conditional kth moment in regression}.
	\textit{Journal of the Royal Statistical Society: Series B (Statistical Methodology)},
	{\bf64(2)} 159-175.
	
	\bibitem[{Yin and Cook (2003)}]{Yin2003}[47]
	Yin, X., and Cook, R. D. (2003) \enquote{Estimating central subspaces via inverse third moments}.
	\textit{Biometrika},
	{\bf90(1)} 113-125.
	
	\bibitem[{Yin et al.(2010)}]{Yin2010}[48]
	Yin, J., Geng, Z., Li, R., and Wang, H. (2010) \enquote{Nonparametric covariance model}.
	\textit{Statistica Sinica},
	{\bf20} 469.

    \bibitem[{Yu (2014)}]{Yu2014}[49]
	Yu, Z., Dong, Y. and Huang, M. (2014) \enquote{General directional regression}.
	\textit{Journal of Multivariate Analysis},
	{\bf124} 94-104.

    \bibitem[{Yu and Dong (2016)}]{Yu2016}[50]
	Yu, Z., and Dong, Y. (2016) \enquote{Model-free coordinate test and variable selection via directional regressionl}.
	\textit{Statistica Sinica},
	{\bf} 1159-1174.
	
	\bibitem[{Zhang et al. (2013)}]{Zhang2013}[51]
	Zhang, R., Zhao, W., and Liu, J. (2013) \enquote{Robust estimation and variable selection for semiparametric partially linear varying coefficient model based on modal regression}.
	\textit{Journal of Nonparametric Statistics},
	{\bf25(2)} 523-544.
	
	\bibitem[{Zhao et al. (1986)}]{Zhao1986}[52]
	Zhao, L. C., Krishnaiah, P. R., and Bai, Z. D. (1986) \enquote{On detection of the number of signals in presence of white noise}.
	\textit{Academic Press},
	{\bf} Inc.
	
\end{thebibliography}
\end{document}